\documentclass[iop,revtex4]{emulateapj}

\usepackage[letterpaper,portrait,left=1.25in,right=0.75in,top=1.5in,bottom=0.5in]{geometry}                
\usepackage{graphicx}
\usepackage{amssymb}
\usepackage{amsmath}
\usepackage{epstopdf}

\usepackage{longtable}
\usepackage{threeparttable}
\usepackage{dcolumn}

\newcolumntype{d}[1]{D{.}{.}{#1} }

\usepackage[backref,breaklinks,colorlinks,citecolor=blue]{hyperref}
\usepackage[all]{hypcap}

\usepackage{natbib}
\usepackage{rotating} 

\defcitealias{2005ApJS..159..141V}{VF05}

\newcommand{\teff}{$T_{\mathrm{eff}}$}
\newcommand{\logg}{$\log g$}
\newcommand{\kepler}{\textit{Kepler}}

\newcommand{\vdw}{van der Waals}
\newcommand{\vald}{VALD-3}
\newcommand{\vbroad}{$v_{\rm broad}$}
\newcommand{\vsini}{$v \sin i$}
\newcommand{\vmac}{$v_{\rm mac}$}
\newcommand{\vmic}{$v_{\rm mic}$}

\begin{document}
\title{Spectral Properties of Cool Stars: Extended Abundance Analysis of 1,617 Planet-Search Stars}

%
\author{John M. Brewer, Debra A. Fischer}
	\affil{Department of Astronomy, Yale University}
	\affil{260 Whitney Avenue, New Haven, CT 06511, USA}
	\email{john.brewer@yale.edu}
	\email{debra.fischer@yale.edu}
	
\author{Jeff A. Valenti}
	\affil{Space Telescope Science Institute}
	\affil{3700 San Martin Drive, Baltimore, MD 21218, USA}
	\email{valenti@stsci.edu}

\and

\author{Nikolai Piskunov}
	\affil{Uppsala University}
	\affil{Department of Physics and Astronomy, Box 516, 75120 Uppsala, Sweden}
	\email{nikolai.piskunov@physics.uu.se}
%


\begin{abstract}
We present a catalog of uniformly determined stellar properties and abundances for 1,617 F, G, and K stars using an automated spectral synthesis modeling procedure.  All stars were observed using the HIRES spectrograph at Keck Observatory.  Our procedure used a single line list to fit model spectra to observations of all stars to determine effective temperature, surface gravity, metallicity, projected rotational velocity, and the abundances of 15 elements (C, N, O, Na, Mg, Al, Si, Ca, Ti, V, Cr, Mn, Fe, Ni, \& Y).  Sixty percent of the sample had \textit{Hipparcos} parallaxes and V-band photometry which we combined with the spectroscopic results to obtain mass, radius, and luminosity.  Additionally, we used the luminosity, effective temperature, metallicity and $\alpha$-element enhancement to interpolate in the Yonsei-Yale isochrones to derive mass, radius, gravity, and age ranges for those stars. Finally, we determined new relations between effective temperature and macroturbulence for dwarfs and subgiants. Our analysis achieved precisions of 25~K in \teff, 0.01~dex in [M/H], 0.028~dex for \logg\ and 0.5~km~s$^{-1}$ in \vsini based on multiple observations of the same stars.  The abundance results were similarly precise, between $\sim 0.01$ and $\sim 0.04$~dex, though trends with respect to \teff\ remained for which we derived empirical corrections.  The trends, though small, were much larger than our uncertainties and are shared with published abundances.  We show that changing our model atmosphere grid accounts for most of the trend in [M/H] between 5000~K and 5500~K indicating a possible problem with the atmosphere models or opacities.
\end{abstract}


\maketitle

%
\section{Introduction}
Abundance analyses of stars through the observation of stellar spectra began with the careful hand measurement of individual line intensities \citep{Payne:1925ei}.  Applying theories of thermodynamics and spectral line formation to the interpretation of these intensities allowed a physical interpretation that required modeling the properties of the stellar atmosphere.  Though detailed abundances would have to wait for better understanding of stellar structure and detailed model atmospheres, the cataloguing of relative abundances in stars had begun. Exacting and time consuming analysis has been a hallmark of the field ever since.

Continuous improvements in understanding and a growing amount of data allowed exploration of a wide variety of astrophysical problems such as galactic chemical evolution, stellar structure, and stellar atmospheres.  The combined advances of well modeled atmospheres, high resolution spectrographs, sensitive CCDs, and computer modeling were needed to finally allow precise abundance analysis of more than a few elements of a handful of stars almost 70 years later \citep{1993A&A...275..101E}.  However, different data sets often had large differences in the derived \teff, \logg\ and metallicity \citep{2001A&A...373..159C} limiting their usefulness in detailed comparisons between them and precluded most comparisons of abundances altogether.

The search for extrasolar planets dedicated a lot of time on several high-quality spectrographs to the study of nearby stars. This resulted in large homogeneous catalogs of spectroscopic properties and abundances all homogeneously observed and analyzed \citep{2005ApJS..159..141V,2012A&A...545A..32A}.  Uniform analyses such as these minimize the relative errors in abundances, furthering our knowledge of planet formation \citep{2004A&A...415.1153S,2005ApJ...622.1102F,2010ApJ...715.1050B} and galactic structure and evolution \citep{Bensby:2014gi} and advancing progress in stellar astrophysics.  Spectroscopic analysis is also used to calibrate the scales for photometric metallicity surveys, though differences between analyses still remain.

As we enter the third decade of exoplanets, we now know that planet formation is ubiquitous \citep{2015ApJ...809....8B}, that hot jupiters are uncommon \citep{2015ApJ...799..229W}, that super-earths are common around other stars \citep{2012ApJS..201...15H,2011arXiv1109.2497M,Howard:2010es}, and that unexpected compact systems of planets close to their host stars are common \citep{Schlaufman:2014gg,2015arXiv151109157B}.  All of these things have led to new and interesting questions.  How do  planetesimals grow from dust, what exactly produces hot jupiters and why do they usually have no close companions, why does the solar system have no super-earths, and what conditions lead to compact multi-planet systems?  The answers to these complex questions will require investigations into many intertwined influences, but a crucial ingredient will be accurate knowledge of the stellar properties as a proxy for the specific compositions of the disks in which the planets formed.

Due to the difficult and time consuming nature of both observation and analysis of high resolution, high S/N spectra there have been relatively few large catalogs. Over the past two decades, a wealth of exoplanets have been discovered and we are now grappling with questions of formation, migration, and composition.  Missions such as Gaia \citep{Perryman:2001cp}, combined with multiple large scale spectroscopic surveys, are beginning to explore the composition and formation history of our galaxy. These low to medium resolution spectroscopic surveys are deriving stellar parameters for tens of thousands to millions of stars with smaller numbers of higher resolution observations for calibration \citep{DeSilva:2015gr,Holtzman:2015ct,Smiljanic:2014ej}. There is a vital need for a homogeneous catalog of accurate stellar parameters and abundances to help in these endeavors.  In addition, such a catalog can provide empirical clues to help resolve problems in our current models of stellar atmospheres.

In this paper we present a uniform spectroscopic analysis for 1,617 stars obtained by the California Planet Survey (CPS) \citep{2010ApJ...721.1467H} using the HIRES spectrograph at Keck Observatory \citep{1994SPIE.2198..362V}.  The stars in the catalog are primarily dwarfs and subgiants spanning the temperature range between 4700~K and 6800~K. We analyzed them with forward modeling using a single line list. The spectroscopically derived parameters include effective temperature, surface gravity, projected rotational velocity, radial velocity, activity, and abundances for 15 elements.  For each element we describe the lines fitted and their sensitivity to abundance changes, and how our analysis compares to other large catalogs.  In the interest of providing a complete resource for the stars in this catalog, we have also included masses, radii, and ages derived using the Yonsei-Yale isochrones with our spectroscopic parameters as a constraint.

%
%
\section{Stellar Sample}

\subsection{Sample Description}
\begin{figure}[ht] 
   \centering
   \includegraphics[width=0.95\columnwidth]{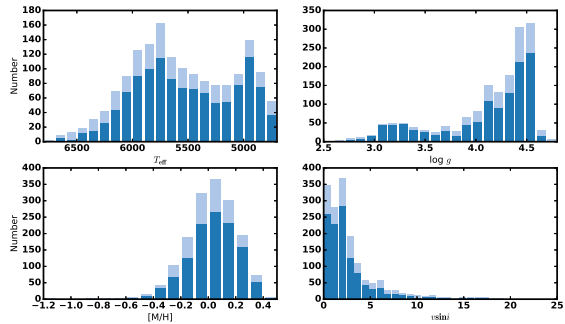} 
   \caption{The distribution of global stellar parameters for all stars in the catalog.  The dark blue bars are those with S/N $\geq 100$ and the light blue bars have S/N $< 100$.  The high-S/N sub-sample was used to examine any trends in the abundance analysis.  The double peaked distribution of temperatures is due to the selection criteria of the various planet-search programs that the spectra are drawn from.}
   \label{fig:global_param_hist}
\end{figure}

The spectra in this study were all collected using the HIRES spectrograph on the Keck I telescope as part of one or more radial velocity planet-search programs under the collaborative umbrella known as the California Planet Survey (CPS).  Each program chooses their stars by a variety of different criteria with an emphasis on bright single F, G, K, and M dwarfs with low projected rotational velocities to ensure sharp lines for precision radial velocities.  There are also a large number of subgiants, a handful of giant stars, members of young clusters, and a few pre-main sequence stars. Table \ref{table:sample_summary} summarizes the min, max, and mean characteristics of the sample and Figure \ref{fig:global_param_hist} shows a histogram of the global properties.

All of the template spectra in this study were taken using HIRES in the red configuration at a resolution of $R \approx 70,000$ with the iodine cell out. The analyzed spectra were all obtained after the HIRES detector upgrade in August 2004 because these spectra have an extended wavelength range that improves the elemental abundance analysis. Typically, template spectra are observed at S/N~$> 200$, limited by their brightness. Recently many faint stars from the \kepler\ mission were observed at lower S/N.  We found that our precision decreased significantly at signal-to-noise ratios below 100, so we divided our sample into S/N~$\geq 100$ (1191 stars) and S/N~$< 100$ (426 stars).

\newcolumntype{d}[1]{D{.}{.}{#1} }

\begin{table}
  \caption{Sample Characteristics}
\begin{center}
\begin{threeparttable}
  \centering
  \begin{tabular}{ l d{1} d{1} d{1} }
\hline 
\hline \\[-1.5ex]
Parameter & \mathrm{Min} & \mathrm{Median} & \mathrm{Max} \\
\\[-1.5ex]
\hline
$T_{\rm{eff}}$ &   4702 &   5600 &   6773 \\
$\log g$ &   2.63 &   4.29 &   4.99 \\
$[M/H]$ &  -1.14 &   0.04 &   0.51 \\
$v\sin i$ &   0.00 &   2.00 &  20.08 \\
  \hline
  \end{tabular} \label{table:sample_summary}
  \end{threeparttable}
  \end{center}
\end{table}

\subsection{Asteroids} \label{sec:asteroids}
In addition to stellar spectra, our data also include 20 spectra of four different asteroids (4 Vesta, 1036 Ganymed, 3 Juno, and 10 Hygiea) from five epochs throughout the 10-year period covered by the observations in our sample.  These spectra provided disk-integrated solar spectra and were obtained to help calibrate our analysis by providing small zero-point offsets for solar parameters and abundances.

%
\section{Spectral Line Parameters} \label{sec:lline_params}
The shapes of spectral lines in LTE in a 1-dimensional stellar spectrum are determined by a combination of the quantum-mechanical parameters for a given atom or molecule in a particular energy state, the global stellar parameters such as surface gravity and effective temperature, and the relative abundances of the element in question and HI (as the most common collisional partner).  Complicating matters in most cases are nearby blends with one or more additional lines.  Precise atomic line parameters are needed to fit observed spectra reasonably well.  However, a good fit does not guarantee accurate stellar parameters.

We use the procedure of \citet{2015ApJ...805..126B} to model 350~\AA\ of the spectrum in specific wavelength segments between 5164~\AA\  and 7800~\AA\ (Table \ref{table:wavelength_regions}).  We expanded the number of spectral lines analyzed by \citet[][hereafter VF05]{2005ApJS..159..141V} to provide stronger constraints on temperature and surface gravity and to increase the number of elements. \citet{2015ApJ...805..126B} demonstrated that the added information provided a robust constraint on \logg, returning gravities consistent with asteroseismology with an RMS scatter of 0.05~dex. We provide all of our line parameters in Table \ref{table:line_parameters}.  The first column designates the element or molecule name and the second the ion in the case of atoms.  The next three columns are the wavelength, excitation potential, and $\log gf$ values for the line.  The $\Gamma_6$ column is the van der Waals broadening coefficient when the value is $< 0$ or the cross section and velocity parameter $\alpha$ from Anstee, Barklem, and O'Mara (ABO) theory when positive \citep{2000A&AS..142..467B}.  Lines in the table that were included in our models but excluded from $\chi^2$ calculations while fitting are marked as ``Masked Out." 

\begin{table}
\caption{Spectral Segments}
\begin{center}
\begin{threeparttable}
\renewcommand{\TPTminimum}{\linewidth}
\centering
    \begin{tabular}{ l l c }
    \hline 
    \hline \\[-1.5ex]
    $\lambda_{start}$ & $\lambda_{end}$ & $\mathrm{VF05}$ \\
    \\[-1.5ex]
    \hline
  5164 &   5190 & x \\
  5190 &   5207 &   \\
  5232 &   5262 &   \\
  6000 &   6015 & x \\
  6015 &   6030 & x \\
  6030 &   6050 & x \\
  6050 &   6070 & x \\
  6100 &   6120 & x \\
  6121 &   6140 & x \\
  6143 &   6160 & x \\
  6160 &   6180 & x \\
  6295 &   6305 &   \\
  6311 &   6320 &   \\
  6579 &   6599 &   \\
  6688 &   6702 &   \\
  6703 &   6711 &   \\
  6711 &   6718 &   \\
  7440 &   7470 &   \\
  7697 &   7702 &   \\
  7769 &   7799 &   \\
    \hline
    \end{tabular} \label{table:wavelength_regions}
\end{threeparttable}
    \begin{tablenotes}
     \item {Note: Segments with a check in the VF05 column cover the same wavelengths as segments from VF05.}
    \end{tablenotes}
\end{center}
\end{table}

\subsection{Initial Line Parameters}
The Vienna Atomic Line Database 3 (\vald) contains both atomic and molecular line data for millions of lines across our spectral regions.  We used the `Extract Stellar' function of the \vald\ website to extract line parameters for a solar type and cooler star represented by the parameters in Table \ref{table:vald_params}.  The stellar parameters bracket the majority of our sample and we set the detection threshold to 0.001.  The total line list contains more than 7,500 lines; almost half of these are molecular lines.

\vald\ is a compilation of the work of many researchers, with parameters for some lines coming from multiple sources.  The largest contributors to information in our line list ($\sim 85$\% of the parameters) were from \citet{KCN,K07,K10,BP09ow,KMGH,K13,K09,RU,KC2,K08,PQWB,BPM,GSGCD,MC,BA-J}.


\begin{table}
\caption{VALD3 Extract Stellar Parameters}
    \begin{center}
    \begin{threeparttable}
    \renewcommand{\TPTminimum}{\linewidth}
    \begin{tabular}{l l l}
        \hline
        \textbf{Parameter}        & \textbf{Solar} & \textbf{Cool Dwarf} \\
        \hline
        \hline
        Detection Threshold  & 0.001  & 0.001 \\
        Microturbulence  & 0.9 km/s & 0.9 km/s \\
        T$_{eff}$  & 5750 K  & 4750 K \\
        log g  & 4.5  & 4.5 \\
        \hline
    \end{tabular} \label{table:vald_params}
    \end{threeparttable}
    \end{center}
\end{table}

\subsection{Empirical Corrections} \label{sec:empirical_corrections}
The line data from \vald\ is compiled from a variety of sources.  We calibrated the atomic line data to the NSO solar flux atlas \citep{2011ApJS..195....6W} to better fit the solar spectrum, adopting an atmospheric model for the Sun (Table~\ref{table:solar_params}) adopting the abundances of \citet{Grevesse:2007cx}. We made empirical corrections to $\log gf$ values, \vdw\ coefficients, and line positions for spectral lines that fit poorly.  Adjustments were required to one or more of these parameters for 20\% of our lines.  However, we visually selected the lines which needed tuning which preferentially selected deeper lines where mismatches between model and atlas were clear.

\begin{table}
    \caption{Solar Parameters}
    \begin{center}
    \begin{threeparttable}
    \renewcommand{\TPTminimum}{\linewidth}
    \centering
    \begin{tabular}{ l c }
    \hline 
    \hline \\[-1.5ex]
Parameter & \multicolumn{1}{r}{Adopted\ Solar\ Value} \\
 \\[-1.5ex]
 \hline
 \teff\  & 5777\ K \\
 \logg\  & 4.44 \\
 $\mathrm{[M/H]}$  & 0.00 \\
 $v \sin i$ & 1.63~km~s$^{-1}$\\
 $v_{\mathrm{mac}}$ & 3.5~km~s$^{-1}$\\
 $v_{\mathrm{mic}}$ & 0.85~km~s$^{-1}$\\
 C &    8.39 \\
N &    7.78 \\
O &    8.66 \\
Na &    6.17 \\
Mg &    7.53 \\
Al &    6.37 \\
Si &    7.51 \\
Ca &    6.31 \\
Ti &    4.90 \\
V &    4.00 \\
Cr &    5.64 \\
Mn &    5.39 \\
Fe &    7.45 \\
Ni &    6.23 \\
Y &    2.21 \\
    \hline
 \end{tabular} \label{table:solar_params}
\end{threeparttable}
 \begin{tablenotes}
 	\item {Note: Parameters used to calibrate atomic line data against the NSO solar atlas.  These are also the zeropoints for the asteroid spectra.}
 \end{tablenotes}
\end{center}
\end{table}

We developed a semiautomated procedure to fit the line parameters using Spectroscopy Made Easy (SME) \citep{1996A&AS..118..595V}.  Within the graphical interface of SME we selected lines that we judged by eye to have a poor fit to the atlas spectrum and marked their $\log gf$ values as free parameters.  We divided the spectral segment into small regions around the marked lines, grouping those lines within 1~\AA\ of each other.  For each group we solved first for $\log gf$ to ensure that the line depths were approximately correct, then adjusted the line positions followed by another iteration to solve for $\log gf$.  Finally, we fit for \vdw\ damping parameters. The fitting was carried out simultaneously for all of the lines in a group.

\begin{figure*} 
   \centering
   \includegraphics[width=\textwidth]{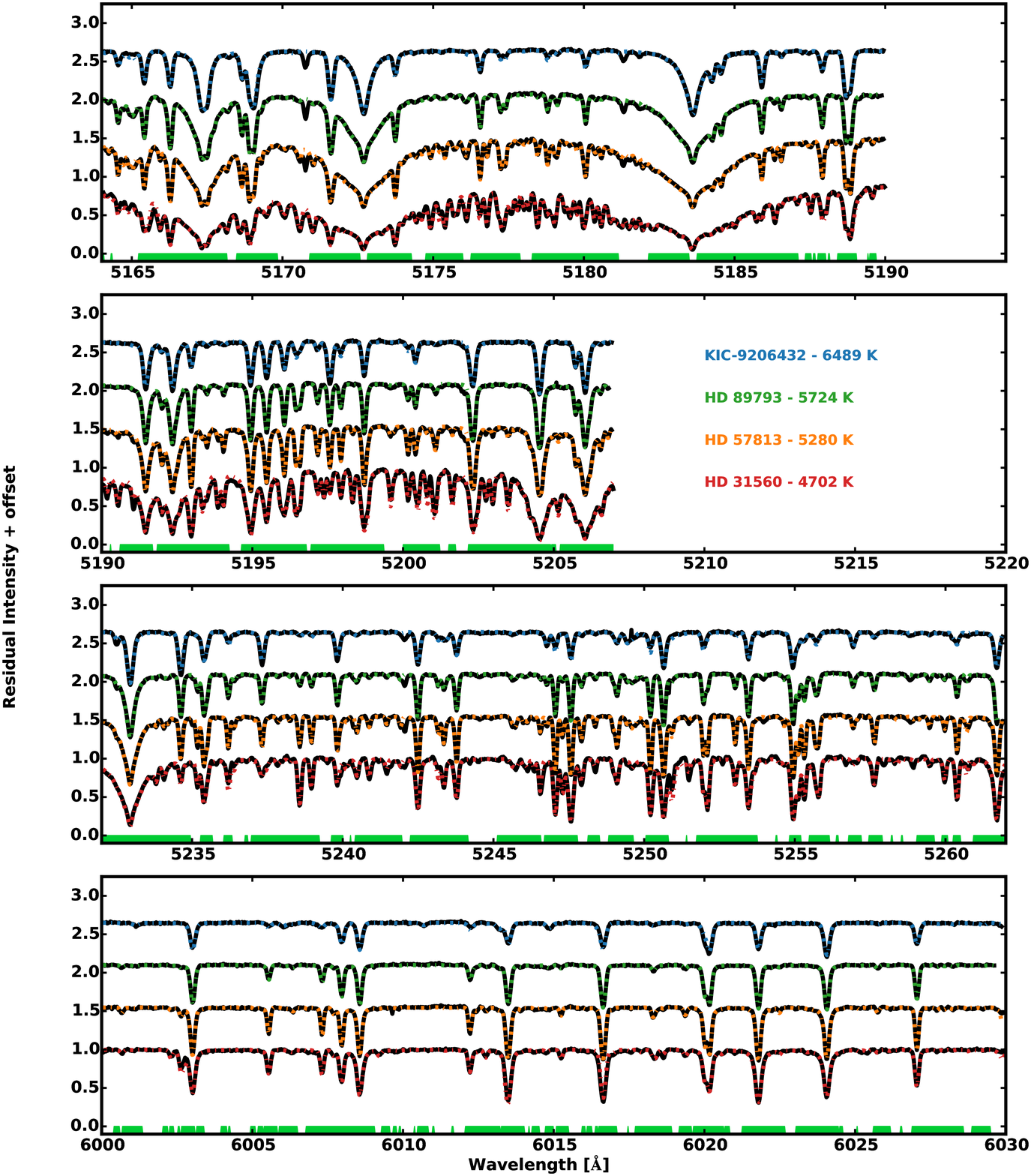} 
   \caption{Representative spectra and our fits across the range of our sample for the first quarter of our wavelength regions.  For each star, the solid black line is the observation, and the dash-dotted colored line is the model.  At the bottom of each subplot, the green solid region indicates pixels included in our line and continuum masks.  Additional regions are removed from this mask based on the positions of telluric lines.  All spectral regions are plotted at the same wavelength scale to facilitate comparisons of spectral features, though this will leave some subpanels relatively empty.}

   \label{fig:spec_regions_1}
\end{figure*}

\begin{figure*} 
   \centering
   \includegraphics[width=\textwidth]{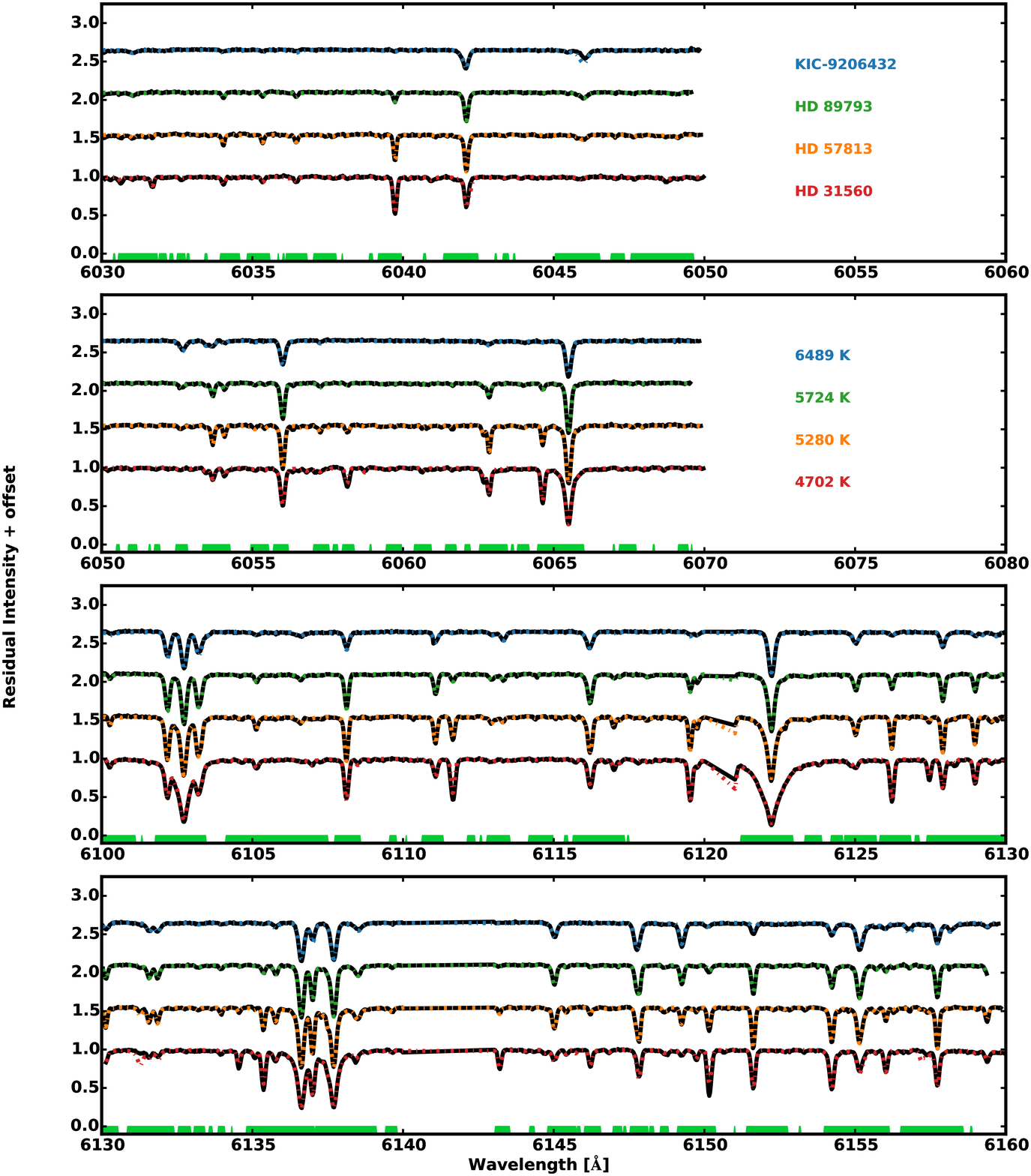} 
   \caption{Same as Figure \ref{fig:spec_regions_1} but for the second quarter of our wavelength regions.}

   \label{fig:spec_regions_2}
\end{figure*}

\begin{figure*} 
   \centering
   \includegraphics[width=\textwidth]{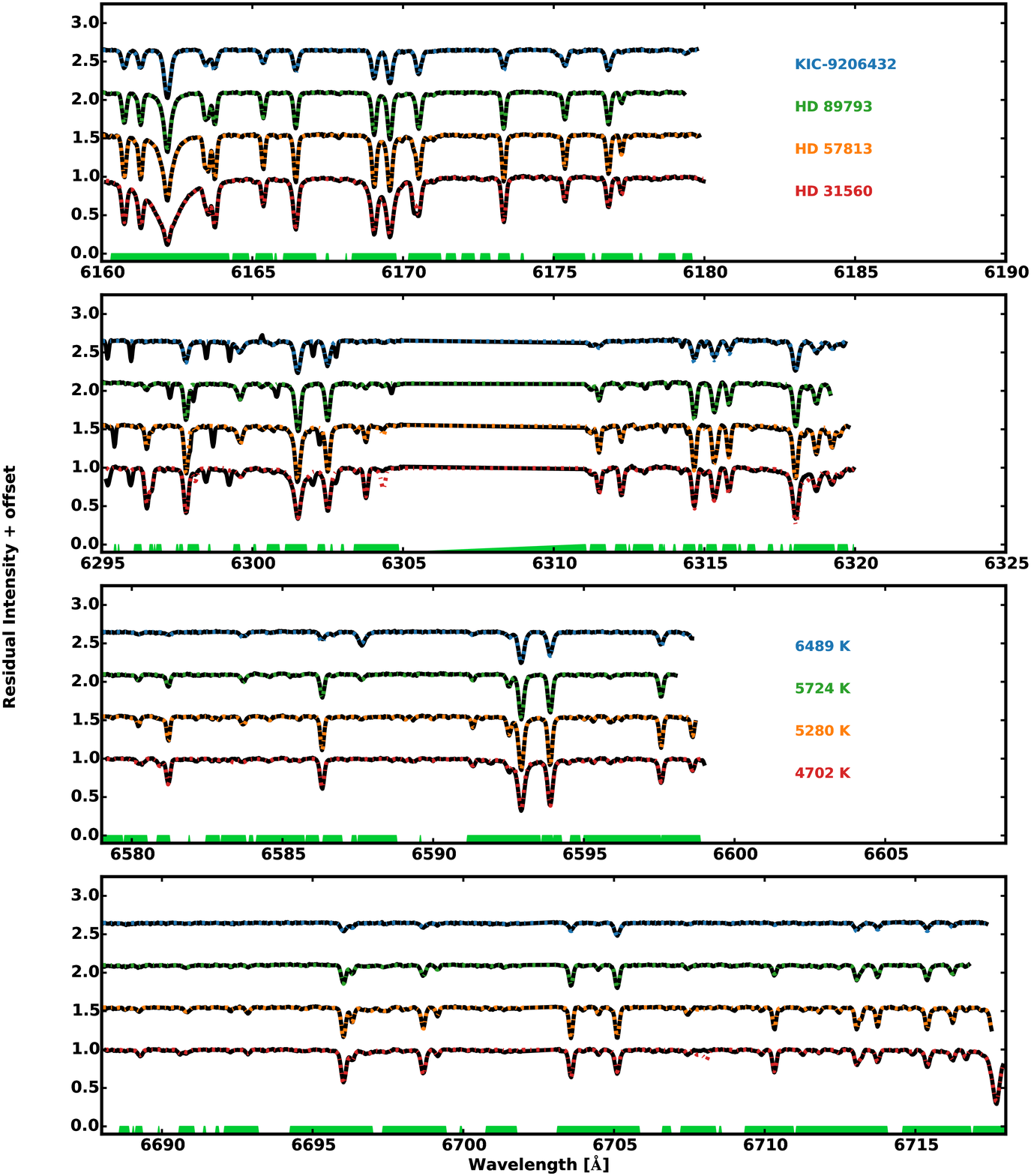} 
   \caption{Same as Figures \ref{fig:spec_regions_1} and \ref{fig:spec_regions_2} but for the third quarter of our wavelength regions.}

   \label{fig:spec_regions_3}
\end{figure*}

\begin{figure*} 
   \centering
   \includegraphics[width=\textwidth]{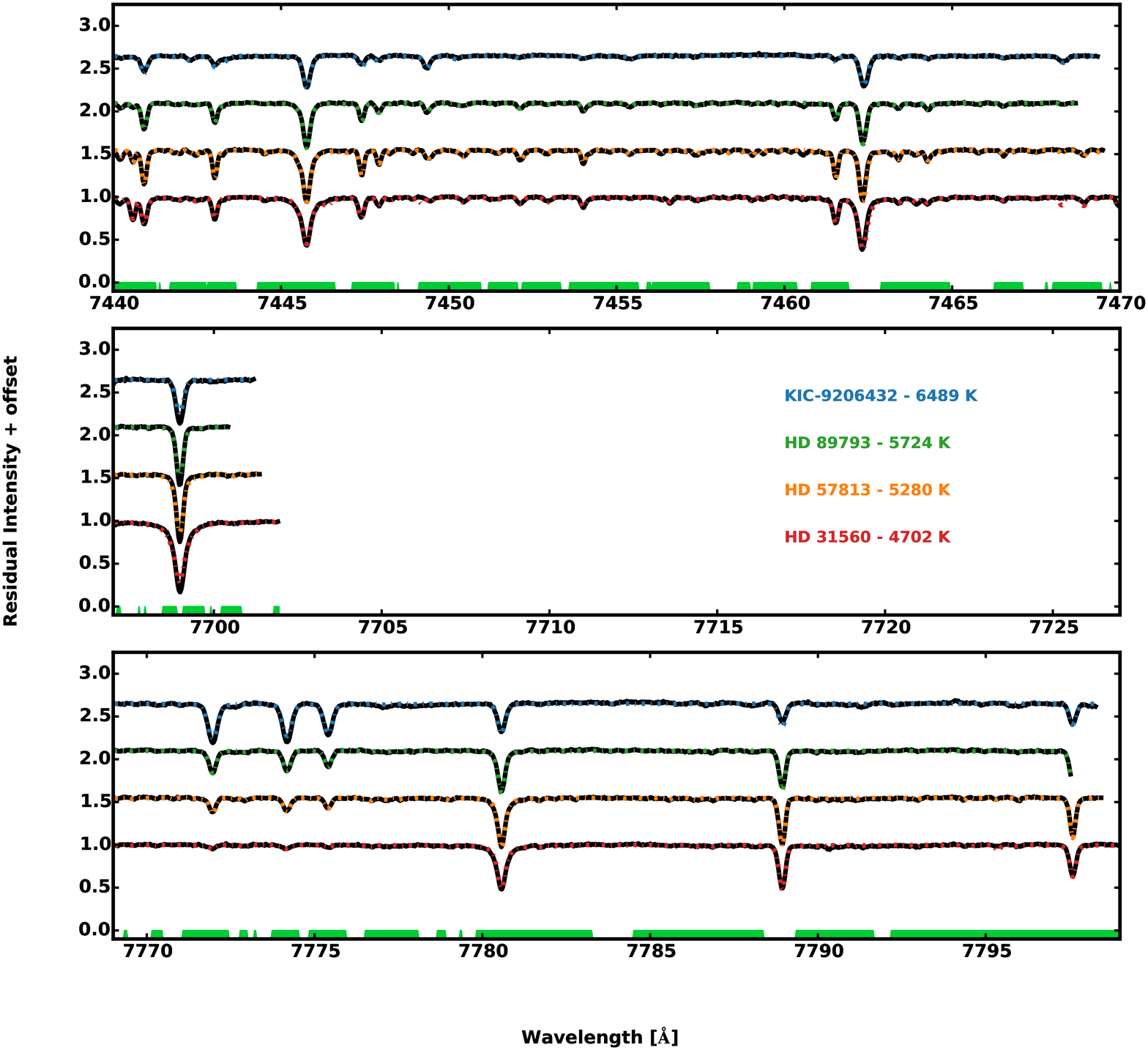} 
   \caption{Same as Figures \ref{fig:spec_regions_1}, \ref{fig:spec_regions_2}, and \ref{fig:spec_regions_3} but for the final quarter of our wavelength regions.}

   \label{fig:spec_regions_4}
\end{figure*}

In principle, the updated parameters should apply across the relatively narrow range of stellar types in our sample, however, in practice some adjustments to atomic parameters compensate for model errors that will manifest as effective temperature deviates from solar.  Lines which are shallow in the solar case but prominent in stars significantly warmer or cooler than the sun might still be poorly tuned.  Some regions of the spectrum were masked out because the line information was missing from \vald\ or because severe line blending gave degenerate results.  The NSO solar flux atlas \citep{2011ApJS..195....6W} we calibrated against had been corrected for telluric contamination; however, lines in our observed spectra that fell directly on telluric features were excluded from our line mask. The mask also marks wavelength regions that are relatively free of spectral lines across a range of temperatures as continuum regions.  Representative spectra can be seen in Figures \ref{fig:spec_regions_1}, \ref{fig:spec_regions_2}, \ref{fig:spec_regions_3} and \ref{fig:spec_regions_4} and the combined line and continuum masks are also shown.  Telluric lines are also dynamically masked out before fitting each spectrum and are not indicated in the figures.

%
\section{Spectral Fitting Procedure} \label{sec:spectral_fitting}
Our analysis consists of fitting observed spectra with one-dimensional (1D)  LTE models in plane-parallel atmospheres from the grid of \citet{Castelli:2004ti}.  We followed the iterative fitting method of \citet{2015ApJ...805..126B} that yielded accurate surface gravities.  First we fit for global free parameters (\teff, \logg, [M/H], \vmac) while assuming a solar abundance pattern except for the alpha elements Ca, Si, and Ti.  We then fix the first set of parameters and fit for the individual elemental abundances.  The next step is iterative, running the global fit again with the updated derived abundance pattern, and then fixing the revised global parameters for a final adjustment of the individual elemental abundances.  We use the same line list and mask as \citet{2015ApJ...805..126B} and are publishing it here for the first time.  The full procedure takes approximately 14 hours of CPU time per spectrum on a modern Core i7 processor.

\subsection{Preparing the Spectra} \label{sec:prepare_spectra}
Nightly thorium-argon calibration spectra were used to obtain a wavelength solution. The spectra were then cross-correlated with the NSO solar atlas \citep{2011ApJS..195....6W} and the barycentric velocity of the Earth at the time of the observation was subtracted to derive a radial velocity to the star.  Before cross-correlating the spectra we use the telluric line list of the atlas to mask out wavelength regions with significant telluric contamination to lessen its impact on the fitting procedure. To create the telluric mask, we select all pixels that are less than 99\% of the continuum in the atlas, then rescale the mask to the resolution of our observation.  The signal to noise for each pixel in the spectrum was stored and the spectra were continuum normalized. We estimated an initial guess for the effective temperature using a B-V color-temperature relation and calculated the absolute visual magnitude to estimate an initial guess for surface gravity.  When these data were missing, the default initial guesses were \teff$=5500$~K and \logg$=4.5$.

\subsection{Instrument Profile Resolution}
The resolution needs to be well known in order to estimate the instrumental line broadening because this will cross-talk with other line broadening parameters (\vsini, \vmac). The resolution can be measured from thorium-argon calibration spectra, however the lamp completely fills the slit.  This can under-estimate the resolution on nights of good seeing where the FWHM of the stellar image is smaller than the slit.  When modeling the asteroid spectra, we adopted the resolution derived from fitting the thorium-argon spectrum.  This yielded macroturbulence and rotational velocities that were smaller than our adopted  solar values (Table \ref{table:solar_params}), implying that our resolution was underestimated (i.e. instrumental line broadening too large) and needed to be revised.

The asteroids in our sample are good proxies for the stars we are observing since their angular sizes are typically smaller than the seeing limit and their apparent magnitudes cover the range of our stellar sample.  Using one asteroid spectrum observed with the B1 decker slit and another from the B5 decker slit we used SME to solve for the resolution with all other parameters fixed to solar values (Table \ref{table:solar_params}).  We used these fits to adjust the resolution for the B1 (70300 to 78409) and  B5 (57000 to 63574) slits.  This yielded a \vsini\  $\simeq 1.6$ km~s$^{-1}$ and \vmac\ $\simeq 3.5$ km~s$^{-1}$ for the SME-modeled asteroid spectra, matching the values used in tuning the solar atlas.  We modeled spectra of the same star obtained with both the B1 and B5 slits using these resolutions.  If the resolution for one or the other slit was incorrect, this would have resulted in a mismatch in the derived broadening parameter.  However, the same rotational broadening was found for both slits, demonstrating that the resolution of the two slits is likely correct for observations of stars on nights of good seeing.

\subsection{Global Stellar Parameters} \label{sec:global_stellar_params}
In the first fitting step, we solve for effective temperature (\teff), surface gravity (\logg), combined rotational (\vsini) and macroturbulent (\vmac) broadening (\vbroad, see \S \ref{sec:rotation_vmac}), and metallicity ([M/H]).  In addition, each spectral segment is allowed to make an independent radial-velocity shift of the entire segment, (v$_{rad}$) to account for inaccuracies in our initial radial velocity determination. This analysis implicitly assumes a solar abundance pattern, which is not accurate. For example, metal poor stars have an $\alpha$/Fe ratio that is larger than the Sun.

Most spectral lines that were analyzed arise from atomic transitions of iron, silicon, titanium and nickel.  The iron and nickel should scale together \citep{Scott:2015hl}.  However, the $\alpha$-elements do not track the iron peak elements.  To improve the fit for non-solar abundance patterns, we also allow the $\alpha$-elements calcium, silicon and titanium to be free parameters.  This allows a more accurate fit to [M/H] which is used in selecting the atmosphere for generating our model spectra.

\subsection{Element Sensitivity} \label{sec:elem_sensitivity}
After solving for the global stellar parameters, we fix them and solve for the elemental abundances of the 15 elements.  These elements have a sufficient number of lines or influence on the spectrum to return the solar abundances with high precision for all of our asteroid spectra (\S \ref{sec:asteroids}).  The abundance precision of individual lines can be estimated from calculating where on the curve of growth the line is from the specific atmospheric parameters.  When using $\chi^2$ to fit a spectrum using a model that spans hundreds of angstroms and thousands of lines, the amount that changing abundances influence the overall model becomes a factor. To decide whether to solve for an element, we developed a metric to quantify the model sensitivity to changes in abundance for each element with lines in our line list.

The metric quantifies the impact the element has over $\chi^2$, which can be influenced by molecule formation in cooler stars (e.g. CN, and OH) as well as chemical equilibrium between elements such as C and O.  Changes in temperature and gravity affect ionization balance and the location of particular lines in curve of growth.  The result is that for a given line list at a given \teff\ and \logg\ there will be a varying sensitivity in a model to changes in abundance.

The values for the metric that we derive here apply only to our model.  Analyses which use different atmospheres, wavelength selections, or line lists will have different values for the metric.  This procedure serves as a model that other studies can follow to quantify their sensitivity to the elemental abundances they derive.

We generated synthetic spectra with solar parameters and abundances for effective temperatures ranging from 4700~K to 6800~K.  At each 100~K step in temperature we varied the abundances of each element ($\epsilon$) by $\pm 0.5$ dex.  We determined the  difference between pairs of models for each element (i.e. model with [O/H]=0.5 - solar model) and the standard deviation of the differences in the line regions served as a sensitivity metric, $S_{\epsilon}$ (Equation \ref{eqn:s_epsilon}).  There were small differences in $S_{\epsilon}$ when increasing as opposed to decreasing the abundance with respect to solar.  However, the relative positions of the elements were generally the same and we use the sensitivity metric from the abundance increases in the rest of our discussion.

\begin{equation}  \label{eqn:s_epsilon}
	S_{\epsilon} = \sigma((Model_1 - Model_2))
\end{equation}

We can also assess the ability to derive abundances for individual elements by modeling the solar spectrum.  We analyzed our asteroid spectra (\S \ref{sec:asteroids}) with up to 29 free abundances for elements we were interested in or which serendipitously had many lines in our wavelength regions.  We eliminated elements more than 1 $\sigma$ away from the solar value and chose a maximum dispersion in the asteroid abundances of $\pm 0.03$ dex which placed nitrogen just inside this limit.  We then used the sensitivity of nitrogen ($S_{N}$) as our cutoff and excluded all elements which had $S_{\epsilon} \leq S_{N}$ at solar \teff\.  The lines from excluded elements are still included in our model, but their abundances are simply solar abundances scaled by overall metallicity.  The remaining 15 elemental abundances  (C, N, O, Na, Mg, Al, Si, Ca, Ti, V, Cr, Mn, Fe, Ni, and Y) are free parameters in our model (\S \ref{sec:elements}).

Some elements had individual lines that varied strongly with abundance changes, but as measured by $S_{\epsilon}$ did not have sufficient sensitivity for precise abundance determinations with our model.  We chose several of these elements (S, K, Co, Zr, and Gd) and tested our ability to recover them using the asteroid spectra.  When these elements were included as additional free parameters in fitting the observed solar spectra, the dispersion in the fitted elemental abundances where high. For stars similar to the sun then, our cutoff at $S_{\epsilon} \leq S_{N}$ seems well placed.

%
\section{Elemental Abundances} \label{sec:elements}
\begin{figure*} 
   \centering
   \includegraphics[width=\textwidth]{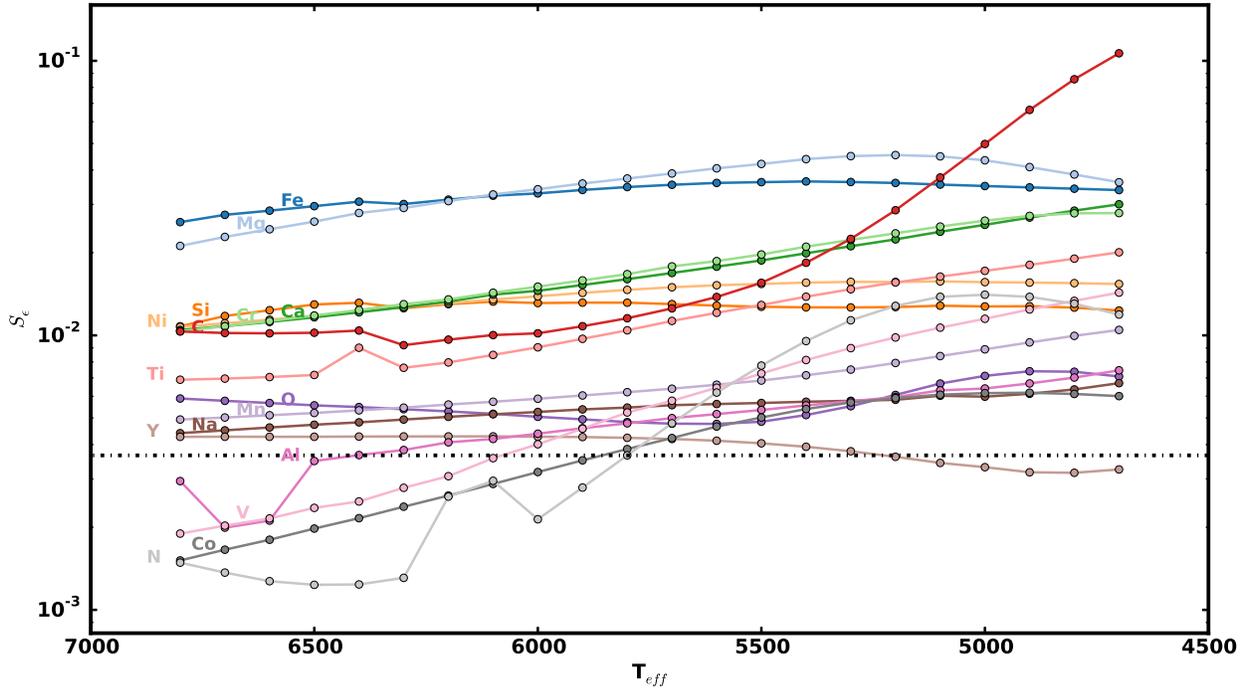} 
   \caption{Relative element sensitivities ($S_{\epsilon}$) for increased abundance for all elements with sensitivities less than that of nitrogen ($S_{\epsilon} \leq S_{N}$) at 5800~K (grey dashed line).  Although cobalt made the cutoff, we found a high dispersion in the abundances returned for the asteroid spectra and decided to exclude it from the analysis for the larger sample.}

   \label{fig:elem_sensitivities}
\end{figure*}
We aren't able to see planet formation in progress, or even the planets themselves in most cases.  Instead, we must derive everything we want to know about a planetary system and its formation from careful study of its host star.  Planets likely form from the same nebular material which gave birth to the star.  Since the composition of the stellar atmosphere undergoes small changes over its main sequence lifetime, the study of its atmospheric composition can tell us about the conditions under which its planets formed.

\begin{figure*} 
   \centering
   \includegraphics[width=\textwidth]{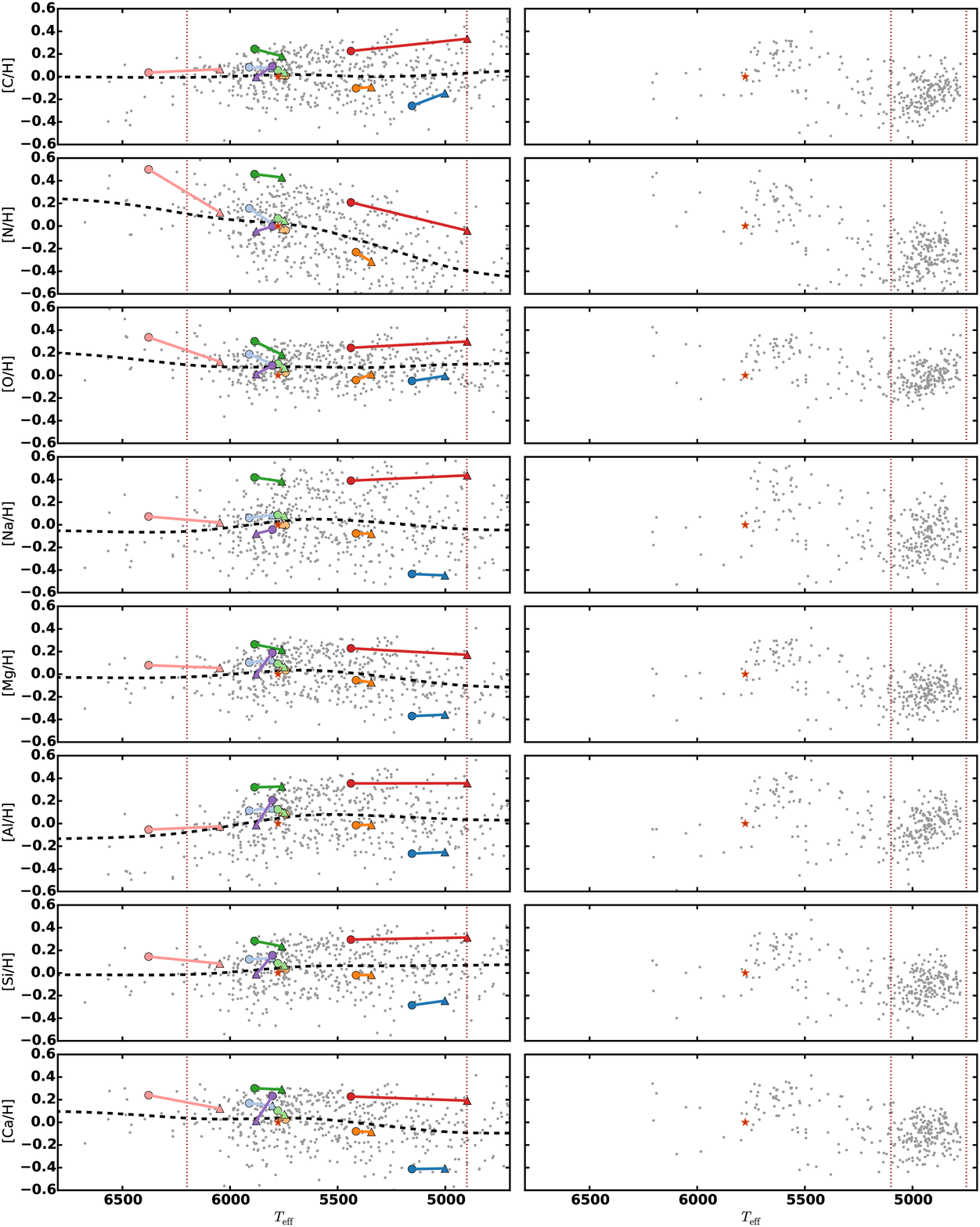} 
   \caption{The left column shows uncorrected Solar relative abundances ([$\epsilon$/H]) for the subset of dwarf stars in our sample with S/N $\geq 100$ and \vsini \ $\leq 3$ km~s$^{-1}$.  The right column shows stars with $3 < \log g < 4$ and the same S/N and \vsini \ cuts.  The black dashed line is a gaussian kernel regression fit to the points and their uncertainties between the vertical dotted lines. The colored points show binary pairs in our sample and their slopes roughly follow the trend line.  Between 6300~K and 5300~K, most elements show no significant trends in abundance with respect to \teff. A notable exceptions is nitrogen which shows a very strong correlations over the entire temperature range.  There were too few points to adequately fit the lower gravity stars, but they generally show the same trends as the dwarfs.}

   \label{fig:raw_light_abunds}
\end{figure*}

\begin{figure*} 
   \centering
   \includegraphics[width=\textwidth]{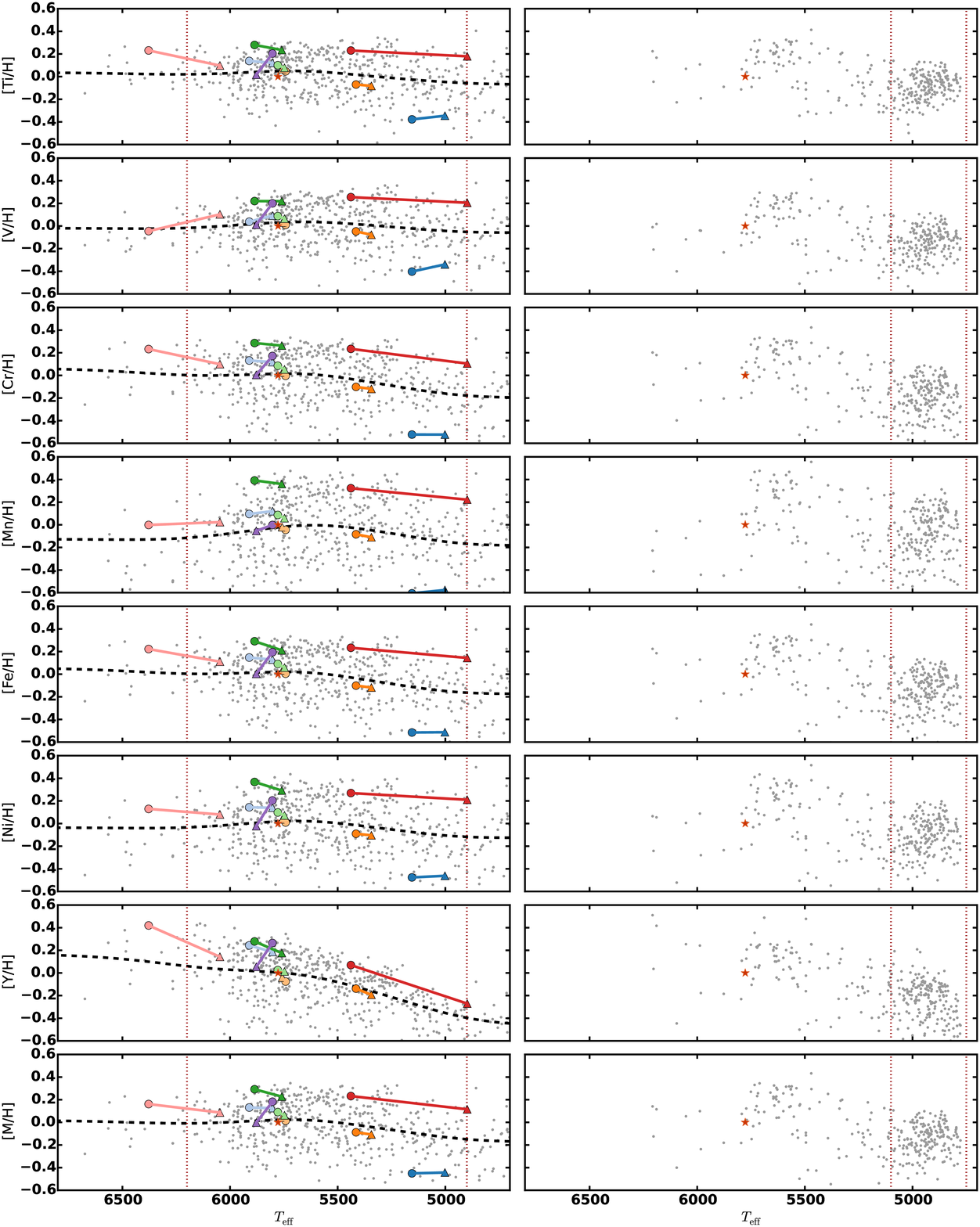} 
   \caption{Same as Figure \ref{fig:raw_light_abunds} but for heavier elements and overall metallicity. The left column shows uncorrected Solar relative abundances ([$\epsilon$/H]) for stars with $\log g \geq 4$ and the right column, stars with $3 \leq \log g < 4$, both with S/N $\geq 100$ and $v \sin i \leq 3.0$ km~s$^{-1}$. The black dashed line is a gaussian kernel regression fit to the points and their uncertainties between the vertical dotted lines. The colored points are binary pairs in our sample and their average slopes roughly follow the trend line.  Between 6300~K and 5300~K, most elements show no significant trends in abundance with respect to \teff, though yttrium shows very strong correlations over the entire temperature range.}

   \label{fig:raw_heavy_abunds}
\end{figure*}

Our models were tuned to solar values and we accurately recover the solar parameters (Table \ref{table:asteroid_summary}).  Although our sensitivity metric quantifies the relative precision in abundances as a function of temperature, it does not tell us apriori whether our results  are accurate across the HR diagram.   Our assumption of LTE, the choice of atmosphere models, and inaccuracies in the atomic line data can all lead to errors, especially for stars that differ from the Sun.  Below, we discuss trends that we observe as a function of \teff in our uncorrected solar relative abundances (Figures \ref{fig:raw_light_abunds} \& \ref{fig:raw_heavy_abunds}) for each of the elements in our spectral synthesis modeling.

%


Except for yttrium and nitrogen, the trends in the derived elemental abundances with respect to temperature are all quite small.  Nevertheless, their presence means that there are probably still systematics unaccounted for in our analysis and we discuss the possible sources when they are apparent.  In general, the scatter in abundances increases with decreasing \teff, and between 5500~K and 5000~K several elements show a marked decrease in the mean.

To find the general shape of these temperature dependent trends, we used gaussian kernel regression. We first found optimal bandwidths using a grid based cross-validation search for each element though the resultant fits sometimes showed more structure than we felt was justified by our understanding of where the trends originated.  Instead, we chose the maximum bandwidth found using this method (150~K) and used this for all elements (Figures \ref{fig:raw_light_abunds} \& \ref{fig:raw_heavy_abunds}).  We have 9 pairs of binaries over this temperature range\footnote{The coolest member of our binary pair sample fell just outside of our goodness of fit cutoff for the catalog.  We included it in the plots to help highlight any trends.} whose abundance trends mostly followed the regression line quite closely for all elements, showing that the trends are likely due to limitations in our analysis.  We discuss our procedure to correct these trends in \S \ref{sec:abundance_trends}.

One feature that is nearly constant is an increase in the dispersion of abundances at lower \teff, which can be most easily seen in the trend-corrected abundances (Figures \ref{fig:final_abundances_light} \& \ref{fig:final_abundances_heavy}).  This is probably due to the increased line crowding in the cooler stars in addition to the systematic problems we appear to have in determining overall metallicity at lower temperatures (\S \ref{sec:atmosphere_grids}).

In the following subsections, we discuss in sequence each of the 15 elements for which we derive abundances.  We provide some astrophysical context, discuss spectral line constraints in unmasked regions of our line list, characterize trends versus temperature in our \textit{uncorrected} abundances (Figures \ref{fig:raw_light_abunds} and \ref{fig:raw_heavy_abunds}), and compare our abundances to selected results from the literature (Figure \ref{fig:raw_elem_comparisons}).  For the trend analysis we use a high S/N subset of stars with S/N $\geq 100$ and \vsini~$ \leq 3.0$. In a later section (\S \ref{sec:final_lit_comparisons}), we compare our final trend-corrected abundances to literature values.

%

\subsection{Carbon} \label{elem:carbon}
Carbon is one of the required elements for life on Earth and has an important role in exoplanet research.   Its abundance relative to oxygen in a protoplanetary disk can affect the interior compositions of the planets in the system \citep{Moriarty:2014cb,2010ApJ...715.1050B} as well as their atmospheric compositions \citep{Konopacky:2013jv}.  Additionally, any differences between abundances in the stellar and planetary atmospheres constrains the formation and migration histories of the planets. Unfortunately, it is also challenging to accurately determine abundance in the optical spectra of cool dwarfs \citep{2012ApJ...747L..27F}.

Our line list contains 63 atomic carbon lines; typically, these are shallow or blended.  However, for cooler stars molecular lines are very sensitive to changes in carbon abundance \ref{fig:elem_sensitivities}.  By 5200~K carbon is the element most sensitive to changes in our model.  Molecular species such as CO and CN can also have large effects on the lines of the species involved outside of the molecular lines themselves.  For instance, an increase in the carbon abundance will increase the probability of combination with the always-plentiful oxygen atoms.  Atomic oxygen lines will then be shallower than expected for a given abundance.  Our models do take this into account as a single abundance has to fit both molecular and atomic lines of both elements.  However, large uncertainties in molecular line strengths could weaken this constraint. Because the molecular lines are very shallow and blended in the Sun there is no way to tune their transition and broadening parameters.

Between 4800 and 6200~K our raw carbon abundances is relatively flat with only a small slope (Figure \ref{fig:raw_light_abunds}) indicating that our carbon abundances are well constrained.  There is a slight increase in the dispersion toward lower temperatures.  For 26 of our sample stars that overlap with \citet{2014A&A...568A..25N}, there is good agreement in both the LTE and NLTE carbon abundances.  These stars only cover the temperature range where our trend in carbon abundances with temperature is small.

\subsection{Nitrogen} \label{elem:nitrogen}
Along with carbon and oxygen, nitrogen is one of the most abundant elements and participates in the CNO cycle which is the dominant path of hydrogen fusion in massive stars.  Molecular nitrogen is also the primary component of the Earth's atmosphere and owing to its abundance, it may be a major constituent of terrestrial exoplanet atmospheres. Exoplanetary nitrogen abundances can not only be used to quantify the bulk properties of their atmospheres but can help discriminate between the biotic and abiotic accumulation of O$_2$ \citep{Schwieterman:2015wg}.

Nitrogen is at the limit or our ability to obtain precise abundances.  For sun-like stars at solar metallicities, nitrogen shows a larger scatter in abundance than all elements other than yttrium and twice the scatter of elements with high $S_{\epsilon}$ values in our analysis.  There is a clear decrease in determined abundance with decreasing temperature, especially below 5500~K.  The $S_{N}$ curve shows that the model is more sensitive to nitrogen abundance at those temperatures, which may mean that the atomic and molecular lines abundance scales are offset, leading to the trend as one dominates over the other.  The CN molecule is prevalent in our spectral regions and although it seems to contribute substantially to our spectral models at lower effective temperatures, it does not help in obtaining trend free nitrogen abundances.  One possible reason for this could be that a large fraction of our molecular lines have poorly determined parameters.

As mentioned above, there is both a large scatter and a trend of $\sim 0.06 dex/100$~K.  This does not preclude the use in comparative analysis in this sample and nitrogen abundances provide a novel look into planet composition and formation.  However, the large scatter and trend make the absolute values less certain and care should be taken in using them as such.

 \subsection{Oxygen} \label{elem:oxygen}
The correlation between giant planet formation and metallicity was one of the earliest results in exoplanet research and confirmation of its primordial origin \citep{2005ApJ...622.1102F,2004A&A...415.1153S} pointed us toward core accretion as the dominant method of planet formation.  Under the assumption of core accretion, elements such as silicon and oxygen should play a much more important role than overall metallicity in generating dust grains to serve as nucleation sites for the rapid accumulation of ices \citep{Robinson:2006cr}.  Confirmation of oxygen's importance though has been hampered by the difficulty in accurately measuring oxygen abundance \citep{2011ApJ...738...97B,2015MNRAS.454L..11A}.  Beyond its role in initial planet formation, free oxygen in terrestrial planet atmospheres will be a primary biomarker in future studies.  Oxygen is a very reactive element, easily combining with most elements, locking it into water and solid minerals and out of the atmosphere unless there is some source of replenishment.  For the first couple billion years of Earth's history, there was very little atmospheric oxygen and only the onset of biological processes pushed it close to current levels \citep{Lyons:2014ii}.

Determining accurate stellar oxygen abundances in the optical has traditionally been hampered by blends and NLTE effects on the most prominent lines\citep{2015MNRAS.454L..11A,2014A&A...568A..25N,Ecuvillon:2005ca}.  Our line list does not include the 6300~\AA\ forbidden line due to the uncertainty introduced by the nickel blend.  It does contain the triplet at 7771~\AA\ which is known to have large NLTE effects and tends to overestimate oxygen abundance compared to 3D NLTE determinations \citep{2015MNRAS.454L..11A}.  We also include many molecular lines, including OH lines which yield abundances similar to the 6300~\AA\ line \citep{Ecuvillon:2005ca}.

The [O/H] abundance for our high S/N sample shows a slight trend toward lower abundances for lower \teff\ but a strong trend toward higher abundances for stars above 6000~K.  The 56 overlapping stars from \citet{Bensby:2014gi} show generally good agreement, and the few common stars hotter than solar show higher abundances in our analysis.  There are few hot stars in common with the \citet{2014A&A...568A..25N} sample.  They determined oxygen abundance using both the forbidden line at 6300~\AA\ and the triplet at 7771~\AA, the latter in both LTE and non-LTE.  This allowed us to explore where our model limitations may lie.  Our abundances match well with those they derive from the triplet in LTE.  However, at temperatures above solar, our abundances are higher than either their 6300~\AA\ abundances or those they derive from the triplet including NLTE effects. Though there are only 5 stars in common, this may indicate that our LTE assumption affects our oxygen abundances in hotter stars.

\subsection{Sodium} \label{elem:sodium}
Planetesimal formation mechanisms and timescales are still poorly understood, though recent progress has been made on both \citep{2015ApJ...809...94M,2012ApJ...745...79L,2012Sci...338..651C}.  One line of evidence which seems to support the direct and rapid formation of planetesimals from tiny grains comes from the surprising abundance of the volatile element sodium in chondrules which formed at high temperatures in the early solar system.  For sodium to have remained in grains at the low pressures in protoplanetary disks, the dust to gas ratio had to be extremely high.  This in turn leads to densities high enough to collapse directly into planetesimals \citep{Alexander:2008ih}.  Though an important clue towards understanding planet formation, it is based on the analysis of rocks instead of abundances in stars or protoplanetary disks.  Sodium is also useful for estimating interstellar reddening, for studying stellar evolution \citep{Campbell:2013bd} and it has been measured in exoplanet atmospheres \citep{2002ApJ...568..377C}.

There are very few optical sodium lines and we only have 3 included in our line list (at 6069~\AA, 6154~\AA, and 6161~\AA). The $S_{Na}$ curve is smooth, indicating a consistent model sensitivity to sodium across our temperature range.

The [Na/H] abundance for the high S/N dwarf sample is fairly flat between 5000~K and 6500~K, though the average does seem to be higher between 5500~K and 6000~K than for hotter or cooler stars.  Both the scatter and slope in the derived [Na/Fe] vs. \teff\ abundance ratio are small.  Overall, our sodium abundances compare well to the published literature for stars, with some increasing differences at higher temperatures.  We find modest and opposite abundance slopes as a function of temperature relative to \citetalias{2005ApJS..159..141V} and \citet{2012A&A...545A..32A} when examined separately.

\subsection{Magnesium} \label{elem:magnesium}
For stars cooler than $\approx$ 6200~K, the pressure-broadened wings of the Mg I b triplet at 5164~\AA\ provides a sensitive gravity indicator.  However, it is crucial to have accurate magnesium abundances for a robust determination of surface gravity \citep{2015ApJ...805..126B}.  In addition to being important for our global analysis, Mg also  plays an important role in the composition and geology of terrestrial planets.  The Mg/Si ratio controls the specific mineralogy of silicates with distinct behavior for ratios less than 1, between 1 and 2 and greater than 2 \citep{2010ApJ...715.1050B}.  At the temperatures and pressures assumed for the mid-plane of the protoplanetary disk, forsterite (MgSiO$_4$) and enstatite (MgSiO$_3$) condense out just after iron grains and can be the dominant grain-nucleation site for ices and lead to the rapid growth of planetesimals.

In addition to the Mg I b triplet at 5164~\AA, we have added an additional triplet at 6319~\AA, a couple of weaker atomic lines and many MgH molecular lines.  The spread in Mg abundances for the asteroid sample was nearly as small as for Fe abundances, which had 10 times as many atomic lines. There is very little trend in abundances of [Mg/H] across most of the temperature range of our sample.

Despite our best efforts, we see a decrease in average abundance for stars cooler than $\approx 5500$~K.  This will be a common feature for most of the remaining elements.  The Mg trend could reflect a trend in overall metallicity tracked by [Fe/H] (\S \ref{elem:iron}) and indeed, [Mg/Fe] has a much shallower trend across the entire temperature range.  Our [Mg/H] abundances are $\sim 0.06$  dex higher than large samples in the literature \citep{Bensby:2014gi,2012A&A...545A..32A} and have better precision and similar trends versus temperature as those studies.  As we correctly recover solar abundances and this offset applies there as well, our abundances are likely more accurate. 

\subsection{Aluminum} \label{elem:aluminum}
Aluminum is one of the most abundant elements in the Earth's crust after silicon and oxygen \citep{1995JGR...100.9761C} and calcium-aluminum-rich inclusions (CAIs) with very high condensation temperatures are the first particles to condense out of the protoplanetary disk according to the inventory of meteorites.  Additionally, the Earth may have formed inside the ice line and therefore been largely dry, requiring water rich planetesimals from further out to deliver its water. These planitesimals in turn were partially dried through radiogenic heating, primarily by the short-lived isotope $^{26}$Al  \citep{2011M&PS...46..903M,2006M&PS...41...95H}.  The ratio of $^{26}$Al/$^{27}$Al, which is about $10^{-5}$ in the solar system \citep{2011M&PS...46..903M,2006M&PS...41...95H}, can influence the habitability of earth-like planets. Though this research will not be able to resolve the relative ratios of $^{26}$Al/$^{27}$Al or $^{26}$Al/$^{26}$Mg, aluminum is still an important element in the study of planet formation and composition.

Our line list contains 9 Al lines (6 neutral, 3 singly ionized), most of them very weak and blended in the Sun with the ionized lines too weak to measure across our entire temperature range. However, the lines at 6696.0~\AA\ and 6698.7~\AA\ are moderately deep and relatively unblended.  Figure \ref{fig:elem_sensitivities} shows good sensitivity to aluminum abundance changes in cool stars, but sensitivity decreases as temperature increases, nearing our cutoff sensitivity for stars hotter than the Sun.  Between 6100~K and 5000~K the [Al/H] abundance is relatively flat for our selection of spectra with S/N $\geq 100$. Below 5000~K there aren't enough of these high S/N spectra to say much about trends but it continues to look flat.  For stars above 6100~K our inferred abundance decreases with increasing \teff.  It is unlikely that all of the warmer stars in our sample have low aluminum abundances so our models are likely deficient, for example due to neglect of NLTE effects. \citet{2012AstBu..67..294M} found a steep increase in the NLTE abundance differences  in the same sense as our trend for aluminum in stars $> 6000$~K.

\begin{figure*} 
   \centering
   \includegraphics[width=\textwidth]{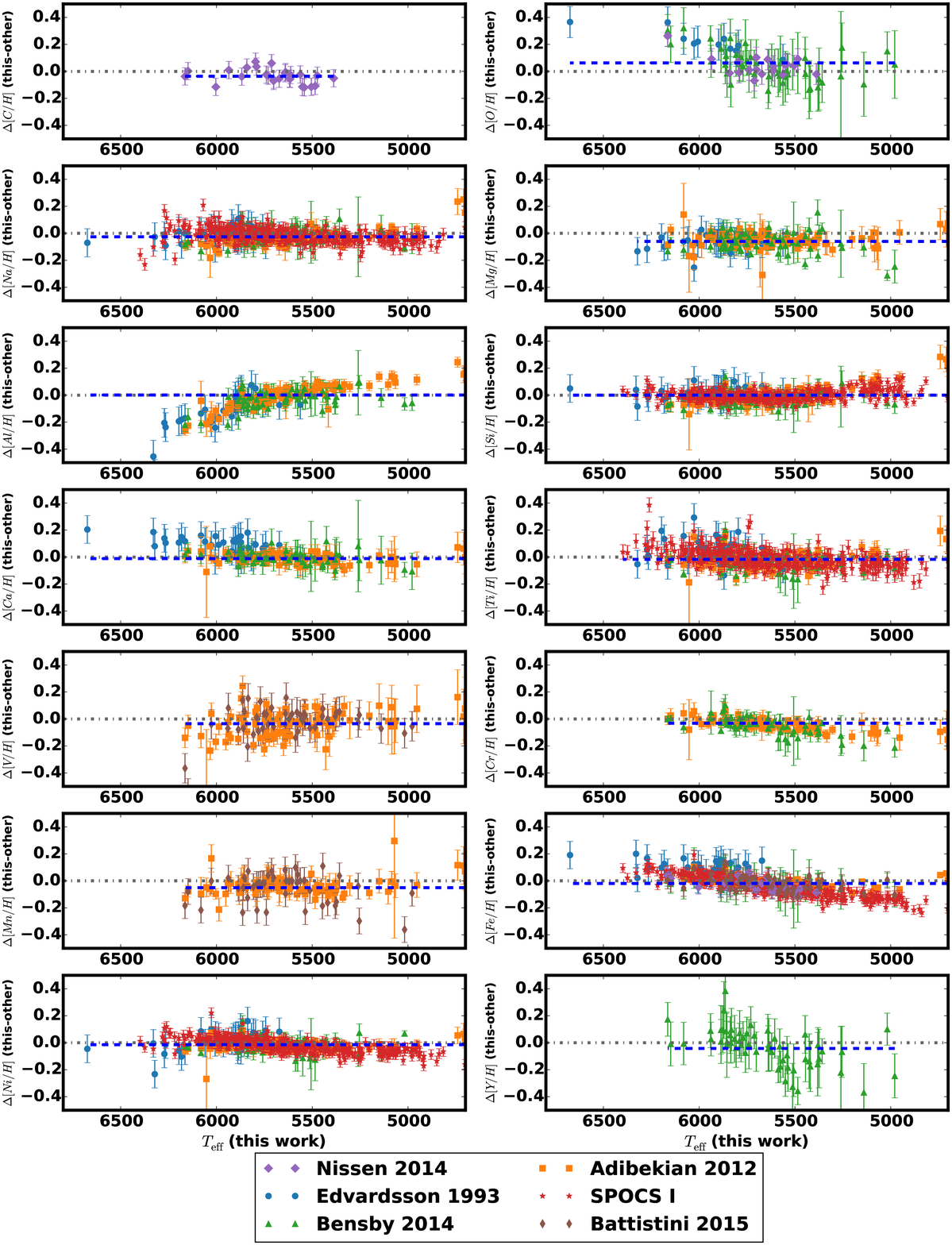} 
   \caption{Comparisons between our uncorrected solar relative abundances and those from literature values.  The grey dash-dot line at zero is for reference, and the blue dashed line shows a fit to the mean weighted offset.  Higher abundances at temperatures above solar are generally related to our improved gravity determinations.  While our uncorrected abundances show a temperature dependent trend toward under-abundance at lower temperatures, comparisons here with literature values generally show good agreement.  This indicates that other studies are also systematically underestimating abundances for lower temperature stars, probably due to errors in the stellar atmosphere models.}

   \label{fig:raw_elem_comparisons}
\end{figure*}

Figure \ref{fig:raw_elem_comparisons} shows that below 5700~K our aluminum abundances have no trends in the mean and are consistent with \citet{Bensby:2014gi} but not \citet{2012A&A...545A..32A}. Above 5700~K, our aluminum abundances are increasingly underestimated at higher temperatures.  Caution should be used in combining our abundances with those from other studies outside of the 5000~K to 6000~K range, but relative [Al/H] abundances in our study can be useful across the entire temperature range after our corrections.

\subsection{Silicon} \label{elem:silicon}
The rapid formation of giant planet cores before the dispersal of the protoplanetary gas disk requires the formation of ices which make up the bulk of the mass.  The ices in turn depend on dust grains to serve as reaction sites for molecular formation such as H$_2$, H$_2$O, and CO.  The density of dust grains, predominantly silicates and graphite, in the cool mid-plane of the disk is then a limiting factor in the formation of ices and giant planets \citep{2011ApJ...726...29F}.  Silicates are also the most abundant minerals in the Earth's crust and probably the dominant component of terrestrial planet crusts in most systems (those with C/O ratios $< 1$) \citep{2015ApJ...804...40G,2012ApJ...747L..27F,2010ApJ...725.2349D}.  The precise Mg/Si ratio in such planets can lead to range of crust compositions as the formation rates of MgSiO$_3$, MgSi$_2$O$_4$, MgO and MgS varies \citep{CarterBond:2012dma}.  Although the exact chemistry of exoplanet crusts depends on formation times and location in the protoplanetary disk \citep{Moriarty:2014cb,CarterBond:2012dma,2010ApJ...715.1050B}, inferring the primordial Mg/Si ratio from the star will give us a good start on determining habitability.

Silicon lines are relatively plentiful in the visible. We have a total of 84 lines in our mask, though all but 3 of them are neutral.  The model sensitivity as measured by $S_{Si}$ is relatively constant across our temperature range, decreasing only slightly as we go to cooler temperatures.   The dispersion in measured [Si/H] in the asteroid sample was relatively low, but the offset from solar was larger than most at 0.043 dex.

In the high S/N dwarf sample, our uncorrected [Si/H] is fairly flat across our temperature range, decreasing slightly for stars warmer than $\sim6200$~K.  The dispersion is also roughly constant, consistent with the sensitivity metric.  Comparing with other catalogs, there is virtually zero offset and a scatter of only 0.040 dex implying very small mutual uncertainties of $\sim 0.028$~dex.  The dispersion in differences does seem to increase at temperatures $\lesssim 5200$~K, though it is still small.  Our silicon abundances should be useful for both relative abundance comparisons and in absolute comparisons with values from other spectroscopic analysis.

\subsection{Calcium} \label{elem:calcium}
In our own solar system, calcium-aluminum-rich inclusions (CAIs) in chondritic meteors are an important tracer of early planet formation.  Due to their high condensation temperature, these grains can condense out almost anywhere in the protoplanetary disk.  Thus calcium is an important element in defining a complete range of chemistry for exoplanet interiors, especially the most interesting terrestrial planets.

There are more than 200 calcium lines in our line list though, similar to silicon, all but 4 of them are neutral. The ionized lines contribute no information at the temperatures and gravities in our sample. Though there are calcium lines throughout our spectral regions, the majority of them lie in the reddest segments.  Like silicon, calcium is a pure alpha element and we expect it to behave similarly. Our model sensitivity as measured through $S_{Ca}$ increases at lower temperatures.  The asteroid sample shows a small offset and low dispersion.

Despite the increasing model sensitivity at low temperatures, the average [Ca/H] of the high S/N dwarf sample decreases slightly at cooler temperatures and the dispersion seems to increase as well.  This mirrors the magnesium (\ref{elem:magnesium}) abundances and could signal that increasing line blends and un-modeled molecular lines at these low temperatures are adding noise and confusing the fit.  The average [Ca/H] for stars warmer than $\sim 5500$~K is constant.  Abundances from \citet{1993A&A...275..101E} showed an offset of $\gtrsim 0.15$~dex from other surveys but the differences showed no trends relative to our analysis. The remaining matching stars showed virtually no offset (-0.017~dex) and no trends. The lack of trends in comparison to published abundances where we see a decrease in our mean [Ca/H] (below 5500~K) may be the result of common problems in our analysis.

%

\subsection{Titanium} \label{elem:titanium}
Titanium is an alpha element, but its precise nucleosynthesis channel is still debated and galactic chemical evolution models fail to reproduce the proper titanium abundance or its observed trend with iron \citep{Hughes:2008hd}.  In cooler stars, titanium oxide becomes a growing source of continuum opacity as well as a confounding factor in model comparisons due to the large number of unknown lines.  The TiO lines may also be present in the atmospheres of hot jupiters \citep{Desert:2008ei} and have become a popular molecule for exoplanet atmosphere studies.

The ratio of ionized to neutral titanium lines is low, but there are still 28 \ion{Ti}{2} lines out of the 254 Ti lines in our mask.  In addition, there are $> 450$ TiO lines.  The influence of titanium in our models grows with decreasing temperature (Figure \ref{fig:elem_sensitivities}), which is likely due to the growing influence of TiO in the spectra of cooler stars.

As with many of our abundances, the mean [Ti/H] abundance for our high S/N dwarf sample is flat at temperatures above 5500~K.  The mean relative abundance decreases at lower temperatures though to a lesser extent than elements like calcium and iron.  This could be related to the increased influence of molecular lines, such as TiO, and their less well known oscillator strengths, positions, and broadening parameters.  There is a negligible offset (-0.017~dex) between the entire comparison sample and our abundances, no appreciable trend and an RMS scatter of only 0.061~dex. When looking at the individual catalogs, there are trends of opposite sign between the \citetalias{2005ApJS..159..141V} and \citet{2012A&A...545A..32A} comparison stars. This leads to an increased scatter at temperatures hotter than solar and $\lesssim 5500$~K our [Ti/H] seems to be systematically lower than those in the \citetalias{2005ApJS..159..141V} catalog stars that make up most of the cooler comparisons.

\subsection{Vanadium} \label{elem:vanadium}
Though vanadium lies on the edge of the iron peak, it is also an odd Z element and so in general has a lower abundance than titanium (\ref{elem:titanium}) and chromium (\ref{elem:chromium}) and tends not to track iron abundance. Instead [V/H] follows a more $\alpha$-element trend, increasing in abundance with decreasing metallicity.  Like TiO, VO is present in the atmospheres of cooler stars and may be a feature of hot jupiter atmospheres, though no strong detections have yet been made.

Almost 20\% of the 208 vanadium lines in our mask are \ion{V}{2} lines which should help mitigate NLTE effects known to bias abundances from neutral lines.  For higher temperature stars, the sensitivity indicator $S_{V}$ dips down below our cutoff though the general trend follows that of titanium, increasing in importance at cooler temperatures.  

The overall trend in average abundance is also very similar to titanium, though there is less deviation from a flat average, dropping noticeably only around 5200~K.  The comparison sample has a small offset of -0.032~dex, but a large RMS scatter of 0.11~dex.  Some of this scatter may be related to the large uncertainties in the comparison catalogs, though some may due to the limited power of our model at high \teff.

\subsection{Chromium} \label{elem:chromium}
Chromium is an iron peak element which generally tracks iron abundance and has the same nucleosynthesis channel.  Its siderophile nature makes it of interest in analyzing the history of differentiation in the earth and has also been used to quantify the extent of differentiation in HgMn stars.  It has not been used directly in exoplanet research, but it is moderately refractory which makes it useful in exploring abundance trends with condensation temperature and their possible connection to planet formation.

Second only to iron, chromium has 326 lines in our mask with more than 1/3 of them \ion{Cr}{2} lines. The sensitivity indicator $S_{Cr}$ shows that sensitivity to model changes in chromium increases at lower temperatures. However, \citet{Tsantaki:2013dc} showed that for stars below 5200~K, errors in temperature can lead to a divergence in the abundances derived from neutral vs. ionized lines.  If the temperature is too high, then the abundance derived from neutral chromium will be systematically higher than that derived from \ion{Cr}{2}.  

The trend in [Cr/H] in our overall sample is nearly identical to that of [Fe/H].  At temperatures $\lesssim 5400$~K, there is a steady decrease in mean abundance.  As we should expect weaker \ion{Cr}{2} lines at lower temperatures, this should give more weight to \ion{Cr}{1} which does not fit with the picture of increased abundance seen by \citet{Tsantaki:2013dc}.  Since this effect assumed overestimated \teff, this is likely not the reason for our trend.  At hotter temperatures, the mean of the chromium abundances in our sample seem relatively flat.

\subsection{Manganese} \label{elem:manganese}
Unlike other odd Z iron peak elements such as vanadium (\ref{elem:vanadium}), manganese does not follow an $\alpha$-element trend.  However, it also does not track iron or overall metallicity.  Instead it seems to stay relatively flat, possibly decreasing slightly, with decreasing [Fe/H].  At metallicities below -0.4, NLTE effects become important and seem to change any downward trend into a plateau \citep{Battistini:2015jy,2008A&A...492..823B}.  Uncertainty in these trends has led to uncertainty in the production sites for Mn, though silicon shells with incomplete burning in SNIa are the suspected primary location \citep{2005ApJ...619..427U}.

There are 152 Mn lines in our mask, 41 of them \ion{Mn}{2} lines.  As mentioned above, at [M/H] $\lesssim -0.4$, NLTE effects begin to become important and LTE analysis can result in systematically lower abundances and higher temperatures \citep{Battistini:2015jy,2008A&A...492..823B}.  Since the $S_{Mn}$ curve shows decreasing sensitivity at higher temperatures, this could also effect the scatter in [Mn/H] values in the same region.

There is a small downward trend in the mean [Mn/H] with increasing temperature in addition to the low temperature trend shared with other elements.  The combined effects of decreased model sensitivity and NLTE effects are the most likely causes.  This is confirmed in our comparison with common stars from \citet{Battistini:2015jy} where our abundances roughly match their LTE abundances but show a decreasing trend when compared with their NLTE [Mn/H] values.  At the low temperature end, we see the same decrease in our sample mean [Mn/H] that we see in most of our other elements.  There are only a few low temperature stars in common with \citet{Battistini:2015jy} and they are all coincidentally lower metallicity stars.  Comparison with \citep{2012A&A...545A..32A} in this cooler region shows no trends, implying that they have the same issues as we do at lower temperatures.

\subsection{Iron} \label{elem:iron}
The large number of possible transitions in the iron atom combined with its relatively large abundance mean that optical spectra are dominated by iron lines.  The high number of lines and ease in determining [Fe/H] has led to its frequent use as a proxy for overall metallicity.  However, although decreasing iron abundance typically signals decreasing metallicity, $\alpha$-elements can make up an increasingly large fraction of total metals (by number of atoms) at sub-solar metallicities with the exact ratios a function of both age and location. The occurrence rate of massive planets increases with the square of [Fe/H] \citep{2005ApJ...622.1102F,2004A&A...415.1153S,1997MNRAS.285..403G}, indicating that the abundance of solids in the protoplanetary disk influences planet formation.  More recent analysis has shown that this dependence is weaker for smaller planets but still exists \citep{2015AJ....149...14W,2014Natur.509..593B}.

Twelve percent or 901 of the 7,500 lines in our model are iron lines and almost 1/3 of those are \ion{Fe}{2}, which provides a strong additional constraint on gravity via ionization balance.  The high number of lines in our model also leads to a nearly constant, and high relative, $S_{Fe}$.  At high temperatures, low metallicities, and low gravities, LTE analysis results in lower [Fe/H] than if NLTE effects are accounted for \citep{2014arXiv1403.3088B}, though for the majority of our sample, these effects will be quite small.

Above 5500~K, the mean iron abundance of our sample is relatively flat with only a mild decrease at the highest temperatures.  However, there is a clear decrease in [Fe/H] with temperature between 5500~K and 5000~K before it levels out again.  Despite the large number of iron lines in our spectrum, we are seeing the same trend in the cooler stars as other spectroscopic analyses \citep{Bensby:2014gi,2012A&A...545A..32A,2005ApJS..159..141V}.  In our comparisons of matching stars, most show very similar [Fe/H] with only a small offset (-0.020 dex) and no trends.  However, the \citetalias{2005ApJS..159..141V} abundance differences show a clear trend with temperature.  This is likely due to the differences induced by our new \logg\ determinations which would necessitate changes in both \teff\ and [Fe/H] to ensure a good model fit.

\subsection{Nickel} \label{elem:nickel}
The primary astrophysical interest in nickel is in measuring the energy output and progenitor mass of supernovae, particulary Type Ia (SNe Ia) \citep{1982ApJ...253..785A,2013FrPhy...8..116H}.  At early times, the decay of ${}^{56}\mathrm{Ni}$ is the primary energy source of the expanding ejecta until about 12 days in when the decay of its product, ${}^{56}\mathrm{Co}$, begins to eclipse it.  The measurement of ${}^{56}\mathrm{Ni}$ though comes from the lines of the iron decay product or through using the peak brightness in ``Arnett's law" \citep{1982ApJ...253..785A}.  Also of interest are the iron and nickel cores of the more massive terrestrial planets as the metals provide heat from friction and free electrons which can power dynamos for global magnetic fields.  Unfortunately, we will not have measurements of these properties outside of our solar system in the foreseeable future.  Nickel abundances are still useful in examining trends in abundance with condensation temperature which could provide insights into gas-dust segregation in molecular clouds or the temperature history of protoplanetary disks.

More than half of the 223 nickel lines in our mask are \ion{Ni}{2} lines, though only a handful of those make a noticeable contribution to the spectrum for our stars. Like Fe, Ni has a relatively constant $S_{Ni}$ across the temperature range of our sample.  The neutral nickel lines at low metallicities are susceptible to NLTE effects \citep{2014arXiv1403.3088B} though the effect should be minimal since our sample contains few low metallicity stars.

Though the average abundances are somewhat flatter, the trends in [Ni/H] with \teff\ mimic those of [Fe/H].  There is a slight decrease in abundances at super solar temperatures, and there is a decline in the mean between 5500~K and 5000~K.  The differences between matching stars from other catalogs are smaller than with iron with an offset of only -0.015 dex.  The comparisons to the \citetalias{2005ApJS..159..141V} stars do show a slight trend, though less significant than for iron.  This again is probably caused by the differing global parameters for these stars.

\subsection{Yttrium} \label{elem:yttrium}
The majority of yttrium nucleosynthesis happens in the weak s-process in massive stars; most of the remainder of Y nucleosynthesis taking place in the r-process \citep{Busso:1999ig,Pignatari:2010ir,Maiorca:2011ey}.  Because the strong-s process in AGB stars is responsible for more massive s-process elements, the yttrium abundance can help identify the specific evolutionary pathway which led to a given population.  Though this may not translate directly into planet formation, the environmental effects of massive stars is of interest for understanding the evolution of protoplanetary disks and the impact on planet formation.

There are 35 yttrium lines in our model, though 17 of them are \ion{Y}{2}.  Though the number of clean lines was comparable to other light s-process elements like copper and zirconium, we were able to more reliably recover the asteroid abundance of yttrium.  Our sensitivity metric, $S_{Y}$, is low but relatively flat over the whole temperature range, dipping to below our cutoff value at about 5200~K.

Though $S_{Y} > S_{N}$ for temperatures above solar, [Y/H] seems to have a very similar overall trend in abundance.  At lower temperatures, there is a clear decrease in mean abundance similar to several other elements.  However, because the trend in [Y/H] is roughly constant across our temperature range, it is likely that our line parameters are incorrectly tuned.  The only comparison sample that we have from the literature is \citet{Bensby:2014gi} which shows good agreement above about 5600~K and then a decreasing trend in our abundance at cooler temperatures.  As this matches closely with the trend we see in our mean [Y/H] values, it reinforces our confidence in the corrections we applied based on those trends.

%
\section{Macroturbulence and Projected Rotational Velocity} \label{sec:rotation_vmac}
\begin{figure}[ht] 
   \centering
   \includegraphics[width=0.95\columnwidth]{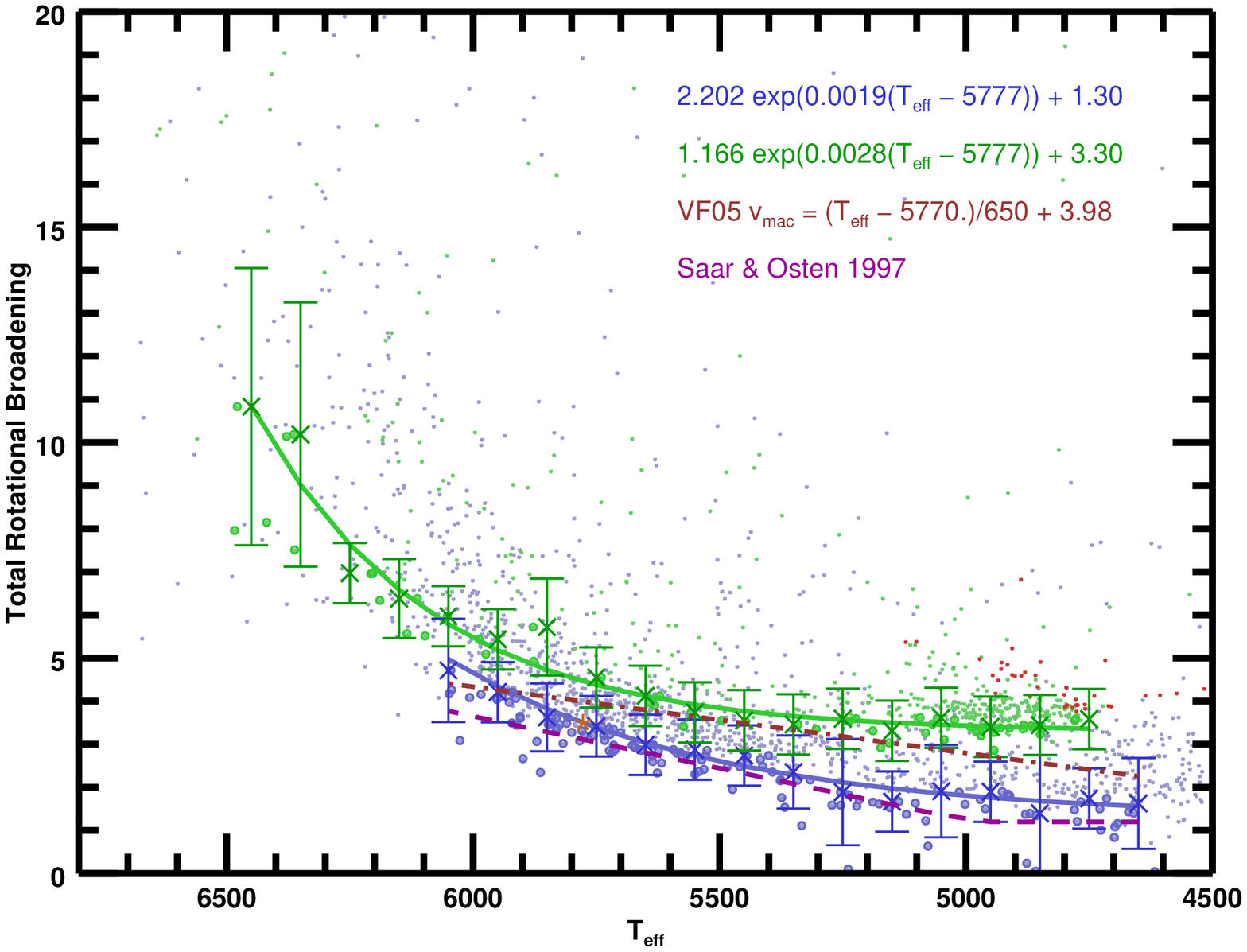} 
   \caption{Fit to the bottom 10th percentile of total rotational broadening (\vsini\ and \vmac) for stars with S/N $\geq 100$ in the sample.  The relations of \citetalias{2005ApJS..159..141V} and \citet{Saar:1997wt} are overplotted in dash-dot and dashed lines respectively.  Blue points are stars with \logg\ $\geq 4$, green have $4 > \log g \geq 3$, and red have \logg\ $< 3$.  Heavier points with error bars represent bin centers and uncertainties used in fitting the relations. Neither of the previous studies separated dwarfs from subgiants, but our separate relations coincide with those of the previous studies.}
   \label{fig:vmac_relation}
\end{figure}

Surface granulation strengthens and asymmetrically broadens spectral lines \citep{2008oasp.book.....G,2000A&A...359..729A}. In hydrostatic models, granulation effects are commonly approximated by a microturbulence parameter (\vmic) that selectively increases line strength and a macroturbulence parameter (\vmac) that symmetrically broadens lines without affecting line strength. This approximation yields symmetric model line profiles that significantly disagree with asymmetric observed line profiles. Bulk rotation of the star is well modeled by a projected equatorial rotation velocity parameter (\vsini), but this formulation neglects surface differential rotation, which can be significant. Given these limitations of the hydrostatic model employed by SME, we made some simplifying assumptions.

As discussed in \citet{2015ApJ...805..126B}, we fix \vmic\ at 0.85 km~s$^{-1}$. For our analysis procedure, this yields spectroscopic \logg\ values that agree with asteroseismic surface gravities as a function of \teff.  Our empirical atomic line parameters (\S \ref{sec:empirical_corrections}) compensate to some extent for errors in the microturbulence approximation and our particular choice of \vmic.

Limitations of the hydrostatic model employed by SME make it difficult to disentangle the remaining line broadening parameters, \vmac\ and \vsini, especially in the common case when rotational and macroturbulent broadening are comparable. Therefore when fitting observed spectra, we adopted a single line broadening parameter. We fixed \vsini$=0$ and solved only for \vmac\ (\S \ref{sec:global_stellar_params}). With this approach, our derived value of the SME parameter \vmac\ is actually the total line broadening (\vbroad) due to both rotation and macroturbulence. This additional approximation is no worse the hydrostatic approximation when rotational broadening is comparable to macroturbulent broadening.

\subsection{A Macroturbulence Relation, \vmac} \label{sec:fitting_vmac}
We obtained \vbroad\ for each star in our sample by fitting observed spectra (\S \ref{sec:global_stellar_params}). Following \citetalias{2005ApJS..159..141V}, we then assume that stars with the lowest values of \vbroad\ have negligible rotational broadening. For these stars, we interpret \vbroad\ as \vmac. We interpret any additional broadening above this level as rotational broadening.

Several previous efforts have derived relations between \vmac\ and effective temperature or color \citep{2005ApJS..159..141V,Saar:1997wt,1988lsla.book.....G}.  Those relations have then been used to disentangle \vsini\ and \vmac\ under the assumption that the two, added in quadrature, comprise \vbroad.  We found that using these relations during our analysis resulted in the temperature (and hence gravity) being artificially pulled toward that defined by the inverse of the relation.  Instead, we solved only for the total broadening and derived new macroturbulence relations for both dwarfs and subgiants based on our sample.  We then apply those relations after the analysis to provide values for \vmac\ and use it to fit for \vsini.  At both low \vbroad\ (where the proportional uncertainty is large) and high \teff\ (where \vmac\ as a function of temperature is highly non-linear), \vbroad\ is a more precisely determined quantity than either the macroturbulence or projected rotational velocity.

To derive our relations between \teff\ and \vmac, we separated our sample into dwarfs (\logg\ $\geq 4.0$), subgiants ($4.0 > \log g \geq 3.0$) and giants ($\log g < 3.0$).  The naming of these broad \logg\ bins is a slight over-simplification but in our sample, results in very little contamination compared to a more careful separation by evolutionary state.  The number of giants was too small to derive a statistically meaningful result, but as can be seen in Figure \ref{fig:vmac_relation} they do seem to have a higher minimum \vbroad\ than dwarfs or subgiants at the same \teff.  Within each group, we bin all the stars in 100~K temperature bins and discard any stars with $\chi^2 > 3 \sigma$ above the mean for that bin.  Since the effects of \vsini\ on line profiles are moderated not only by rotation, but orientation of the rotation along our line of site, some fraction of our stars with minimal \vbroad\ will have nearly zero \vsini. We fit an exponential decay function to the median of the \vbroad\ in the bottom 10\% of the stars in each 100~K bin with more than 10 stars adopting uncertainties for each bin based on the standard deviation of points selected divided by the square root of the number of points (Figure \ref{fig:vmac_relation}).  The combined relation is in Equation \ref{eqn:vmac} with the addition of a constant \vmac\ for giant stars.

{\footnotesize
\begin{equation} \label{eqn:vmac}
	v_{mac} =
	\begin{cases} 
	2.202 e^{0.0019(T_{eff} - 5777)} + 1.30		&	\log g \ge 4.0 \\
	1.166 e^{0.0028(T_{eff} - 5777)} + 3.30	&	4.0 > \log g \ge 3.0  \\
	4.0								& 	\log g < 3.0
	\end{cases}
\end{equation}}

\subsection{Rotational Broadening, \vsini} \label{sec:determining_vsini}
When fitting our relation for macruturbulence, we made the assumption that stars on the floor of the distribution would have \vsini~$\approx 0$ and so all of \vbroad\ would be \vmac.  As \vsini\ increases and dominates over macroturbulence, \vbroad\ will not necessarily be a simple quadrature sum of the two.  To  ensure self consistent broadening in our models, we used Equation \ref{eqn:vmac} to set \vmac\ and solved for \vsini\ for each spectrum.

%
\section{Error Analysis} \label{sec:error_analysis}

The sources of errors in our derived parameters can be grouped broadly into three categories: uncertainties in the observed spectra (i.e. S/N and stellar variability), the inability of our model to accurately and consistently reproduce the spectra, and degeneracies between parameters which yield similar spectral shapes.  Observational uncertainties, variations in photon counts in any wavelength bin, are the easiest to quantify and the high S/N of a majority of our spectra make this a negligible contribution to the overall parameter error budget.  The other two are both tightly linked and much more difficult to quantify.  We discuss below our efforts to normalize all of our quantities to solar, characterize the systematics which arise from our modeling, and quantify the contributions to our uncertainties from random observational errors.

\subsection{Solar Zero Points}

\begin{table}
    \caption{Asteroid Sample}
    \begin{center}
    \begin{threeparttable}
    \renewcommand{\TPTminimum}{\linewidth}
    \centering
    \begin{tabular}{ l d{3} d{3} }
    \hline 
    \hline \\[-1.5ex]
    Parameter & \mathrm{Mean} & \mathrm{Std Dev} 	\\
    \\[-1.5ex]
    \hline
	$T_{\rm{eff}}$	&	5779.2\ \mathrm{K}	&	4.7\ \mathrm{K} \\
	$\log g$		&	4.458		&	0.007 \\
	$v_{\rm{broad}}$	&	4.024~\mathrm{km/s}		&	0.190~\mathrm{km/s}\\
	{[}C/H]				&	0.025		&	0.007 \\
	{[}N/H]				&	0.016		&	0.027 \\
	{[}O/H]				&	-0.015		&	0.016 \\
	{[}Na/H]			&	0.032		&	0.011 \\
	{[}Mg/H]			&	0.030		&	0.003 \\
	{[}Al/H]			&	0.004		&	0.009 \\
	{[}Si/H]			&	0.044		&	0.007 \\
	{[}Ca/H]			&	0.035		&	0.007 \\
	{[}Ti/H]			&	0.040		&	0.005 \\
	{[}V/H]				&	0.049		&	0.012 \\
	{[}Cr/H			&	0.038		&	0.005 \\
	{[}Mn/H]			&	0.025		&	0.007 \\
	{[}Fe/H]			&	0.024		&	0.002 \\
	{[}Ni/H]			&	0.025		&	0.004 \\
	{[}Y/H]				&	0.020		&	0.011 \\
    \hline
    \end{tabular} \label{table:asteroid_summary}
    \end{threeparttable}
    \begin{tablenotes}
        \item {Note: Values are the $\chi^2$ weighted means of the parameters for 20 asteroid
            spectra and their standard deviations.  Square bracketed notation designates standard log of solar relative abundances.}
    \end{tablenotes}
    \end{center}
\end{table}

\begin{figure}[ht] 
   \centering
   \includegraphics[width=0.95\columnwidth]{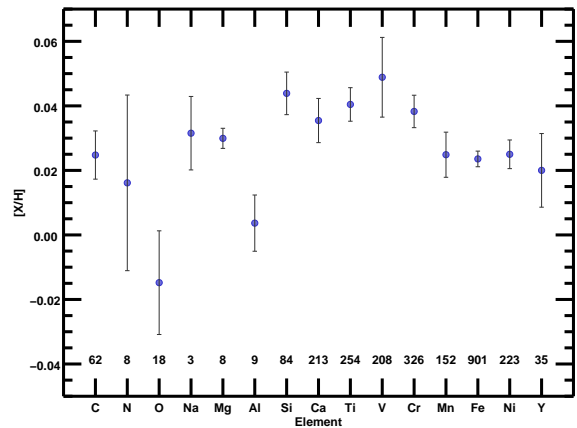} 
   \caption{We analyzed the solar spectrum using 20 asteroid spectra (\S \ref{sec:asteroids}) and recovered the solar abundances to within $\sim 0.03$~dex for most elements, and dispersions (error bars) $< 0.03$~dex.  The number of lines in our mask for each element is indicated at the bottom of the plot.}
   \label{fig:asteroid_abunds}
\end{figure}

Our analysis of the solar spectrum is based on 20 asteroid spectra (\S \ref{sec:asteroids}) taken throughout the 10 years of observations. This gives us both a sense of the accuracy in returning the parameters for stars similar to the Sun, and in the precision with which our procedure returns those values for multiple spectra.  We recover the parameters we used when tuning the atomic line data (Table \ref{table:asteroid_summary}).  The typical offset for most abundances is only $\sim 0.03$ dex and the dispersions in results are small (Figure \ref{fig:asteroid_abunds}) despite the simplifying assumptions made in our models (1D, static atmospheres, LTE, fixed microturbulence).  However, as we move away from the solar values, those simplifications can lead to larger uncertainties that won't be captured by either the covariance matrix of the fit or this analysis of the Sun.

In our standard fitting procedure, we also vary the initial temperatures of the models in the two stages where we fit for the global parameters.  This can tell us how sensitive our fitting procedure is to the choice of initial parameters.  We looked at the distribution of parameters returned for the asteroid spectra for these perturbed models to examine the dependence of our final parameters on the starting point.  \teff\ was the only parameter where the variation in the starting point had an effect (3.3~K) comparable to the difference between final parameters between different observations (4.7~K).

\subsection{Random Errors} \label{sec:random_errors}
From a single spectrum, we are able to get an estimate of the uncertainties in the fit for each parameter, but this does not capture systematic errors from simplifications in our model, nor does it include the effects of stellar variability or observational errors.  However, almost 30\% of the high S/N stars in our sample have multiple observations and 14\% have 3 or more.  We can use these repeat observations to build up a statistical error distribution for each parameter.

\begin{table}
  \caption[]{Statistical Uncertainties}
  \begin{center}
    \begin{threeparttable}
    \renewcommand{\TPTminimum}{\linewidth}
    \centering
    \begin{tabular}{ l d{2} d{2} }
\hline 
\hline \\[-1.5ex]
Parameter & \multicolumn{1}{c}{Std Dev} & \multicolumn{1}{c}{Adopted Unc} 	\\
\\[-1.5ex]
\hline
	$T_{\rm{eff}}$	&	12.4\	\mathrm{K}	&	25\ \mathrm{K} \\
	$\log g$		&	0.014		&	0.028 \\
	$[M/H]$			&   0.005		&	0.010 \\
	$v_{\rm{broad}}$	&	0.35		&	0.7~\mathrm{km/s}   \\
	C				&	0.013		&	0.026 \\
	N				&	0.021		&	0.042 \\
	O				&	0.018		&	0.036 \\
	Na				&	0.007		&	0.014 \\
	Mg				&	0.006		&	0.012 \\
	Al				&	0.014		&	0.028 \\
	Si				&	0.004		&	0.008 \\
	Ca				&	0.007		&	0.014 \\
	Ti				&	0.006		&	0.012 \\
	V				&	0.017		&	0.034 \\
	Cr				&	0.007		&	0.014 \\
	Mn				&	0.010		&	0.020 \\
	Fe				&	0.005		&	0.010 \\
	Ni				&	0.006		&	0.012 \\
	Y				&	0.015		&	0.030 \\
\hline
    \end{tabular} \label{table:random_errors}
    \end{threeparttable}
    \begin{tablenotes}
        \item {Note: Standard deviations derived from parameter differences in observations
        of the same star.  Statistical uncertainties have been multiplied by a factor of 2.}
    \end{tablenotes}
    \end{center}
\end{table} 

We used the procedure outlined in \citetalias{2005ApJS..159..141V} to compare all pairs of observations of single stars, weighting them by the inverse of the number of repeated observations.  Plotting histograms of these differences for each free parameter and taking the 68th percentile region gives us statistical uncertainties for each parameter.  We have tabulated the results for all of our free parameters in Table \ref{table:random_errors}.  Most of the distributions had broader wings than a standard gaussian distribution, though all of the distributions were narrow. 

These statistical uncertainties are roughly 1/4 those from \citetalias{2005ApJS..159..141V}, which indicates that the new version of SME and our new procedure are doing a better job at converging to the same solution.  However, our parameter comparisons to matching stars in other catalogs indicate that these are probably too small by a factor of 2 or more.  In \citet{2015ApJ...805..126B}, we found that the differences between our surface gravity and that of asteroseismology was 0.05 dex, or almost 3 times larger than what we determined here.  At the same time, the standard deviations in our asteroid parameters are smaller by a factor of 2 or more, which says that near solar, we probably are doing much better.  To give a more reasonable average statistical error, we will multiply the standard deviations determined here by 2 to obtain the statistical uncertainties for our parameters (Table \ref{table:random_errors}).  We assumed approximately equal contributions to errors in overall line broadening from \vsini\ and \vmac. This yields uncertainties in each of 0.5~km~s$^{-1}$ from the overall \vbroad\ uncertainty of 0.7~km~s$^{-1}$ listed in the table.

\subsection{Modeling Limitations} \label{sec:modeling_limitations}
The sensitivity metric we developed (\S \ref{sec:elem_sensitivity}) is meant to capture the sensitivity of our \textit{model} to changes in abundances.  We can perform a similar analysis of how well the model is able to fit the data.  This method is described in detail in (Piskunov \& Valenti, 2016 in prep) but we will discuss it briefly here.  At every point in the line regions of an individual spectrum, we obtain the difference between model and observation and the partial derivative with respect to the parameter of interest.  Because not every point in the spectrum may be effected by changes in that parameter, we exclude all points which have partial derivatives less than the median for all spectral points.  We then divide the differences by the partial derivatives to find the amount the parameter would need to change to place that pixel at the correct location.  We calculate the cumulative distribution of all pixels, and the middle 68\% show us the uncertainty range for that parameter.

This method has the benefit of giving parameter uncertainties for individual observations, but unfortunately does not capture the degeneracies between the parameters which can often help to constrain the range of acceptable values.  It is also overstating the influence of each parameter since each individual uncertainty estimate assumes that only that parameter is responsible for the mismatch between observation and model, then makes the same assumption for the next parameter.  Nevertheless, calculating these uncertainties for our global parameters (those which have influence over most of the spectrum) can give us an upper limit on what to expect from our uncertainties.

We calculated these uncertainty limits for \teff, \logg, [M/H], and $v_{\mathrm{broad}}$ for every spectrum in our sample and then calculated the mean and standard deviation in the distribution of returned limits.  The limits we found were $\Delta T_{\mathrm{eff}} = 60 \pm 37$, $\Delta \log g = 0.15 \pm 0.083$, $\Delta [M/H] = 0.06 \pm 0.025$, and $\Delta v_{\mathrm{broad}} = 1.4 \pm 7.1$.  Considering the accuracy with which we recovered the solar parameters and the precision with which we matched literature values from other studies, these uncertainties are probably too large by a factor of 2 or so in most cases.  In fact, in comparison to the statistical errors calculated in \S \ref{sec:random_errors} they are a factor of 4 to 10 times larger.

\subsection{NLTE Effects} \label{sec:nlte_effects}

Our models assume local thermodynamic equilibrium and have a large number of lines of varying excitation potentials, ionization states, and formation depths.  This tends to smooth out some of the NLTE effects which tend to affect deep lines and those in the minor ionization state \citep{2001A&A...366..981G,2014arXiv1403.3088B}.  It is still instructive to look at the NLTE corrections when available to examine how much our models might be biased.  It is not possible to calculate a straight-forward abundance adjustment without the correction data for every line in our line list.  We can, however, look at the mean corrections for lines with available NLTE data and see how those corrections would effect the relative abundances of binary pairs of stars in our sample.

The INSPECT \footnote{Data obtained from the INSPECT database, version 1.0 (inspect.coolstars19.com)} tool provides a web interface to look up NLTE corrections based on abundance, temperature, gravity, metallicity, and microturbulence for a number of lines for seven elements.  We obtained corrections for lines in common with our line list of oxygen \citep{2015MNRAS.454L..11A}, sodium \citep{Lind:2011ho}, magnesium \citep{2015A&A...579A..53O}, titanium \citep{Bergemann:2011kc}, and iron \citep{2012MNRAS.427...27B,Lind:2012gc}.  This ranged between two (sodium) and 16 (iron) lines, just a small fraction of the lines for all but sodium.  Only one pair had stellar parameters within the databases limits for titanium, and only two pairs were in the range for magnesium (that had zero correction) so we dropped these elements from this analysis.  For each star, we averaged the corrections from all lines of a given element weighted by their equivalent widths and applied the correction to our abundance.  We then compared the abundances of the binary pairs to see if the abundances were brought more in line with one another. There may be innate differences in abundances in some binaries \citep{2015ApJ...801L..10T,2015ApJ...808...13R}, but they should not be correlated with temperature. Although the corrections were derived for only a small fraction of our lines for oxygen and iron, we applied the averaged adjustments for this test assuming that all of our other lines would behave in a similar manner.

There were no appreciable changes in the offsets between the iron abundances of the binary pairs after applying the NLTE corrections. The lack of agreement with the iron line changes is probably due to our large line list which includes both neutral and ionized lines as well as many weak and high excitation lines which are not very susceptible to NLTE effects.

The changes in sodium abundances were all very similar, averaging -0.078~dex for the hotter components and -0.083~dex for the cooler component. This tended to make the differences between the pairs larger rather than smaller.  The two sodium lines for which we had NLTE corrections are the primary source of information for our sodium abundances.  However, in hotter stars where NLTE effects can play a stronger role, these lines are weaker and so are less susceptible to NLTE effects.

The NLTE corrections to the oxygen abundances varied much more and typically decreased the differences in abundance between the binary components.  The typical corrections were $\approx -0.22$ dex for the hotter component and $\approx -0.15$ dex for the cooler component and tended to lower the abundances overall, especially at super-solar temperatures. This corresponds to the trend that we see in our raw oxygen abundances.  At higher temperatures, most of the oxygen information in our spectral regions comes from the triplet at 7771~\AA\ which get deeper in hotter stars.  These lines form high in the atmosphere and are subject to greater NLTE effects.  It seems likely that our trend in oxygen abundance at higher temperatures is due to our assumption of LTE.

%
\section{Results}
\subsection{Spectroscopic Properties} \label{sec:spectroscopic_properties}
The parameters determined from fitting observed spectra with synthetic spectra are divided into two tables. The global stellar parameters are in Table \ref{table:final_params} and the final trend-corrected abundances in Table \ref{table:final_abundances}.

If the reader wishes to recover our original solar relative abundances,  we provide a table of the abundance corrections applied over the temperature range of our sample (Table \ref{table:dwarf_abund_trend_offsets}).  Each column indicates the correction applied at that temperature for a given element.  Subtracting this value (or an interpolated value in between) based on the final temperature of the star as given in Table \ref{table:final_params} will recover the uncorrected solar relative abundance.

Column (1) of Tables \ref{table:final_params} and \ref{table:final_abundances} is the SPOCS ID.  The SPOCS ID numbers are shared with the \citetalias{2005ApJS..159..141V} catalog for stars in common between that work and this.  Stars new to this catalog have an ID~$\geq 2000$.  Also shared between the two tables is the star name in Column (2).  Columns (6) and (7) of Table \ref{table:final_params} are the activity metrics described in \S \ref{sec:activity}.  Columns (8)-(10) are the final line broadening values (\S \ref{sec:rotation_vmac}) for the stars and Column (11) is the barycentric radial velocity at the time of observation.  The S/N ratio reported in Column (12) is calculated at the blaze peak near 6000~\AA.  The final columns contain the RMS scatter in continuum (Column (13)) and line (Column (14)) regions of the spectra and the number of spectra fit for that star (Column (15)).  Columns (16)-(30) of Table \ref{table:final_abundances} are our final trend corrected abundances.

\subsection{Derived Properties}  \label{sec:derived_properties}
About 60\% of our stars had measured distances, which allowed us to compute absolute visual magnitudes.  We used these, combined with our spectroscopic parameters, to interpolate in the Yonsei-Yale isochrone grid to obtain mass, radius, and age for these stars (Table \ref{table:derived_props}).  Stars without distances are also included in this table as it also includes the location for easier matching to other catalogs.

Columns (1) and (2) in Table \ref{table:derived_props} are the SPOCS ID and star name shared with the stellar parameters and abundances tables.  Columns (31)-(34) are literature values from the SIMBAD database \citep{Wenger:2000ef} and the Hipparcos catalog \citep{1997ESASP1200.....E}.  Using those parameters, the bolometric correction of \citet{2003AJ....126..778V} and our values of \teff\ and \logg\ we calculated the luminosity, radius, mass and their uncertainties in Columns (35)-(37).  The isochrone based values in Columns (38)-(42) were derived by combining the luminosity with our final \teff, [Fe/H], and [Si/H] and interpolating in a grid of Yonsei-Yale isochrones \citep{2004ApJS..155..667D}.  For all of the values in Columns (35)-(42) we followed the procedures detailed in \citetalias{2005ApJS..159..141V}.

%
\section{Surface Gravity Comparisons} \label{sec:surface_gravity}
\begin{figure*} 
   \centering
   \includegraphics[width=\textwidth]{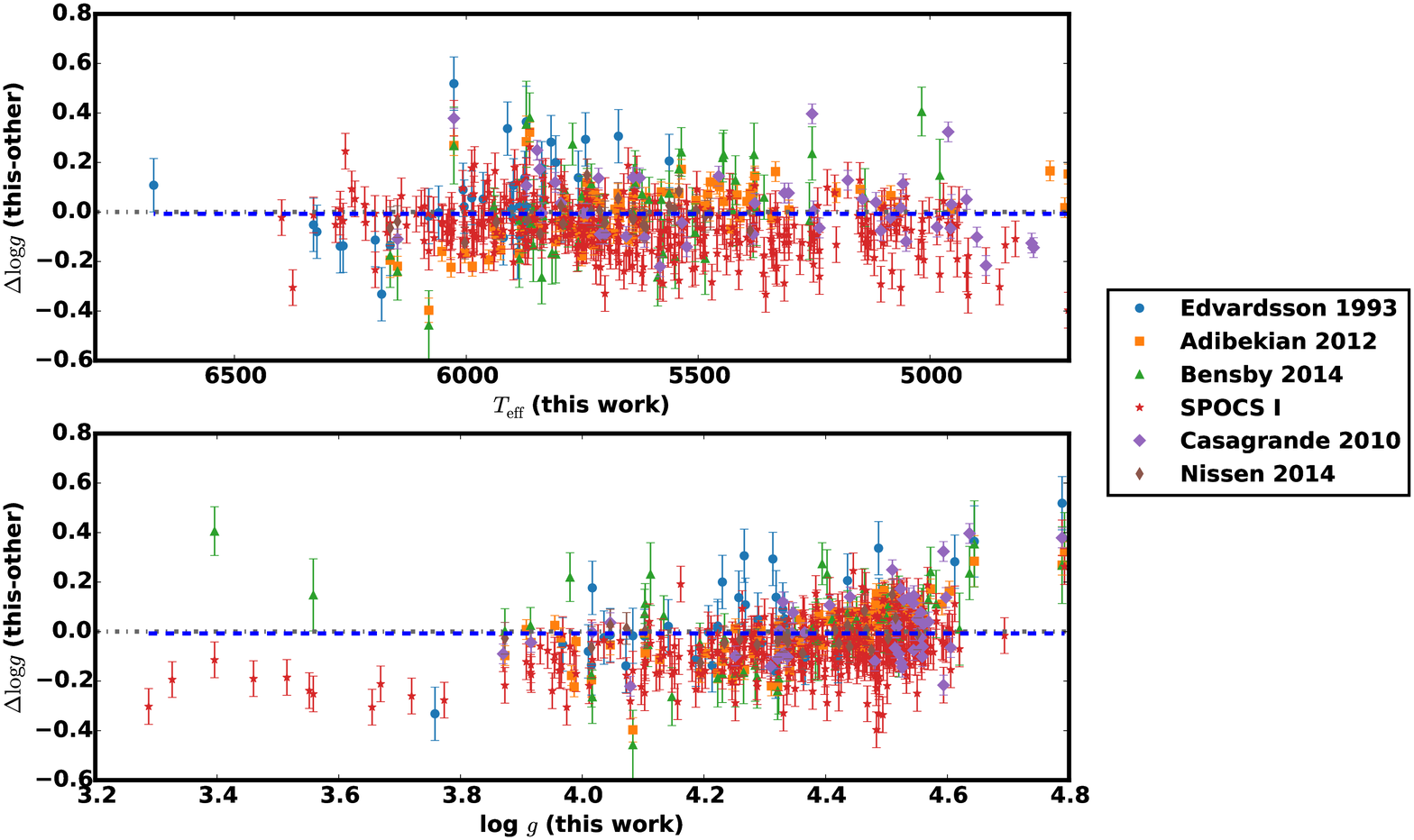} 
   \caption{The bulk of the comparison stars are from the SPOCS catalog \citepalias{2005ApJS..159..141V} which uses a very similar analysis technique to ours, but that was shown to have problems recovering accurate gravities for hotter and lower gravity stars \citep{2015ApJ...805..126B, 2012ApJ...757..161T}.  At gravities below solar, we obtain consistently lower \logg, especially with respect to the SPOCS stars. The RMS scatter in $\Delta\log g$ of 0.12 dex about the fitted offset of -0.007 dex (blue dashed line) is more than double the scatter for our comparison with asteroseismically determined gravities.  An offset was preferred over a sloping fit by both AIC and BIC tests.}
   \label{fig:surface_gravity}
\end{figure*}

\citet{2015ApJ...805..126B} demonstrated that our procedure recovered surface gravities consistent with those derived using asteroseismology.  Due to the time requirements of asteroseismology, most of the stars in that study were hotter and had lower surface gravities than the bulk of the stars in this study.  However, those were the stars which have been hardest to accurately characterize with spectral analysis \citep{2012ApJ...757..161T}.  The stars in the comparisons here (Figure \ref{fig:surface_gravity}) come from other spectroscopic catalogs using both curve of growth analysis and spectral modeling and the majority of them are of cool, high surface gravity dwarfs.

When compared against all literature values there is virtually no offset (-0.007 dex) and a scatter of only 0.12 dex in $\Delta\log g$ suggesting only a small underestimate in the combined uncertainties.  Since our procedure is an extension of that used in \citetalias{2005ApJS..159..141V}, we also examined only these stars, and the rest of the matches excluding these stars.  There was virtually no change in the offset or scatter when comparing the gravities of stars other than those in the SPOCS catalog.  There was a slight decrease in scatter (0.10 dex) and increase in offset (-0.06 dex) when looking at only the SPOCS stars.  The offset is likely directly related to the improvements in our analysis procedure which yield more accurate gravities.

We can expect that stars with large offsets in gravity between this paper and other forward modeling studies will also have significant differences in temperature and metallicity.  In fitting modeled spectra to observations, changes in surface gravity necessitate changes in temperature and metallicity to continue to provide a close fit.  However, the parameters are not completely degenerate.  At different temperatures and gravities, lines of different excitation potentials and ionization states can grow in opposite directions with changes in \teff\ and \logg.    The large number of lines in our model constrains how much of a tradeoff can be made and still provide a good fit to the observation.

%
\section{Effective Temperature Comparisons} \label{sec:effective_temperature}
Determining effective temperature from high resolution spectra has typically yielded precise temperatures but offsets between scales using different analysis techniques remain.  We showed in \citet{2015ApJ...805..126B} that our procedure yields accurate gravities consistent with those from asteroseismology.  We would like to make the same kind of comparison for effective temperature.  There are a growing number of stars with interferometrically determined angular diameters \citep{2013ApJ...771...40B} but the coverage of dwarfs is still small.  To augment our comparisons, we will compare our effective temperatures to those derived using the infrared flux method \citep{2010A&A...512A..54C} and other spectroscopic surveys.

\subsection{\teff\ from Angular Diameters}
Interferometrically derived effective temperatures rely on very few model dependencies and can be obtained using only the limb darkened angular diameter $\Theta_{\mathrm{LD}}$ and the bolometric flux $F_{\mathrm{BOL}}$ \citep{2013ApJ...771...40B}.  This should result in the most accurate determination of effective temperature and the systematic uncertainties reported are correspondingly small.  However stars with measurements by different groups still show differences in radius, and hence temperature, that are larger than the reported uncertainties.  

\begin{figure*} 
   \centering
   \includegraphics[width=\textwidth]{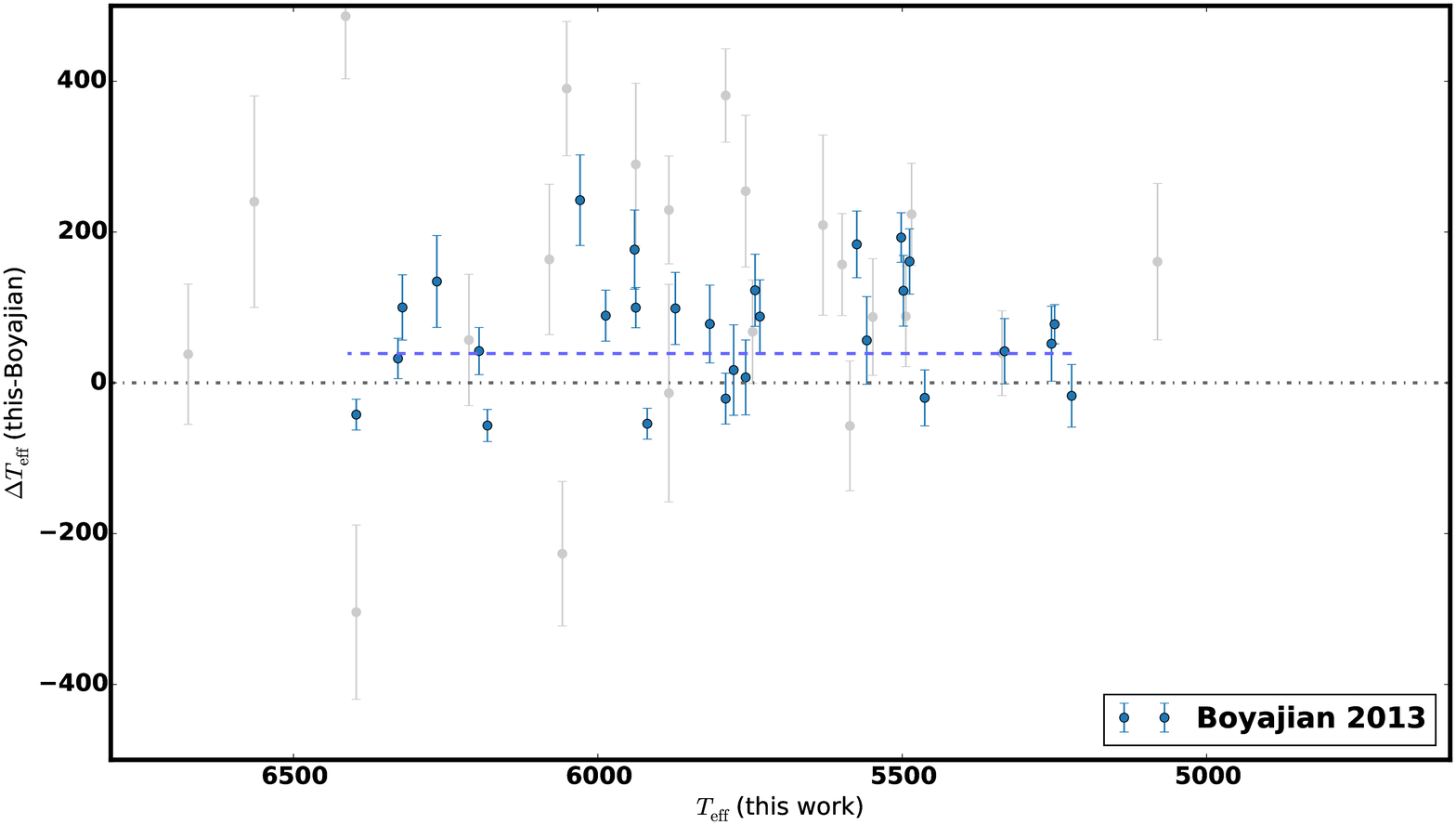} 
   \caption{Comparisons to temperatures from interferometric diameter measurements had a scatter several times larger than the combined errors suggesting that the errors were underestimated.  The blue points above are only those points with reported errors in \teff\ from diameters  $< 1\%$ which seemed to match their scatter.  These tend to be cooler than our spectroscopically determined temperatures with an offset of 39~K (blue dashed line) and an RMS scatter of 82~K.  Including all matching stars (both blue and grey points) resulted in an offset of 52~K and a scatter of 121~K which appears to increase at higher effective temperatures, though there are few overlapping measurements above 6200~K.}

   \label{fig:interferometric_teff}
\end{figure*}

Our high S/N sample has 45 stars in common with 50 unique measurements in the catalog of \citet{2013ApJ...771...40B}.  We compared the differences between \teff\ for these stars and fit an offset using the quadrature sum of our uncertainties and theirs (Figure \ref{fig:interferometric_teff}).  The spectroscopically determined temperatures are 52~K hotter than those from interferometry and have an RMS scatter of 121~K which is much greater than the combined uncertainties.  We removed all stars with reported uncertainties larger than 1\% from the interferometric sample and found that the remaining stars were not only closer to ours (only 39~K cooler) but also had significantly reduced RMS scatter of only 82~K.  This implies that the uncertainties of the angular diameter temperatures may be underestimated.  Although the number of points is small, it also appears that the scatter increases with increasing \teff\ and there is better agreement for stars cooler than the sun.  There is no significant trend with \teff, \logg, or [M/H].  This result agrees with \citet{2013ApJ...771...40B} which finds that spectroscopic \teff\ tends to be overestimated with respect to interferometrically derived temperatures.

\subsection{\teff\ from Infrared Flux}

\begin{figure*} 
   \centering
   \includegraphics[width=\textwidth]{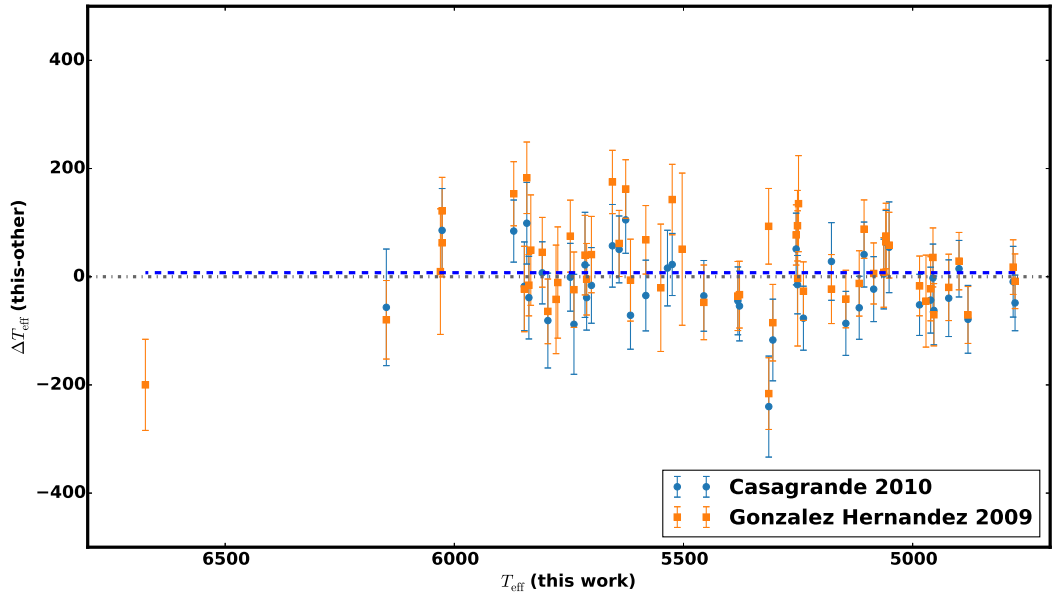} 
   \caption{Differences with temperatures determined from infrared flux measurements with uncertainties added in quadrature.  The two different IRFM methods had an offset of $\sim 35$~K from each other, which when combined led to an offset 7~K hotter than our temperatures with an RMS scatter of 72~K.  For clarity, the plot above has removed 16 points from \citet{2009A&A...497..497G} with uncertainties larger than 3\% which leaves the offset unchanged and the scatter at a nearly identical 71~K.  Separating the two shows that the temperatures of \citet{2010A&A...512A..54C} are 11~K hotter than ours, and those of \citet{2009A&A...497..497G} are 24~K cooler.}

   \label{fig:irfm_teff}
\end{figure*}

The infrared flux method (IRFM) relies on the same basic principle as angular diameters.  Rather than measuring angular diameters, the ratio between bolometric and monochromatic infrared flux is compared to those from stellar atmosphere models \citep{2009A&A...497..497G}.  The bolometric flux and IR flux depend equally on diameter, but the IR flux is only weakly dependent on temperature, which allows the diameter to cancel in the ratio.  The typical uncertainties are similar to those from high resolution spectroscopy and their zero points tend to lie in between those of spectroscopy and direct angular diameter measurements.  We compared our temperatures to two large catalogs \citep{2010A&A...512A..54C,2009A&A...497..497G} which had $> 50$ stars apiece in common with ours (Figure \ref{fig:irfm_teff}).  We found generally good agreement with our temperatures when compared to the whole set.  On average they were only 7~K hotter with an RMS scatter of 72~K.  However, when looking at the two catalogs individually, there was a 35~K difference between them.  Several of the IRFM temperatures from \citet{2009A&A...497..497G} had large offsets, but all of those had uncertainties larger than 3\% and removing them did not significantly reduce the scatter of the sample as a whole.

\subsection{\teff\ Comparisons with other Spectroscopy}

\begin{figure*} 
   \centering
   \includegraphics[width=\textwidth]{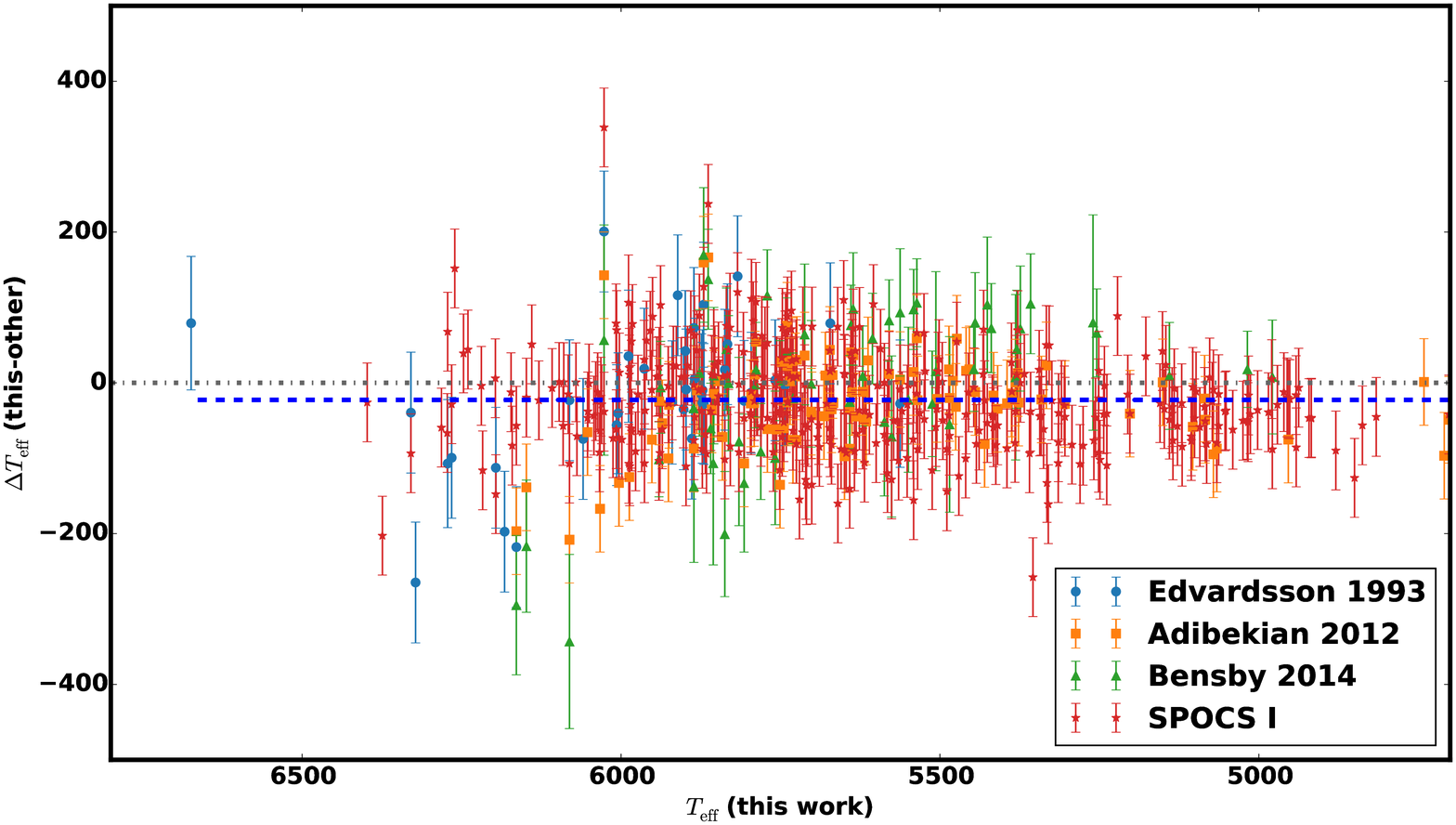} 
   \caption{Differences with temperatures determined from spectroscopic measurements with uncertainties added in quadrature.  For catalogs without reported errors, we assumed an average error of 50~K.  Our temperatures are 22~K cooler (blue dashed line) with an RMS scatter of 67~K implying a mutual precision better than 1\%.  The scatter again increases at higher effective temperatures. We expect this behavior in this context as our new analysis procedure has larger differences in \logg\ (and hence \teff) for hotter stars.}

   \label{fig:spectroscopic_teff}
\end{figure*}

We have a large overlap, $> 500$ stars, with a number of spectroscopic catalogs with great coverage between 5000~K and 6000~K.  Comparing them in the same manner as for angular diameters and IRFM above, we found that our temperatures are $\sim 22$~K cooler with an RMS scatter of 67~K (Figure \ref{fig:spectroscopic_teff}).  The scatter does seem to increase for hotter stars with very tight clustering at the cooler end of the scale.  Some of this scatter is due to offsets between the different comparison catalogs, though there is agreement at the 1\% level among the different spectroscopic temperature scales.  Our new analysis procedure had the largest effect on the gravities of stars hotter than the sun.  Adjusting the gravity necessitates new temperatures and metallicities to fit model to spectrum, which is the likely reason for an increase in the differences with temperatures from other forward modeling techniques such as that used in \citetalias{2005ApJS..159..141V}.

%
\section{Abundance Determinations} \label{sec:abundance}
As the Sun is the only star where we can compare our spectroscopic abundance determinations to actual rocks (and even there, get things only mostly correct), there is frequently large disagreement between different authors, especially for elements with few or blended lines.  In order to ensure that we have the most precise relative abundances, we have ensured that our global parameters are as accurate as possible.  Here we discuss systematics which may still be affecting our abundances and derive uncertainty estimates as a function of surface gravity, effective temperature, and overall metallicity.  We also note any abundances which show evidence of trends or deviations from literature comparisons and discuss their likely origins.

\subsection{Temperature Dependent Metallicity} \label{sec:abundance_teff_metals}
\begin{figure}[ht] 
   \centering
   \includegraphics[width=0.95\columnwidth]{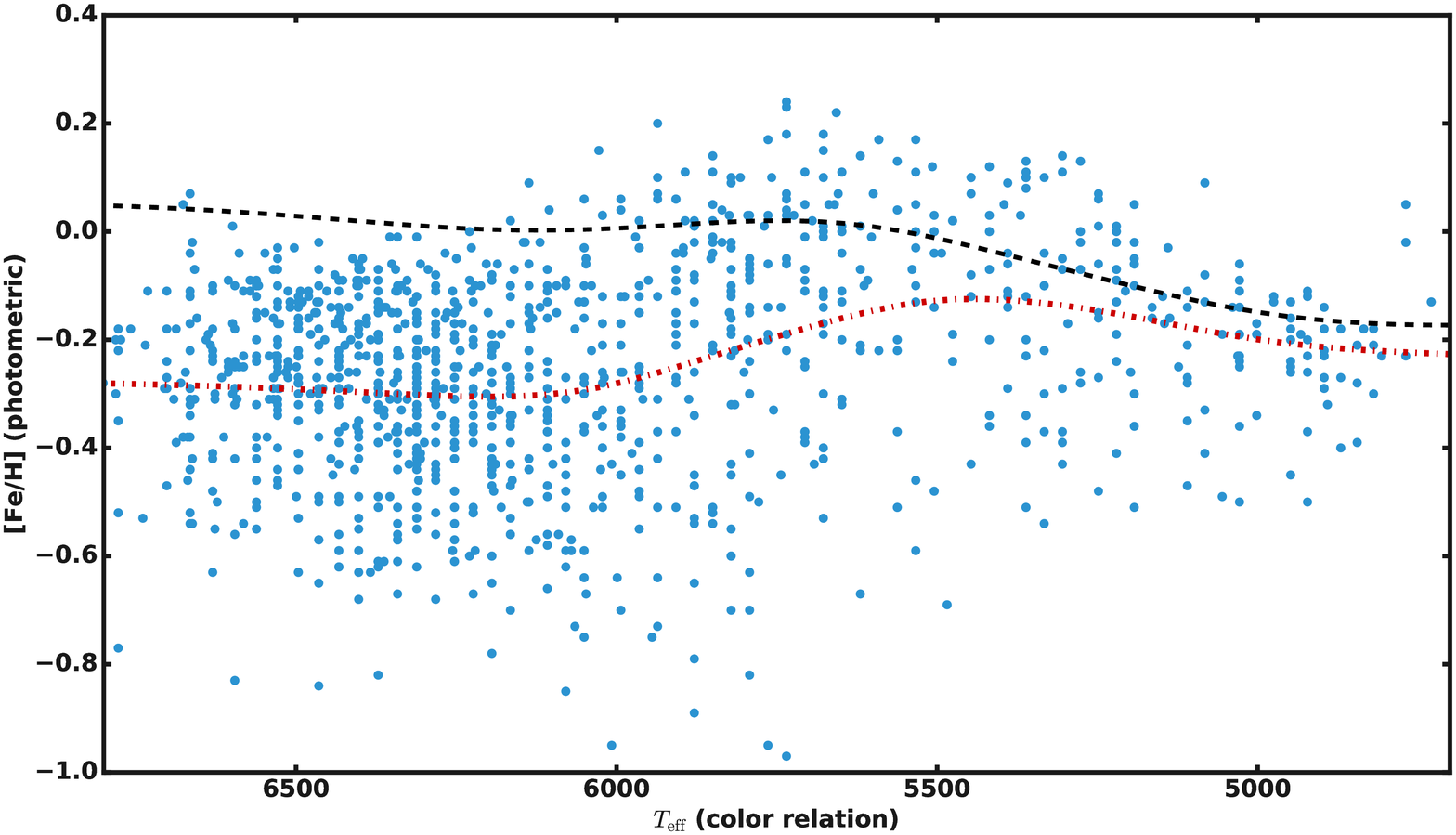} 
   \caption{The distribution of photometric metallicities of a selection of nearby stars from \citet{2002A&A...394..927I}.  The temperatures were calculated from the B-V color relation of \citet{2013ApJ...771...40B}. In black (dashed line) we have overplotted the fit to our [Fe/H] data and in red (dot-dashed line) a fit to this data using the same gaussian kernel regression method. These stars show the same lack of high metallicity stars at low \teff\ as our sample.  This trend is most likely a systematic induced by errors in the models and extends to photometric metallicities which are calibrated against those from spectra.}
   \label{fig:sample_metallicity_bias}
\end{figure}

Our sample shows a clear temperature dependent trend in overall metallicity which is also evident in comparison to other spectroscopic surveys. The trend can also be seen in literature values for photometrically determined metallicities \citep{2002A&A...394..927I} (Figure \ref{fig:sample_metallicity_bias}). The agreement across techniques suggests that the trend may be either astrophysical or a selection bias.  However, a set of binary pairs in our sample shows a very similar trend with the cooler member of each pair always being more metal poor, which seems to indicate a systematic problem that is common across different analysis techniques.

\subsubsection{Stellar Evolution} \label{sec:stellar_evolution}
Stars of different mass which begin with the same initial metallicity will evolve to have different abundances in the atmospheres.  The combined effects of temperature, surface gravity, and convective zone depth will effect the amount of gravitational settling, or diffusion, of the heavier elements out of the photosphere but the effect is generally small.  We used YREC \citep{2008Ap&SS.316...31D} to create representative models of two stars similar to one of the binary pairs in our sample, separated by $\sim$450~K, with a $\Delta$[M/H] of 0.11~dex.

To model the two stars, we first found their ages and masses by interpolating into the Yonsei-Yale isochrone grids using our final parameters for \teff, \logg, and [M/H]. The warmer star has \teff~=5440~K, \logg~=4.40 and [M/H]~=0.23.  The cooler star has \teff~=4900~K, \logg~=4.45, and [M/H]~=0.12.  These values yielded masses of 0.98 M$_{\odot}$ and 0.82 M$_{\odot}$ and ages ranging from 4.45~Gyr to 8.5~Gyr.  We chose 3 different initial metallicities (0.14, 0.23, and 0.35) and created individual zero-aged main sequence models with both helium (LDIFY) and gravitational settling (LDIFZ) enabled with tolerances set at $2.0e^{-7}$ as recommended in the YREC documentation\footnote{\url{http://www.ap.stmarys.ca/~guenther/evolution/yrec7d.pdf}}.  We then evolved all six models to 6~Gyrs with the same diffusion and settling settings.

The cooler models had systematically lower metallicity at the end of 6~Gyrs than the warmer models by 0.023~$\pm$~0.003~dex.  This is far shy of the 0.11~dex difference we see in our binary pair but does indicate that stellar evolution will effect overall trends in metallicity.  Because the gravitational settling time is a function of the mass of the particles, we can expect this to have a larger effect on heavier elements than lighter elements.  In addition, older stars should show stronger differences than younger stars.  This effect should be kept in mind when comparing abundances of stars with different masses or ages.

\subsubsection{Observational Bias}
As stars of a given mass get more metal rich, their effective temperatures decrease and their overall brightness decreases.  However, the effect is very small.  Using the 6~Gyr models we created in \S\ref{sec:stellar_evolution}, we find that a change of 0.12 dex in [M/H] would decrease the brightness of the star by only 0.2~mag.  This might effect the reddest edge of our sample to some extent, pushing more metal rich stars below the magnitude cutoff required to get high S/N for fainter stars.  However, the more metal rich models are also $\sim$200~K cooler which would tend to erase this effect as more massive stars would simply take the place of less massive ones which were lost.

\subsubsection{Continuum Fitting} \label{sec:continuum}
One possible cause for the metallicity trend could be poor continuum fitting for cooler stars.  We chose spectra of Vesta and several of the cooler components of our binary pairs and divided the observation by the model at our chosen continuum points.  We fit a third order polynomial to these points for each spectral order and applied this correction to our normalized spectra.  We then re-ran these adjusted spectra using our normal procedure.

The global parameters for the sample were largely unchanged, with differences in \logg\ of less than 0.02 dex.  The changes in [Fe/H] were also small, but tended to decrease with decreasing temperature with the coolest star changing by -0.025 dex.  This is too small and in the opposite sense needed to account for the [Fe/H] vs. \teff\ trend and shows that our analysis is not very sensitive to errors in continuum fitting.

\subsection{Microturbulence, \vmic} \label{sec:microturbulence}
The microturbulence parameter is meant to account for Doppler line broadening by velocity fields on length scales smaller than the mean free path of photons in the photosphere (i.e. granulation). The upflows in granules combined with the downflows in intergranular lanes results in line broadening \citep{2000A&A...359..729A}.  It is common practice in equivalent width analysis to adjust the microturbulence parameter to remove correlations between iron abundance and equivalent width. Empirical relations have been derived to calculate \vmic\ based on temperature and surface gravity \citep{2013ApJ...764...78R}.

Our analysis procedure simultaneously fits thousands of lines of varying strengths with a single abundance pattern, while yielding accurate surface gravities. We found that varying microturbulence with position in the HR diagram or making it a free parameter resulted in strong trends in \logg\ with respect to temperature, motivating us to instead use a fixed value \citep{2015ApJ...805..126B}.  Here we revisited this analysis to determine if that decision could be the cause of trends in our abundances.

\begin{figure}[ht] 
   \centering
   \includegraphics[width=0.95\columnwidth]{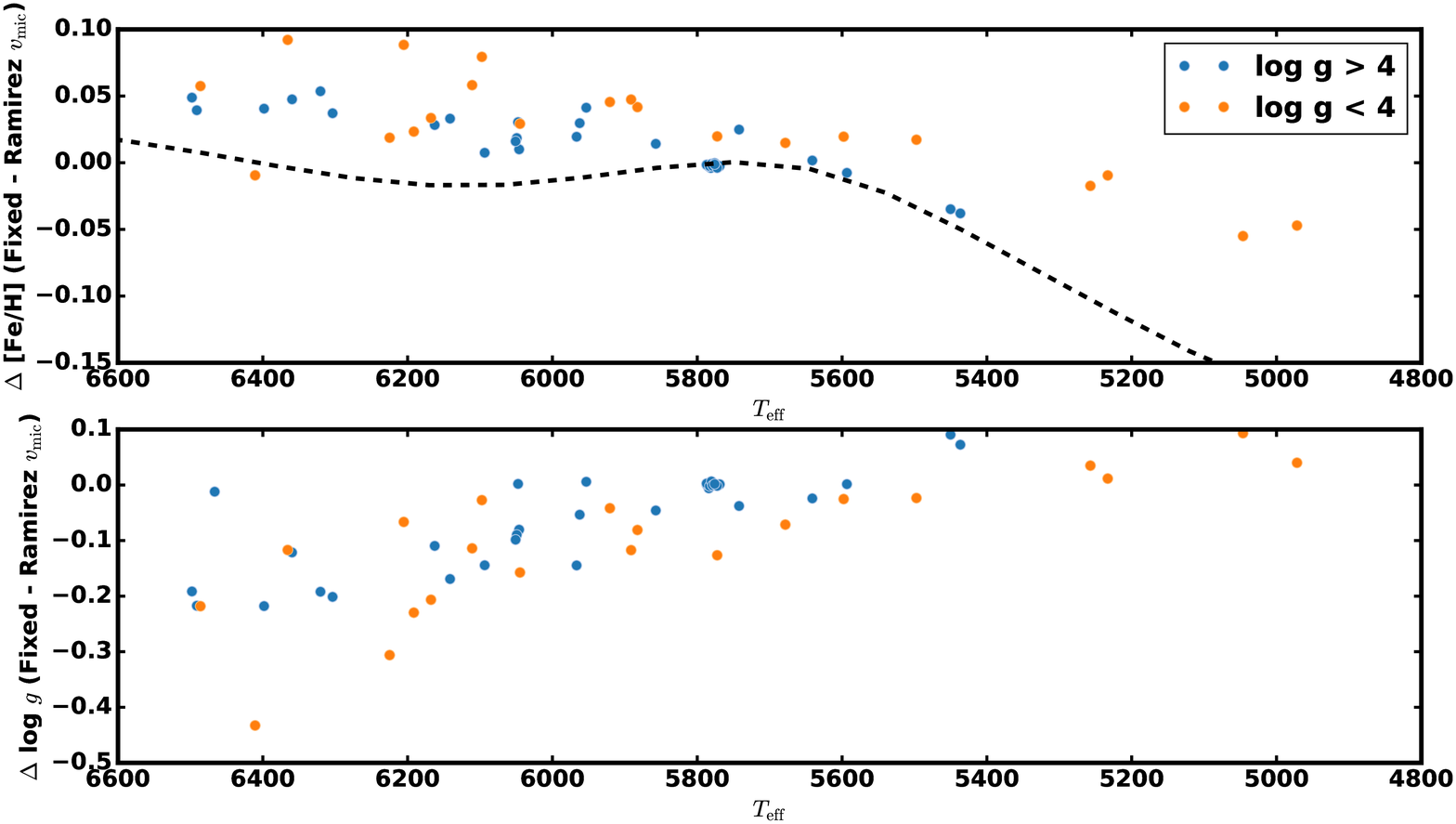} 
   \caption{A comparison of [Fe/H] and \logg\ derived with microturbulence fixed to 0.85~km~s$^{-1}$ or set using the empirical relation of \citet{2013ApJ...764...78R}.  In orange are stars with \logg~$< 4$, and in blue those with \logg~$> 4$.  The dashed black line in the top panel is the fit to the mean [Fe/H] trend of our sample discussed in \S \ref{sec:elements}.  The differences in [Fe/H] don't correspond closely to our trend.  The differences in \logg\ are of much larger magnitude and allowing the microturbulence to vary returns gravities inconsistent with those of asteroseismology.}
   \label{fig:vmic_compare}
\end{figure}

We modified our analysis procedure to set microturbulence according to the empirical formula of \citet{2013ApJ...764...78R} minus 0.22~km~s$^{-1}$ so that the value for the Sun matched the 0.85~km~s$^{-1}$ that we used to tune our line parameters.  We then analyzed the asteroid spectra along with a sample of 44 stars with asteroseismically determined surface gravities.  The stars have $3.7 <$~\logg~$< 4.5$ covering the bulk of our sample, though excludes more evolved stars.   We then compared the new abundances to those determined using a fixed microturbulence of 0.85~km~s$^{-1}$.  In Figure \ref{fig:vmic_compare} we have plotted the differences for [Fe/H] with the trend we found in our overall sample as well as the differences in gravity.

For most elements, the magnitude of the differences were less than 0.1~dex and differences in the remaining 5 (N, Na, Al, V, Mn) were less than 0.15 dex.  In some cases, the trends were in the same sense as the trends that we saw in the overall sample and in other cases opposite.  As we tuned our lines using a fixed value of microturbulence, our updated line parameters likely absorbed some of the broadening that would otherwise be taken care of by the \vmic\ parameter.  However, from these tests it also appears that a single parameter for all elements is an inadequate proxy for the asymmetric broadening of convection.  More importantly, the surface gravity is inaccurate at higher temperatures when using the microturbulence relation (Figure \ref{fig:vmic_compare}).  Heavily blended lines would offer one possible explanation for the abundance trends seen, but the large number of absorption lines makes it more likely that the change in gravity would introduce the trends.  We also ran the same test with \vmic\ a free parameter in our model and found nearly identical results.

\subsubsection{Atmosphere Grids} \label{sec:atmosphere_grids}
Literature metallicities are derived using a variety of different analysis methods including broadband colors.  Our analysis and those of other large spectroscopic samples tend to use the \citet{Castelli:2004ti} atmospheres in their analysis.  If the T-$\tau$ relationship in the atmosphere grids are incorrect at low effective temperatures \citep{Trampedach:2014jb}, model lines could be generated at higher column densities for a given \teff\ than they should for a given abundance, leading to the systematic under-estimation of abundances that we see.  This might also effect surface gravities, leading to lower \logg\ than expected at these low effective temperatures.  Broadband colors are all calibrated against spectral samples which will lead to them showing the same bias.

The trend in our mean raw [M/H] values for our high S/N sample is strongest between 5500~K and 5000~K where it decreases by almost 0.15 dex.  We performed two tests to evaluate the extent to which the atmosphere grid might be responsible for the trend.  We selected two stars with temperatures close to 5000~K and altered our atmosphere grid interpolation code to return a model atmosphere 0.1 dex lower in metallicity and then ran them through our standard procedure.  The [M/H] of the final models increased by 0.023 and 0.066 dex.  Although hardly definitive, it encouraged us to test the effects of using a different atmosphere grid.

\begin{figure}[ht] 
   \centering
   \includegraphics[width=0.95\columnwidth]{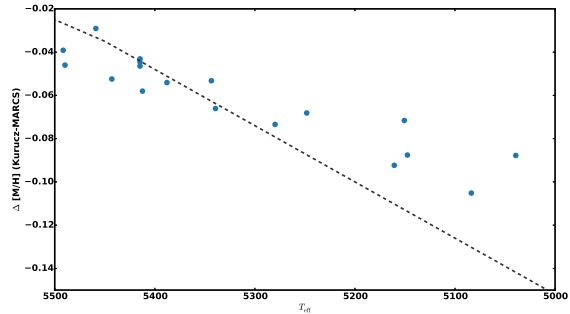} 
   \caption{We substituted MARCS atmosphere models for ATLAS models in our analysis procedure and compared the [M/H] values for a sample of stars between 5500~K and 5000~K.  There is an increasing offset in [M/H] using the different model atmospheres which corresponds to the trend we see in the mean raw [M/H] for our high S/N sample (black dashed line for reference). In this temperature range, the mean raw [M/H] values of our sample as a whole decrease by about 0.15 dex, so the change in atmosphere models alone corrects for more than half the trend we see.}
   \label{fig:monh_marcs_kurucz}
\end{figure}

We selected 20 spectra of 18 stars with [M/H] between -0.2 and +0.2 with temperatures between 5500~K and 5000~K in addition to the 20 asteroid spectra and solved for their parameters using our standard procedure, but using the MARCS atmosphere grid of \citet{Gustafsson:2008df} instead of the ATLAS grid of \citet{Castelli:2004ti}.  Using the $\chi^2$ weighted mean results of the asteroids as our solar values we obtained solar relative metallicities for the spectra and compared them to the solar relative (but uncorrected) metallicities of the spectra from our sample (Figure \ref{fig:monh_marcs_kurucz}).  The [M/H] values derived using the MARCS atmospheres were increasingly higher at lower \teff.  Although the difference was not enough to erase the entire trend, it was consistent with our hypothesis that there is a problem with the ATLAS grid.  It is possible that with proper tuning of our line list using the MARCS atmospheres could bring further improvements.

\subsection{Fitting out Systematic Trends} \label{sec:abundance_trends}
The ideal method for identifying systematic abundance errors would be to examine the abundances of binary stars which have large temperature differences.  Assuming that all of these stars are coeval and have only very small differences due to varying convection zone depths and evolutionary states, they should each have the same abundance pattern.  Unfortunately, we only have 9 pairs of binaries where both have \teff\ $> 4700$~K.  Below this temperature, the growing importance of molecular lines in both continuum and line blends limits the effectiveness of our analysis technique.

Another method that we can use to identify any trends is to compare each element with a reference element as a function of temperature for our entire sample.  For un-evolved dwarf stars we should not expect to find any temperature dependent trends in one element with respect to another.  However, this requires that we have a suitable reference element with no trend with respect to temperature.  Although we have some elements with very small trends, there are none that are trend free.  Other studies have used overall metallicity, as their reference, but a close examination of their data shows that they also had trends in [M/H] which could bias their corrections.

None of the elements we have analyzed should contain any intrinsic abundance trends with respect to \teff.  Assuming that we also have no significant age or location bias with respect to \teff, then the mean abundance of our entire sample should be constant with only the dispersion changing as our ability to recover a given abundance changes.  As mentioned in \S \ref{sec:elements}, we used Nadaraya-Watson regression with a gaussian kernel to fit the mean of the abundances in the subset of our sample having S/N~$\geq 100$ and \vsini~$ \leq 3$ km~s$^{-1}$, which should have the most precisely determined abundances (Figures \ref{fig:raw_light_abunds} \& \ref{fig:raw_heavy_abunds}).  We determined the optimum bandwidth for each element using a cross-validation grid search but found that there was more structure in the fit than we felt was justified by our knowledge of the source of the trends.  Instead, we used the maximum bandwidth found of 150 for all elements.

\begin{figure*} 
   \centering
   \includegraphics[width=\textwidth]{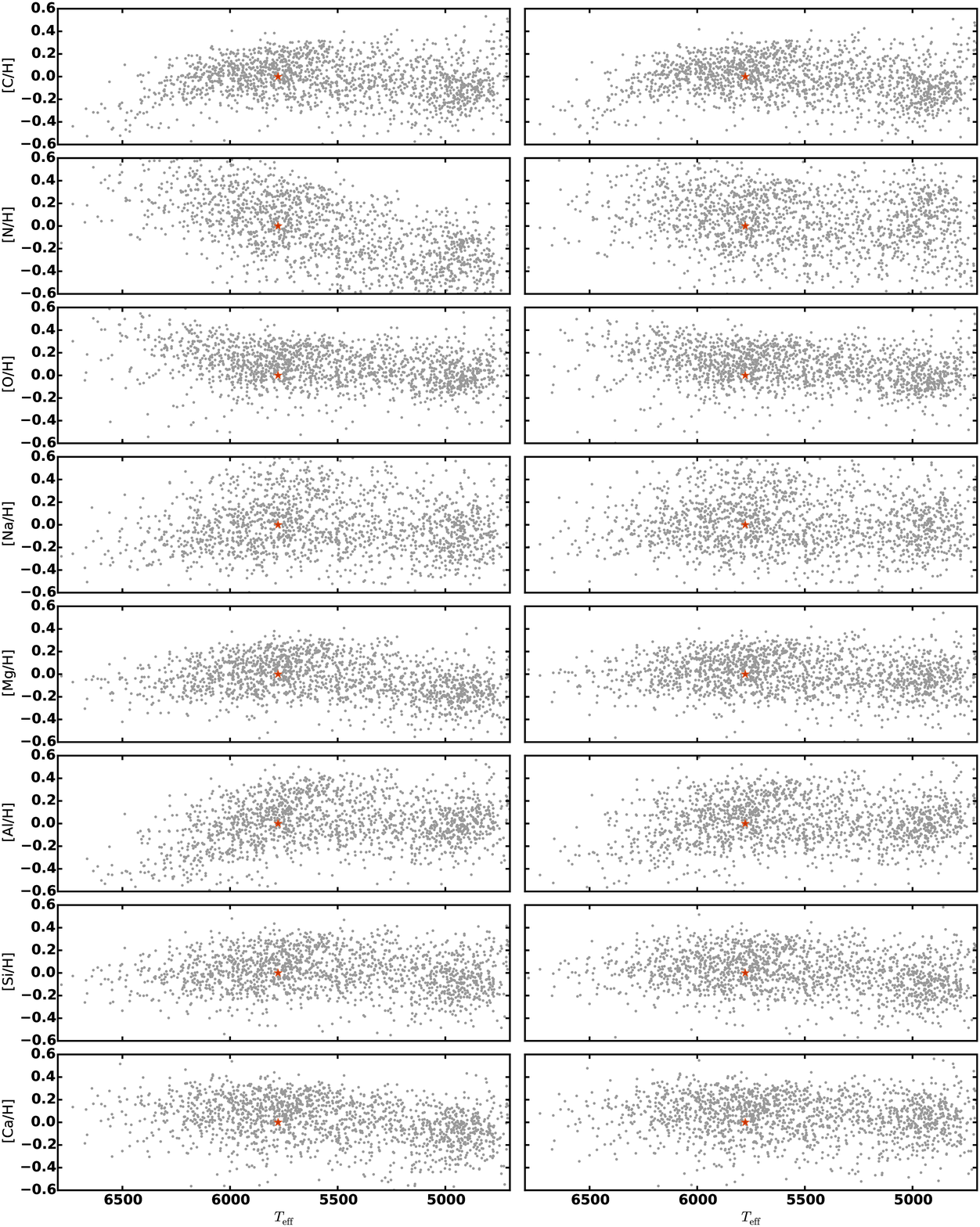} 
   \caption{After setting the offset at solar temperatures to zero, we subtracted our fits from the solar relative abundances of all stars to arrive at our final abundances.  Stars with \logg\ $< 3.0$ were not corrected as there were too few stars in this range in our sample to determine any trends.  The left column shows the full, uncorrected sample and the right shows the sample after trend corrections were applied to dwarfs and subgiants based on the slowly rotating, high S/N members of each group.}

   \label{fig:final_abundances_light}
\end{figure*}

\begin{figure*} 
   \centering
   \includegraphics[width=\textwidth]{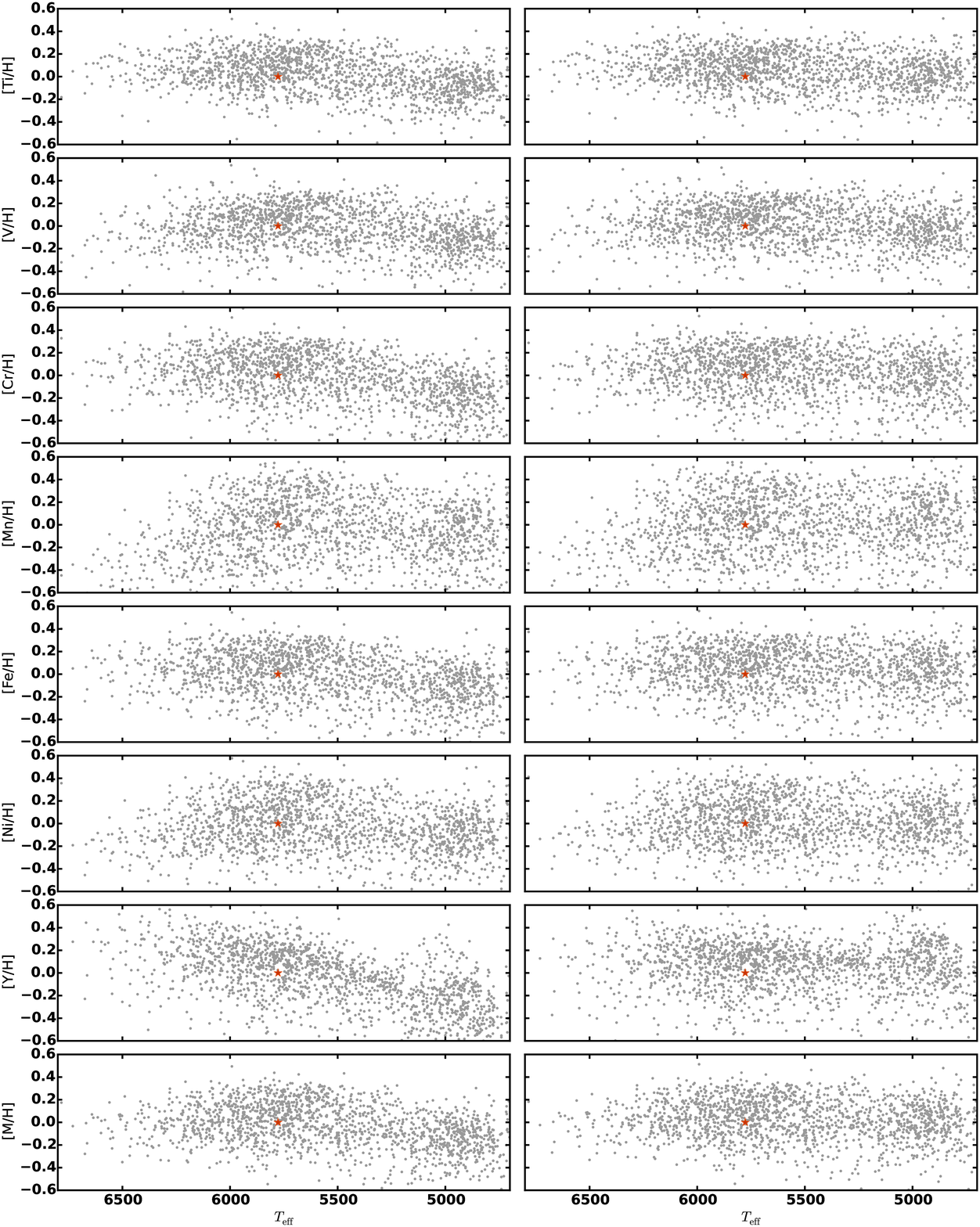} 
   \caption{Same as Figure \ref{fig:final_abundances_light} but for the heavier elements and overall metallicity.}

   \label{fig:final_abundances_heavy}
\end{figure*}

To correct for the trends, we adjust the fit to return no offset for our asteroid spectra (since there is no reason to suspect that the sun is an average star in our sample), and subtract the trend (Figures \ref{fig:final_abundances_light} \& \ref{fig:final_abundances_heavy}).  We are providing both the corrected solar differential abundances and the offsets calculated at 150~K intervals for each element so that the raw values can be recovered (Table \ref{table:dwarf_abund_trend_offsets}). Care should be taken in understanding the systematics involved and where the abundances are most reliable. No correction was applied to the handful of stars with \logg\ $< 3.0$.

\subsection{Comparisons to Literature Values} \label{sec:final_lit_comparisons}
\begin{figure*}
   \centering
   \includegraphics[width=\textwidth]{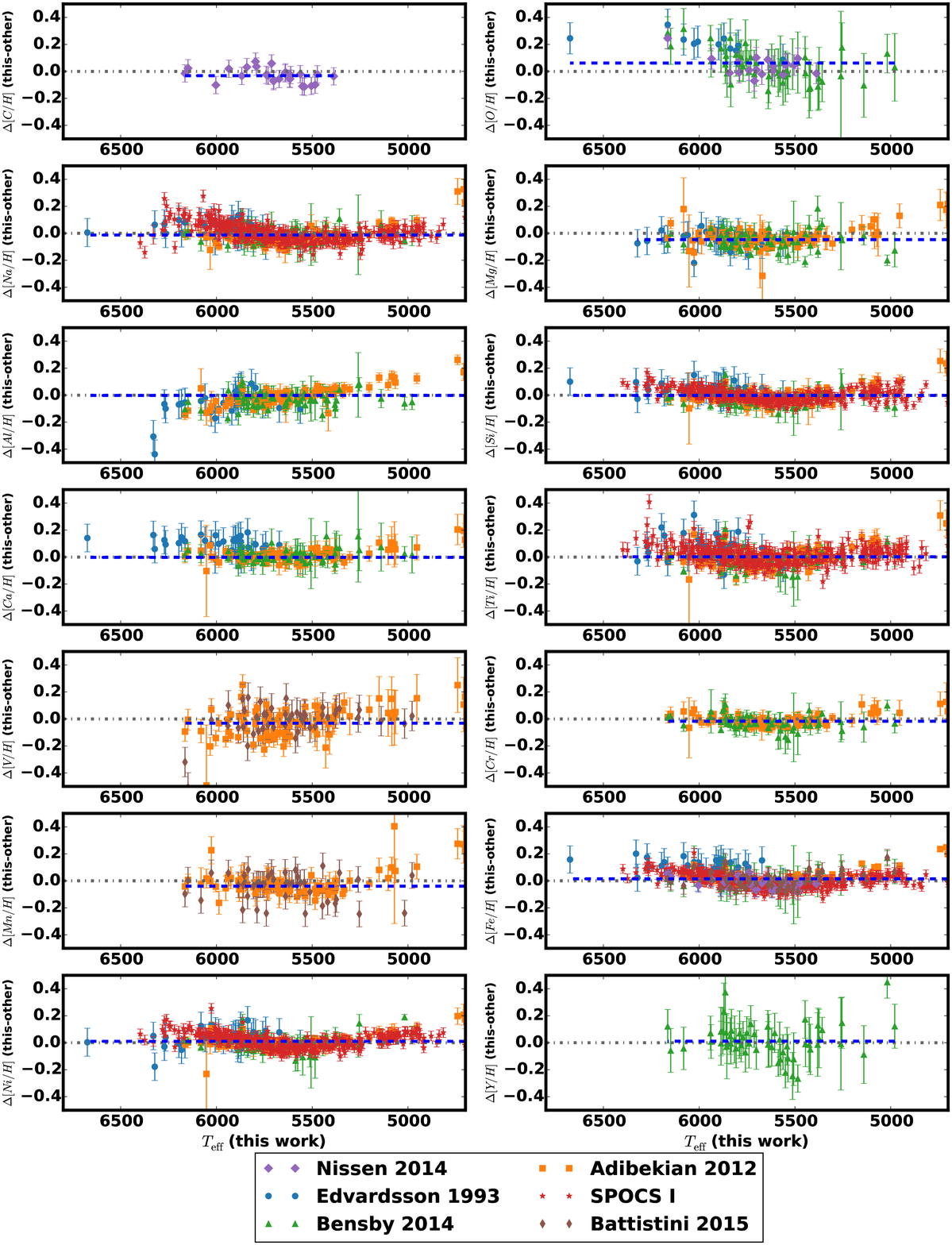} 
   \caption{After correcting for trends in our abundance determinations, we again compare our abundances to literature values and find that there are now trends with some elements at lower temperatures which were not there in our uncorrected abundances.  The trends we subtracted out may be due to incorrect opacities or temperature/density structures in the stellar atmosphere grids, a model feature we share with other studies.}
   \label{fig:trend_corrected_comparisons}
\end{figure*}

For several elements, our comparisons with literature values look worse after we apply our corrections (Figure \ref{fig:trend_corrected_comparisons}) despite the now relatively trend free distribution of our sample.  At the cool temperature end there is often now a slope between 5500~K and 5000~K as seen in comparisons of Mg, Al, Cr, and Ni.  This may be due to shared reliance on the \citet{Castelli:2004ti} atmosphere grids at these temperatures or be a consequence of the differing methods of correcting trends.  At higher temperatures we also see new disagreement with Na and Si.  It is possible that we have overcorrected, subtracting out true astrophysical signal at these temperatures.  However, the differences are quite small, appearing slightly magnified due to the larger changes at the low temperature end.

%
\subsection{Activity} \label{sec:activity}
Our spectra included the calcium H \& K lines which are typically used to evaluate stellar activity.  We calculated both the $\log R^{'}_{HK}$ and $S_{HK}$ values for all stars where we also had broadband colors following the procedure outlined in \citet{Isaacson:2010gk}.  The $\log R^{'}_{HK}$ values were only calibrated between $0.4 < B-V < 1.0$, and we have indicated those which fall outside of that range in the table. Activity may have subtle effects on the photospheric lines and hence our derived parameters \citep{Valenti:1994uv}.

%
\subsection{Abundance Ratios} \label{sec:abundance_ratios}
\begin{figure*} 
   \centering
   \includegraphics[width=\textwidth]{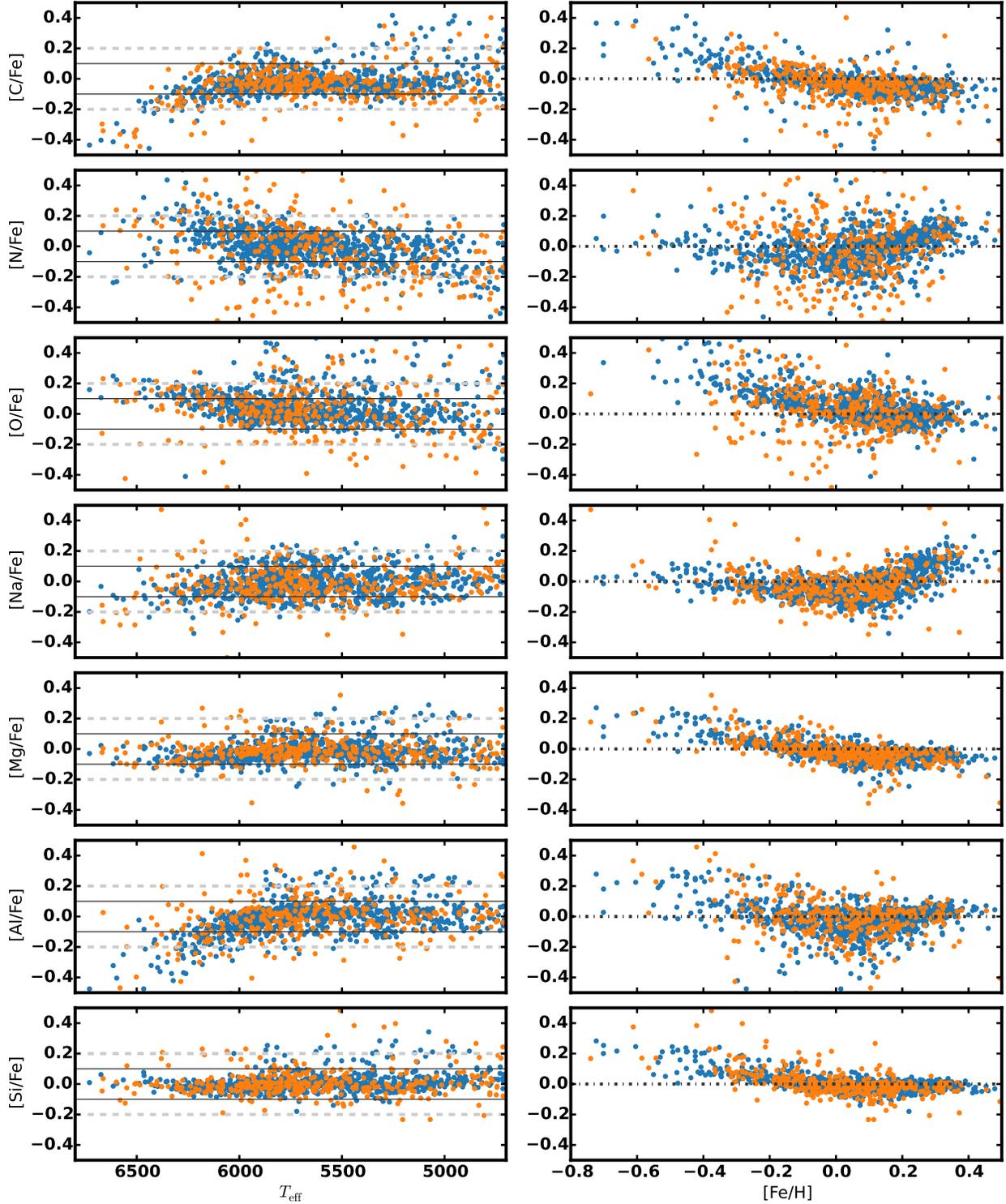} 
   \caption{Abundance ratios for stars with \logg~$ > 4.0$ after applying our abundance corrections.  Blue points are those stars with S/N~$ \geq 100$ and orange have S/N~$ < 100$.  In the left column of panels, we plot the ratio of [$\epsilon$/Fe] vs. \teff\ to show any residual trends we may have after applying our abundance corrections.  Lines mark the $\pm 0.1$ and $\pm 0.2$~dex regions for reference.  In the right column, we then show the behavior of that ratio with respect to changing [Fe/H] with a line at zero for reference.}

   \label{fig:abund_ratios_light}
\end{figure*}

\begin{figure*} 
   \centering
   \includegraphics[width=\textwidth]{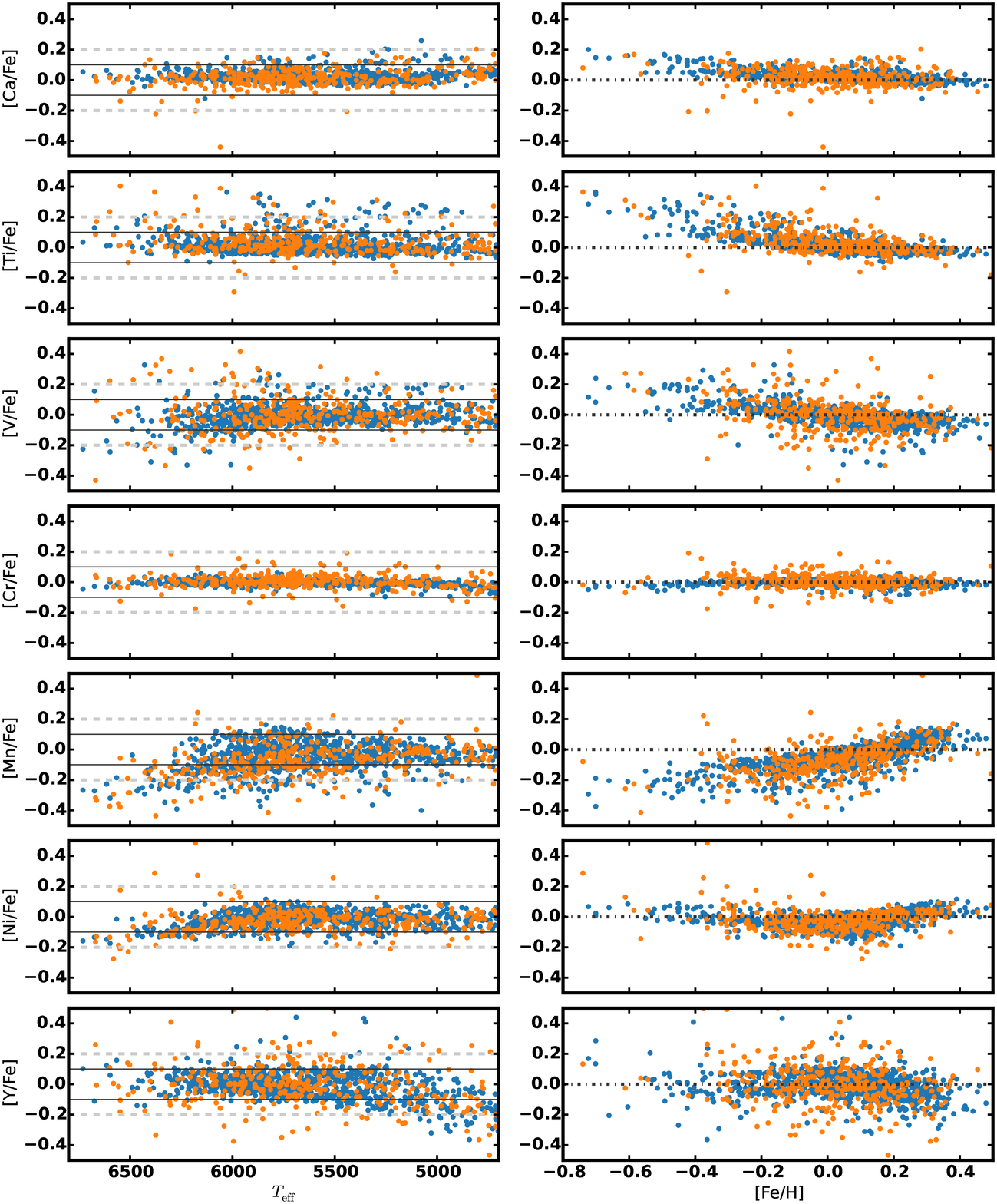} 
   \caption{Same as Figure \ref{fig:abund_ratios_light} but for the heavier elements.  The [Cr/Fe] ratios demonstrate both the high precision we achieve in our abundances and the effects of S/N ratio on their determination.  Points in orange (S/N~$< 100$) have more than twice the RMS scatter as those in blue (S/N~$\geq 100$).  We have excluded iron from the comparisons since this would result in an uninformative flat line.}

   \label{fig:abund_ratios_heavy}
\end{figure*}

In looking for astrophysical trends in abundances, it is useful to remove the effects of overall metallicity to reduce scatter.  This can help in seeing if we retain any residual trends in abundance with respect to \teff\ as well.  Most studies choose to use [Fe/H] as a proxy for [M/H] and so we do the same here to facilitate comparisons with other work.  In Figures \ref{fig:abund_ratios_light} and \ref{fig:abund_ratios_heavy}, we plot the final corrected abundances of stars with \logg~$ > 4.0$ and separate the sample into those with S/N~$ \geq 100$ and those with S/N~$ < 100$.

For most elements, the behavior of the abundance ratios between $\sim$ 5000~K and $\sim$ 6200~K is relatively flat.  There is also a very tight correlation with iron, with most elements having an RMS scatter in [$\epsilon$/Fe]~$< 0.25$~dex. If we look only at the high S/N sample all elements have RMS scatters $< 0.12$ dex. Chromium stands out for its exceedingly tight correlation and also illustrates the limits of our procedure due to low S/N.  The scatter of low S/N points (0.05~dex) is more than double those of stars with S/N~$ \geq 100$ (0.02~dex).  The precision of nitrogen, sodium, and vanadium are particularly hard hit by the decrease in S/N with the scatter increasing by more than an order of magnitude.

Turning to the trends in [$\epsilon$/Fe] with respect to [Fe/H] we see the expected increase in the relative abundance of $\alpha$ elements with respect to iron at sub-solar metallicities \citep{2015MNRAS.454L..11A,1993A&A...275..101E}.  Unfortunately, our sample of predominantly nearby stars means that we have very few stars at [Fe/H]~$< -0.4$.  Below [Fe/H] of -0.2 we also begin to see evidence of a separate population of more $\alpha$-rich stars, particularly in [Mg/Fe], [Si/Fe] and [Ti/Fe].

%
\section{Discussion}  \label{sec:discussion}
The homogeneous nature of this catalog and its large size provide a wealth of information not just for our use in the study of exoplanet formation, but will also enable a more detailed exploration of many other fields of astrophysics.  It may, however, be most useful in illustrating the current limitations in spectral modeling and the derivation of accurate elemental abundances.  

Great strides have been made in developing 3D model atmospheres and calculating NLTE departure coefficients.  However, the computational power required for both means that we will continue to rely on 1D LTE analysis for most tasks.  It is important then to understand what the current limitations of these models are and what, if anything, can be done to improve them.

\subsection{Precision vs. Accuracy} \label{sec:precision_accuracy}
We have shown that our procedure, with its extended line list and iterative fitting, are able to obtain very \textit{precise} results (\S \ref{sec:error_analysis}.  The precision is evident regardless of starting position and holds for multiple observations of the same star over multiple epochs.  This leads to uncertainties which are smaller than our confidence in deriving the \textit{correct} stellar parameters.

In this work we have obtained accurate gravities (\S \ref{sec:surface_gravity}) and temperatures (\S \ref{sec:effective_temperature}), and derived abundances for 15 elements with very high \textit{precision} (\S \ref{sec:random_errors}).  We labored to obtain accurate and self consistent \logg\ and \teff with the expectation that this would lead to accurate abundances.  However, we still see trends in those abundances with respect to temperature that are much larger than our formal uncertainties.  This is not then a problem with how well we are fitting our models to our observations, rather, it is a limitation in our models.

One solution to the problems of accuracy in spectroscopic analyses has been to limit the analysis to a narrow range of stars whose properties are very like the Sun.  The benefit of a homogeneously derived spectroscopic catalog such as this one is that relative comparisons can be made across a broader range of spectral types.  It should be understood though that the very high precision obtained in these efforts does not necessarily translate to accurate abundances.

\subsection{Robustness to NLTE Effects}
Our analysis procedure involves modeling 350~\AA\ of stellar spectrum containing 7,500 atomic and molecular lines.  For each element, this means that we will be simultaneously fitting lines with a variety of excitation potentials, ionization states, spectral region and physical location in the atmosphere.  We showed that this generally mitigates NLTE effects which are prone to effect stronger lines and those which form higher in the atmosphere (\S \ref{sec:nlte_effects}).  The one exception to this seemed to be oxygen which showed a trend of increasing abundance at super-solar temperatures as expected in LTE analysis of the oxygen triplet at $\sim 7770$~\AA.  These lines cover more spectral range than the remaining smaller oxygen lines in our line list and increase in strength at higher temperatures, so their effects dominate in determining the abundance.  This exception though serves to highlight the benefit of simultaneously modeling large numbers of lines for a given element when available.

\subsection{Model Atmosphere Grids}
At lower temperatures we expect little to no contribution from NLTE effects, yet between 5500~K and 5000~K we saw a substantial decrease in metallicity of our sample as a whole, which was echoed in many of the individual abundances.  A review of literature values uncovered the same trend in spectroscopic analyses performed by a variety of techniques, leading us to search for commonalities which might explain this unexpected result.  One trait shared with most studies was the ATLAS model atmosphere grid of \citet{Castelli:2004ti}.  We re-analyzed a subset of stars in this temperature range using a MARCS atmosphere grid  \citep{Gustafsson:2008df} and found that almost all of the trend was eliminated (\S \ref{sec:atmosphere_grids}).  The opacities or density structure of the atmosphere models could be incorrect at these lower temperatures and a more detailed analysis should be performed.  This could have implications not only for spectral analysis, but could lead to improvements in stellar models and a better understanding of stellar structure.

\subsection{Conclusions}
We have performed a uniform analysis of 2,644 spectra of 1,617 stars using spectral synthesis modeling.  For each star we determined solar relative effective temperature, surface gravity, projected rotational velocity, macroturbulence, metallicity, and abundances for 15 elements. We find that our gravities and temperatures are both precise and accurate, in general agreeing well with published results.  At higher temperatures, discrepancies arise that may be due to our new analysis procedure, shown to accurately recover asteroseismic gravities.

For the metallicity and the abundances we also analyzed trends in our sample and derived empirical corrections to the abundances which are likely due to systematics in our analysis stemming from errors in the stellar atmosphere grids we used.  These errors will continue to limit the accuracy of spectroscopic abundance analysis until the models can be improved. Care should be taken in understanding these trends when comparing these abundances to those of other works.

%
\acknowledgments
The authors thank Ulrike Heiter for helpful discussions about NLTE effects.  We would also like to acknowledge the work by all of the people who contributed data to the VALD database that were included in our line list, only a fraction of whom could be directly cited in this paper. We thank the anonymous referee for their careful reading that led to a better paper and also their advocacy for the unsung work of theoretical line parameter calculation.  Though our line data came from more than 100 papers, the bulk was from the tireless work of one man, Bob Kurucz.  This work would not have been possible without the thousands of spectra observed by programs led by Gaspar Bakos, Debra Fischer, Lynne Hillenbrand, Andrew Howard, John Johnson, Geoff Marcy, Phil Muirhead, and Jason Wright. Data presented herein were obtained at the W. M. Keck Observatory from telescope time allocated to the National Aeronautics and Space Administration through the agency's scientific partnership with the California Institute of Technology and the University of California. The Observatory was made possible by the generous financial support of the W. M. Keck Foundation. This research has made use of the SIMBAD database, operated at CDS, Strasbourg, France.

The authors wish to recognize and acknowledge the very significant cultural role and reverence that the summit of Mauna Kea has always had within the indigenous Hawaiian community. We are most fortunate to have the opportunity to conduct observations from this mountain.

JMB and DAF thank NASA grant NNX12AC01G and funding from NASA ADAP program NNX15AF02G.

%
\bibliography{ms,lineparams}

\clearpage
\newgeometry{top=1.5in, bottom=0.5in, left=1.25in, right=0.75in}

\begin{sidewaystable}
\begin{center}
{\footnotesize 
  \begin{threeparttable}
  \renewcommand{\TPTminimum}{\\linewidth}
  \begin{longtable}{ l c d{4} d{2} d{3} d{3} c }
  \caption[]{Atomic and Molecular Line Parameters}  \label{table:line_parameters} \\\\
\hline 
\hline \\[-1.5ex]
Element & Ion & \multicolumn{1}{c}{Wavelength} & \multicolumn{1}{c}{Excitation} & \multicolumn{1}{c}{log gf} & \multicolumn{1}{c}{$\mathrm{\Gamma_6}$} & \multicolumn{1}{c}{Masked Out} \\
\\[-1.5ex]
\hline
  Cr &  1 & 5164.0314 &     3.09 &   -3.6630 &   -7.7900 & x \\
  C2 &    & 5164.0722 &     0.20 &   -0.7850 &       &   \\
  C2 &    & 5164.0722 &     0.20 &   -0.7340 &       &   \\
  Fe &  1 & 5164.1125 &     4.19 &   -2.4710 &   -7.5200 &   \\
  Dy &  2 & 5164.1150 &     2.94 &    0.1540 &       & x \\
  Fe &  1 & 5164.1634 &     3.42 &   -4.7450 &   -7.7600 & x \\
  C2 &    & 5164.2536 &     0.21 &   -0.6370 &       &   \\
  C2 &    & 5164.2536 &     0.27 &   -0.3260 &       &   \\
  C2 &    & 5164.2536 &     0.27 &   -0.3480 &       &   \\
  Nb &  1 & 5164.3660 &     0.27 &   -1.0400 &       &   \\
  \hline
\end{longtable}
\begin{tablenotes}[para,flushleft]
\item{\textbf{Note:}~Table \ref{table:line_parameters} is published in its entirety in the electronic edition of this article. A portion is shown here for guidance regarding its form and content.}
\end{tablenotes}
\end{threeparttable}
}
\end{center}

\begin{center}
{\footnotesize 
\begin{threeparttable}
\renewcommand{\TPTminimum}{\linewidth}
\begin{longtable}{ r l c r r r r r r r r r r r r }

\caption{Spectroscopically Determined Stellar Properties} \label{table:final_params} \\
\hline 
\hline \\[-1.5ex]

 & & \multicolumn{1}{c}{$T_{eff}$} & \multicolumn{1}{c}{$log g$} & & & &  \multicolumn{1}{c}{$v_{\mathrm{broad}}$} & \multicolumn{1}{c}{$v sin i$} & \multicolumn{1}{c}{$v_{\mathrm{mac}}$} & \multicolumn{1}{c}{$v_{\mathrm{rad}}$} & \multicolumn{1}{c}{SNR} & & & \\
\multicolumn{1}{c}{ID} & \multicolumn{1}{c}{Name} & \multicolumn{1}{c}{(K)} & \multicolumn{1}{c}{(cm s$^{-2}$)} & \multicolumn{1}{c}{[M/H]} & \multicolumn{1}{c}{$S_{HK}$} & \multicolumn{1}{c}{log $R^{'}_{HK}$} & \multicolumn{1}{c}{(km s$^{-1}$)} & \multicolumn{1}{c}{(km s$^{-1}$)} & \multicolumn{1}{c}{(km s$^{-1}$)} & \multicolumn{1}{c}{(km s$^{-1}$)} & \multicolumn{1}{c}{6000~\AA} & \multicolumn{1}{c}{C-rms} & \multicolumn{1}{c}{L-rms} & \multicolumn{1}{c}{\#} \\
\multicolumn{1}{c}{(1)} & \multicolumn{1}{c}{(2)} & \multicolumn{1}{c}{(3)} & \multicolumn{1}{c}{(4)} & \multicolumn{1}{c}{(5)} & \multicolumn{1}{c}{(6)} & \multicolumn{1}{c}{(7)} & \multicolumn{1}{c}{(8)} & \multicolumn{1}{c}{(9)} & \multicolumn{1}{c}{(10)} & \multicolumn{1}{c}{(11)} & \multicolumn{1}{c}{(12)} & \multicolumn{1}{c}{(13)} & \multicolumn{1}{c}{(14)} & \multicolumn{1}{c}{(15)}\\

\\[-1.5ex]
\hline
\endfirsthead
\hline
\hline \\[-8pt]

 & & \multicolumn{1}{c}{$T_{eff}$} & \multicolumn{1}{c}{$log g$} & & & &  \multicolumn{1}{c}{$v_{\mathrm{broad}}$} & \multicolumn{1}{c}{$v sin i$} & \multicolumn{1}{c}{$v_{\mathrm{mac}}$} & \multicolumn{1}{c}{$v_{\mathrm{rad}}$} & \multicolumn{1}{c}{SNR} & & & \\
\multicolumn{1}{c}{ID} & \multicolumn{1}{c}{Name} & \multicolumn{1}{c}{(K)} & \multicolumn{1}{c}{(cm s$^{-2}$)} & \multicolumn{1}{c}{[M/H]} & \multicolumn{1}{c}{$S_{HK}$} & \multicolumn{1}{c}{log $R^{'}_{HK}$} & \multicolumn{1}{c}{(km s$^{-1}$)} & \multicolumn{1}{c}{(km s$^{-1}$)} & \multicolumn{1}{c}{(km s$^{-1}$)} & \multicolumn{1}{c}{(km s$^{-1}$)} & \multicolumn{1}{c}{6000~\AA} & \multicolumn{1}{c}{C-rms} & \multicolumn{1}{c}{L-rms} & \multicolumn{1}{c}{\#} \\
\multicolumn{1}{c}{(1)} & \multicolumn{1}{c}{(2)} & \multicolumn{1}{c}{(3)} & \multicolumn{1}{c}{(4)} & \multicolumn{1}{c}{(5)} & \multicolumn{1}{c}{(6)} & \multicolumn{1}{c}{(7)} & \multicolumn{1}{c}{(8)} & \multicolumn{1}{c}{(9)} & \multicolumn{1}{c}{(10)} & \multicolumn{1}{c}{(11)} & \multicolumn{1}{c}{(12)} & \multicolumn{1}{c}{(13)} & \multicolumn{1}{c}{(14)} & \multicolumn{1}{c}{(15)}\\

\\[-1.5ex]
\hline
\endhead
    2 &           HD 105 &  6033 &   4.53 &   0.04 &       0.38 &      -4.35 & 
 17.8 &  14.6 &   4.9 &   10.2 &  247 &  0.02 &  0.01 &  2 \\ 
    4 &           HD 166 &  5489 &   4.51 &   0.07 &       0.42 &      -4.40 & 
  5.0 &   4.1 &   2.6 &   -7.2 &  357 &  0.01 &  0.01 &  1 \\ 
    6 &           HD 377 &  5895 &   4.46 &   0.12 &       0.38 &      -4.37 & 
 17.5 &  14.6 &   4.1 &    0.9 &  248 &  0.01 &  0.02 &  1 \\ 
   10 &           HD 691 &  5489 &   4.48 &   0.18 &       0.56 &      -4.25 & 
  6.4 &   5.6 &   2.6 &   -3.2 &  258 &  0.01 &  0.02 &  1 \\ 
   12 &          HD 1388 &  5924 &   4.32 &   0.03 &       0.16 &      -4.97 & 
  4.9 &   2.3 &   4.2 &   29.2 &  250 &  0.01 &  0.01 &  2 \\ 
   13 &          HD 1461 &  5739 &   4.34 &   0.16 &       0.16 &      -5.01 & 
  3.9 &   1.8 &   3.3 &  -11.6 &  317 &  0.01 &  0.01 &  3 \\ 
   22 &          HD 2589 &  5062 &   3.65 &  -0.02 &       0.15 &  $\cdots$  & 
  3.6 &   0.6 &   3.5 &   12.9 &  393 &  0.01 &  0.01 &  1 \\ 
   30 &          HD 3795 &  5379 &   4.11 &  -0.49 &       0.18 &      -4.93 & 
  3.1 &   1.9 &   2.3 &  -43.5 &  250 &  0.01 &  0.01 &  1 \\ 
   33 &          HD 3861 &  6219 &   4.29 &   0.13 &       0.18 &      -4.81 & 
  4.8 &   0.1 &   6.4 &  -15.9 &  293 &  0.01 &  0.01 &  3 \\ 
   34 &          HD 4208 &  5639 &   4.50 &  -0.26 &       0.19 &      -4.83 & 
  3.1 &   0.1 &   3.0 &   56.5 &  246 &  0.01 &  0.01 &  2 \\ 
\hline
\end{longtable}
\begin{tablenotes}[para,flushleft]
\item{\textbf{Note:}~Table \ref{table:final_params} is published in its entirety in the electronic edition of this article. A portion is shown here for guidance regarding its form and content.}
\end{tablenotes}
\end{threeparttable}
}
\end{center}

\end{sidewaystable}
\clearpage

\begin{sidewaystable}
\begin{center}
{\footnotesize 
\begin{threeparttable}
\renewcommand{\TPTminimum}{\linewidth}
\begin{longtable}{ r l r r r r r r r r r r r r r r r }

\caption{Spectroscopically Determined Abundances} \label{table:final_abundances} \\
\hline 
\hline \\[-1.5ex]

\multicolumn{1}{c}{ID} & \multicolumn{1}{c}{Name} & \multicolumn{1}{c}{[C$/$H]} & \multicolumn{1}{c}{[N$/$H]} & \multicolumn{1}{c}{[O$/$H]} & \multicolumn{1}{c}{[Na$/$H]} & \multicolumn{1}{c}{[Mg$/$H]} & \multicolumn{1}{c}{[Al$/$H]} & \multicolumn{1}{c}{[Si$/$H]} & \multicolumn{1}{c}{[Ca$/$H]} & \multicolumn{1}{c}{[Ti$/$H]} & \multicolumn{1}{c}{[V$/$H]} & \multicolumn{1}{c}{[Cr$/$H]} & \multicolumn{1}{c}{[Mn$/$H]} & \multicolumn{1}{c}{[Fe$/$H]} & \multicolumn{1}{c}{[Ni$/$H]} & \multicolumn{1}{c}{[Y$/$H]} \\
\multicolumn{1}{c}{(1)} & \multicolumn{1}{c}{(2)} &\multicolumn{1}{c}{(16)} & \multicolumn{1}{c}{(17)} & \multicolumn{1}{c}{(18)} & \multicolumn{1}{c}{(19)} & \multicolumn{1}{c}{(20)} & \multicolumn{1}{c}{(21)} & \multicolumn{1}{c}{(22)} & \multicolumn{1}{c}{(23)} & \multicolumn{1}{c}{(24)} & \multicolumn{1}{c}{(25)} & \multicolumn{1}{c}{(26)} & \multicolumn{1}{c}{(27)} & \multicolumn{1}{c}{(28)} & \multicolumn{1}{c}{(29)} & \multicolumn{1}{c}{(30)} \\ 

\\[-1.5ex]
\hline
\endfirsthead
\hline
\hline \\[-8pt]

\multicolumn{1}{c}{ID} & \multicolumn{1}{c}{Name} & \multicolumn{1}{c}{[C$/$H]} & \multicolumn{1}{c}{[N$/$H]} & \multicolumn{1}{c}{[O$/$H]} & \multicolumn{1}{c}{[Na$/$H]} & \multicolumn{1}{c}{[Mg$/$H]} & \multicolumn{1}{c}{[Al$/$H]} & \multicolumn{1}{c}{[Si$/$H]} & \multicolumn{1}{c}{[Ca$/$H]} & \multicolumn{1}{c}{[Ti$/$H]} & \multicolumn{1}{c}{[V$/$H]} & \multicolumn{1}{c}{[Cr$/$H]} & \multicolumn{1}{c}{[Mn$/$H]} & \multicolumn{1}{c}{[Fe$/$H]} & \multicolumn{1}{c}{[Ni$/$H]} & \multicolumn{1}{c}{[Y$/$H]} \\
\multicolumn{1}{c}{(1)} & \multicolumn{1}{c}{(2)} &\multicolumn{1}{c}{(16)} & \multicolumn{1}{c}{(17)} & \multicolumn{1}{c}{(18)} & \multicolumn{1}{c}{(19)} & \multicolumn{1}{c}{(20)} & \multicolumn{1}{c}{(21)} & \multicolumn{1}{c}{(22)} & \multicolumn{1}{c}{(23)} & \multicolumn{1}{c}{(24)} & \multicolumn{1}{c}{(25)} & \multicolumn{1}{c}{(26)} & \multicolumn{1}{c}{(27)} & \multicolumn{1}{c}{(28)} & \multicolumn{1}{c}{(29)} & \multicolumn{1}{c}{(30)} \\ 

\\[-1.5ex]
\hline
\endhead
    2 &           HD 105 &   0.07 &   0.10 &   0.13 &  -0.09 &  -0.01 &  -0.19
 &   0.05 &   0.11 &   0.11 &   0.07 &   0.13 &  -0.03 &   0.10 &  -0.02 & 
  0.18 \\ 
    4 &           HD 166 &   0.01 &   0.01 &   0.08 &   0.01 &   0.00 &   0.00
 &   0.05 &   0.15 &   0.06 &   0.08 &   0.13 &   0.09 &   0.12 &   0.04 & 
  0.16 \\ 
    6 &           HD 377 &   0.12 &   0.24 &   0.23 &   0.06 &   0.04 &   0.04
 &   0.13 &   0.25 &   0.17 &   0.16 &   0.21 &   0.14 &   0.20 &   0.08 & 
  0.20 \\ 
   10 &           HD 691 &   0.10 &   0.00 &   0.13 &   0.11 &   0.11 &   0.14
 &   0.17 &   0.27 &   0.20 &   0.19 &   0.23 &   0.22 &   0.24 &   0.15 & 
  0.27 \\ 
   12 &          HD 1388 &   0.01 &  -0.02 &  -0.01 &  -0.00 &   0.03 &   0.02
 &   0.01 &   0.04 &   0.05 &   0.05 &   0.03 &  -0.05 &   0.03 &   0.01 & 
  0.01 \\ 
   13 &          HD 1461 &   0.13 &   0.25 &   0.10 &   0.26 &   0.16 &   0.19
 &   0.16 &   0.15 &   0.15 &   0.14 &   0.16 &   0.24 &   0.16 &   0.20 & 
  0.12 \\ 
   22 &          HD 2589 &  -0.05 &  -0.04 &   0.11 &  -0.12 &  -0.00 &   0.05
 &  -0.11 &  -0.02 &   0.02 &  -0.07 &  -0.08 &  -0.07 &  -0.03 &  -0.04 & 
 -0.04 \\ 
   30 &          HD 3795 &  -0.51 &  -0.33 &  -0.19 &  -0.50 &  -0.45 &  -0.35
 &  -0.40 &  -0.41 &  -0.28 &  -0.37 &  -0.57 &  -0.81 &  -0.54 &  -0.51 & 
 -0.33 \\ 
   33 &          HD 3861 &   0.03 &   0.21 &   0.19 &   0.08 &   0.11 &   0.07
 &   0.13 &   0.18 &   0.15 &   0.15 &   0.15 &   0.06 &   0.16 &   0.11 & 
  0.24 \\ 
   34 &          HD 4208 &  -0.23 &  -0.37 &  -0.14 &  -0.32 &  -0.21 &  -0.21
 &  -0.23 &  -0.25 &  -0.23 &  -0.24 &  -0.30 &  -0.39 &  -0.29 &  -0.30 & 
 -0.33 \\ 
\hline
\end{longtable}
\begin{tablenotes}[para,flushleft]
\item{\textbf{Note:}~Table \ref{table:final_abundances} is published in its entirety in the electronic edition of this article. A portion is shown here for guidance regarding its form and content.}
\end{tablenotes}
\end{threeparttable}
}
\end{center}

\begin{center}
{\footnotesize 
\begin{longtable}{ l c c c c c c c c c c c c c c c }

\caption[]{Abundance trend corrections} \\
\hline 
\hline \\[-1.5ex] 

\multicolumn{1}{c}{Element} & \multicolumn{1}{c}{4700.0} & \multicolumn{1}{c}{4850.0} & \multicolumn{1}{c}{5000.0} & \multicolumn{1}{c}{5150.0} & \multicolumn{1}{c}{5300.0} & \multicolumn{1}{c}{5450.0} & \multicolumn{1}{c}{5600.0} & \multicolumn{1}{c}{5750.0} & \multicolumn{1}{c}{5900.0} & \multicolumn{1}{c}{6050.0} & \multicolumn{1}{c}{6200.0} & \multicolumn{1}{c}{6350.0} & \multicolumn{1}{c}{6500.0} & \multicolumn{1}{c}{6650.0} & \multicolumn{1}{c}{6800.0} \\ 
\\[-1.5ex] 
\hline 
\endfirsthead 
\hline 
\hline \\[-8pt] 
\multicolumn{1}{c}{Element} & \multicolumn{1}{c}{4700.0} & \multicolumn{1}{c}{4850.0} & \multicolumn{1}{c}{5000.0} & \multicolumn{1}{c}{5150.0} & \multicolumn{1}{c}{5300.0} & \multicolumn{1}{c}{5450.0} & \multicolumn{1}{c}{5600.0} & \multicolumn{1}{c}{5750.0} & \multicolumn{1}{c}{5900.0} & \multicolumn{1}{c}{6050.0} & \multicolumn{1}{c}{6200.0} & \multicolumn{1}{c}{6350.0} & \multicolumn{1}{c}{6500.0} & \multicolumn{1}{c}{6650.0} & \multicolumn{1}{c}{6800.0} \\ 
\\[-1.5ex] 
\hline 
\endhead 

C & 0.044 & 0.027 & 0.010 & -0.003 & -0.011 & -0.008 & 0.001 & 0.001 & -0.007 & -0.015 & -0.018 & -0.017 & -0.015 & -0.014 & -0.014 \\
N & -0.457 & -0.425 & -0.366 & -0.283 & -0.198 & -0.119 & -0.048 & -0.005 & 0.017 & 0.041 & 0.082 & 0.131 & 0.173 & 0.202 & 0.219 \\
O & 0.038 & 0.033 & 0.026 & 0.017 & 0.009 & 0.004 & 0.003 & 0.001 & -0.002 & 0.003 & 0.023 & 0.054 & 0.085 & 0.109 & 0.126 \\
Na & -0.066 & -0.064 & -0.048 & -0.024 & -0.002 & 0.014 & 0.020 & 0.005 & -0.027 & -0.059 & -0.079 & -0.085 & -0.084 & -0.080 & -0.077 \\
Mg & -0.139 & -0.131 & -0.110 & -0.076 & -0.041 & -0.013 & 0.004 & 0.002 & -0.014 & -0.034 & -0.047 & -0.053 & -0.053 & -0.053 & -0.053 \\
Al & -0.015 & -0.013 & -0.002 & 0.015 & 0.030 & 0.038 & 0.033 & 0.007 & -0.034 & -0.079 & -0.117 & -0.145 & -0.161 & -0.172 & -0.179 \\
Si & 0.039 & 0.033 & 0.030 & 0.030 & 0.027 & 0.024 & 0.020 & 0.004 & -0.020 & -0.040 & -0.051 & -0.054 & -0.054 & -0.054 & -0.056 \\
Ca & -0.127 & -0.127 & -0.115 & -0.090 & -0.059 & -0.030 & -0.006 & 0.000 & -0.004 & -0.005 & 0.006 & 0.026 & 0.044 & 0.057 & 0.064 \\
Ti & -0.112 & -0.106 & -0.091 & -0.069 & -0.046 & -0.023 & -0.003 & 0.001 & -0.008 & -0.019 & -0.023 & -0.021 & -0.017 & -0.013 & -0.012 \\
V & -0.086 & -0.083 & -0.070 & -0.049 & -0.026 & -0.007 & 0.005 & 0.002 & -0.012 & -0.029 & -0.041 & -0.045 & -0.047 & -0.047 & -0.048 \\
Cr & -0.203 & -0.194 & -0.169 & -0.128 & -0.082 & -0.039 & -0.008 & 0.001 & -0.006 & -0.011 & -0.004 & 0.012 & 0.028 & 0.040 & 0.047 \\
Mn & -0.151 & -0.144 & -0.119 & -0.075 & -0.030 & 0.005 & 0.020 & 0.005 & -0.028 & -0.061 & -0.084 & -0.096 & -0.099 & -0.100 & -0.101 \\
Fe & -0.186 & -0.182 & -0.162 & -0.124 & -0.082 & -0.041 & -0.009 & 0.000 & -0.006 & -0.012 & -0.008 & 0.005 & 0.019 & 0.029 & 0.035 \\
Ni & -0.136 & -0.133 & -0.117 & -0.086 & -0.052 & -0.021 & 0.001 & 0.002 & -0.014 & -0.032 & -0.044 & -0.049 & -0.049 & -0.049 & -0.050 \\
Y & -0.446 & -0.413 & -0.353 & -0.270 & -0.184 & -0.107 & -0.042 & -0.004 & 0.014 & 0.032 & 0.060 & 0.095 & 0.126 & 0.146 & 0.156 \\
M & -0.184 & -0.172 & -0.145 & -0.106 & -0.066 & -0.030 & -0.004 & 0.001 & -0.009 & -0.020 & -0.022 & -0.017 & -0.010 & -0.004 & -0.002 \\
\hline  \label{table:dwarf_abund_trend_offsets}
\end{longtable} 
} 
\end{center}

\end{sidewaystable}
\clearpage

\begin{sidewaystable}
\begin{center} 
{\footnotesize 
\begin{threeparttable} 
\renewcommand{\TPTminimum}{\linewidth} 
\begin{longtable}{ l l c r c d{2} r c c c c c d{1} c }

\caption[]{Derived Stellar Parameters} \label{table:derived_props} \\
\hline 
\hline \\[-1.5ex] 

\multicolumn{1}{c}{} & \multicolumn{1}{c}{} & \multicolumn{1}{c}{$\alpha$} & \multicolumn{1}{c}{$\delta$} & \multicolumn{1}{c}{V} & \multicolumn{1}{c}{d} & \multicolumn{1}{c}{log L} & \multicolumn{1}{c}{R} & \multicolumn{1}{c}{M} & \multicolumn{1}{c}{M$_{\mathrm{iso}}$} & \multicolumn{1}{c}{$\Delta$ M$_{\mathrm{iso}}$} & \multicolumn{1}{c}{log $g_{\mathrm{iso}}$} & \multicolumn{1}{c}{Age} & \multicolumn{1}{c}{$\Delta$ Age} \\
\multicolumn{1}{c}{ID} & \multicolumn{1}{c}{Name} & \multicolumn{1}{c}{J2000} & \multicolumn{1}{c}{J2000} & \multicolumn{1}{c}{mag} & \multicolumn{1}{c}{pc} & \multicolumn{1}{c}{L$_{\odot}$} & \multicolumn{1}{c}{R$_{\odot}$} & \multicolumn{1}{c}{M$_{\odot}$} & \multicolumn{1}{c}{M$_{\odot}$} & \multicolumn{1}{c}{M$_{\odot}$} & \multicolumn{1}{c}{cm s$^{-2}$} & \multicolumn{1}{c}{Gyr} & \multicolumn{1}{c}{Gyr} \\
\multicolumn{1}{c}{(1)} & \multicolumn{1}{c}{(2)} & \multicolumn{1}{c}{(31)} & \multicolumn{1}{c}{(32)} & \multicolumn{1}{c}{(33)} & \multicolumn{1}{c}{(34)} & \multicolumn{1}{c}{(35)} & \multicolumn{1}{c}{(36)} & \multicolumn{1}{c}{(37)} & \multicolumn{1}{c}{(38)} & \multicolumn{1}{c}{(39)} & \multicolumn{1}{c}{(40)} & \multicolumn{1}{c}{(41)} & \multicolumn{1}{c}{(42)} \\
\\[-1.5ex] 
\hline 
\endfirsthead 
\hline 
\hline \\[-8pt] 
\multicolumn{1}{c}{} & \multicolumn{1}{c}{} & \multicolumn{1}{c}{$\alpha$} & \multicolumn{1}{c}{$\delta$} & \multicolumn{1}{c}{V} & \multicolumn{1}{c}{d} & \multicolumn{1}{c}{log L} & \multicolumn{1}{c}{R} & \multicolumn{1}{c}{M} & \multicolumn{1}{c}{M$_{\mathrm{iso}}$} & \multicolumn{1}{c}{$\Delta$ M$_{\mathrm{iso}}$} & \multicolumn{1}{c}{log $g_{\mathrm{iso}}$} & \multicolumn{1}{c}{Age} & \multicolumn{1}{c}{$\Delta$ Age} \\
\multicolumn{1}{c}{ID} & \multicolumn{1}{c}{Name} & \multicolumn{1}{c}{J2000} & \multicolumn{1}{c}{J2000} & \multicolumn{1}{c}{mag} & \multicolumn{1}{c}{pc} & \multicolumn{1}{c}{L$_{\odot}$} & \multicolumn{1}{c}{R$_{\odot}$} & \multicolumn{1}{c}{M$_{\odot}$} & \multicolumn{1}{c}{M$_{\odot}$} & \multicolumn{1}{c}{M$_{\odot}$} & \multicolumn{1}{c}{cm s$^{-2}$} & \multicolumn{1}{c}{Gyr} & \multicolumn{1}{c}{Gyr} \\
\multicolumn{1}{c}{(1)} & \multicolumn{1}{c}{(2)} & \multicolumn{1}{c}{(31)} & \multicolumn{1}{c}{(32)} & \multicolumn{1}{c}{(33)} & \multicolumn{1}{c}{(34)} & \multicolumn{1}{c}{(35)} & \multicolumn{1}{c}{(36)} & \multicolumn{1}{c}{(37)} & \multicolumn{1}{c}{(38)} & \multicolumn{1}{c}{(39)} & \multicolumn{1}{c}{(40)} & \multicolumn{1}{c}{(41)} & \multicolumn{1}{c}{(42)} \\
\\[-1.5ex] 
\hline 
\endhead 

   2 & HD 105 & 00 05 52.5 & -41 45 11 &  7.51 & 39.39 & 0.10 $\pm$ 0.03 & 1.02 $\pm$ 0.04 & 1.29 $\pm$ 0.18 & 1.12 & 1.10-1.15 & 4.44 $\pm$ 0.02 & 1.0 & 0.3-2.0 \\
   4 & HD 166 & 00 06 36.8 & +29 01 17 &  6.07 & 13.67 & -0.21 $\pm$ 0.02 & 0.87 $\pm$ 0.03 & 0.89 $\pm$ 0.12 & 0.96 & 0.94-0.98 & 4.53 $\pm$ 0.02 & 2.2 & 0.9-4.4 \\
   6 & HD 377 & 00 08 25.7 & +06 37 00 &  7.59 & 39.08 & 0.06 $\pm$ 0.03 & 1.03 $\pm$ 0.04 & 1.12 $\pm$ 0.16 & 1.12 & 1.09-1.14 & 4.44 $\pm$ 0.02 & 1.3 & 0.5-2.5 \\
  10 & HD 691 & 00 11 22.4 & +30 26 58 &  7.95 & 34.20 & -0.17 $\pm$ 0.03 & 0.91 $\pm$ 0.03 & 0.91 $\pm$ 0.13 & 0.99 & 0.97-1.01 & 4.50 $\pm$ 0.03 & 2.7 & 1.1-4.9 \\
  12 & HD 1388 & 00 17 58.9 & -13 27 20 &  6.51 & 27.22 & 0.18 $\pm$ 0.02 & 1.17 $\pm$ 0.04 & 1.05 $\pm$ 0.14 & 1.09 & 1.06-1.11 & 4.34 $\pm$ 0.03 & 4.3 & 3.2-5.5 \\
  13 & HD 1461 & 00 18 41.9 & -08 03 11 &  6.47 & 23.25 & 0.07 $\pm$ 0.02 & 1.10 $\pm$ 0.03 & 0.97 $\pm$ 0.13 & 1.07 & 1.05-1.09 & 4.39 $\pm$ 0.03 & 4.1 & 2.9-5.3 \\
  22 & HD 2589 & 00 30 55.1 & +77 01 10 &  6.18 & 38.51 & 0.69 $\pm$ 0.02 & 2.89 $\pm$ 0.09 & 1.37 $\pm$ 0.18 & 1.23 & 1.16-1.28 & 3.60 $\pm$ 0.04 & 5.2 & 4.5-6.1 \\
  30 & HD 3795 & 00 40 32.8 & -23 48 18 &  6.14 & 28.89 & 0.43 $\pm$ 0.02 & 1.88 $\pm$ 0.06 & 1.67 $\pm$ 0.22 & 0.94 & 0.90-0.97 & 3.86 $\pm$ 0.03 & 11.0 & 10.1-12.0 \\
  33 & HD 3861 & 00 41 11.9 & +09 21 18 &  6.52 & 33.44 & 0.34 $\pm$ 0.02 & 1.28 $\pm$ 0.04 & 1.16 $\pm$ 0.16 & 1.25 & 1.22-1.27 & 4.33 $\pm$ 0.03 & 1.6 & 1.0-2.2 \\
  34 & HD 4208 & 00 44 26.7 & -26 30 56 &  7.78 & 32.37 & -0.15 $\pm$ 0.03 & 0.88 $\pm$ 0.03 & 0.90 $\pm$ 0.13 & 0.90 & 0.86-0.94 & 4.50 $\pm$ 0.04 & 5.5 & 2.4-8.6 \\
\hline 
\end{longtable} 
\begin{tablenotes}[para,flushleft]\item{\textbf{Note:}~Table \ref{table:derived_props} is published in its entirety in the electronic edition of this article. A portion is shown here for guidance regarding its form and content.}\end{tablenotes} 
\end{threeparttable} 
} 
\end{center}

\clearpage
\end{sidewaystable}
\clearpage

\typeout{get arXiv to do 4 passes: Label(s) may have changed. Rerun}
\end{document}